\documentclass[ms]{informs3}
\OneAndAHalfSpacedXI % current default line spacing
\usepackage{graphicx}
\usepackage[normalem]{ulem}
\usepackage{epstopdf}
\usepackage{endnotes}
\usepackage{subfigure}
\usepackage{mathrsfs}
\usepackage{amsmath}
\usepackage{microtype}
\usepackage{enumitem}
\usepackage{comment}
\usepackage{float}
\usepackage{cases}
\usepackage{setspace}
\usepackage{url}
\usepackage{soul}
\usepackage{booktabs}
\usepackage{hyperref}
\usepackage{bbm}
\usepackage{bm}
\usepackage[title]{appendix}
\usepackage[dvipsnames]{xcolor}
\usepackage{multirow}
\usepackage{wrapfig}
\hypersetup{
	colorlinks   = true, %Colours links instead of ugly boxes
	urlcolor     = black, %Colour for external hyperlinks
	linkcolor    = black, %Colour of internal links
	citecolor   = black %Colour of citations
}

\graphicspath{ {Figures/} }
\usepackage[semicolon,authoryear,sectionbib]{natbib}
\newtheorem{lemma}{Lemma}

\newtheorem{proposition}{Proposition}

\long\def\comment#1{}

\EquationsNumberedThrough    % Default: (1), (2), ...
\begin{document}
	%%%%%%%%%%%%%%%%
	
	%\setcounter{page}{1}
	%\DOI{10.1287/xxxx.0000.0000}%
	
	% \RUNAUTHOR{Mookerjee, Kumar, and Mookerjee}
	
	\RUNTITLE{}
	
	\TITLE{Does Machine Learning Amplify Pricing Errors in the Housing Market? --- The Economics of Machine Learning Feedback Loops}
	

\ARTICLEAUTHORS{%
\AUTHOR{Nikhil Malik$^*$ and Emaad Manzoor$^\dag$}
{\small $^\ddag$USC Marshall School of Business, maliknik@usc.edu}\\
{\small $^\dag$Cornell SC Johnson College of Business, emaadmanzoor@cornell.edu} \\
\vspace{-0.0in}
}

	\ABSTRACT{%
		\looseness -1 \looseness -1
		% {\centering \textbf{Abstract}}\\
      \begin{center}
      \textbf{Abstract}\\
      \end{center}
    Machine learning algorithms are increasingly employed to price or value homes for sale, properties for rent, rides for hire, and various other goods and services. Machine learning-based prices are typically generated by complex algorithms trained on historical sales data. However, displaying these prices to consumers anchors the realized sales prices, which will in turn become training samples for future iterations of the algorithms. The economic implications of this machine learning ``feedback loop'' --- an indirect human-algorithm interaction  --- remain relatively unexplored. In this work, we develop an analytical model of machine learning feedback loops in the context of the housing market.
    % \emaad{TODO: highlights and challenges addressed}
    % findings
    We show that feedback loops lead machine learning algorithms to become overconfident in their own accuracy (by underestimating its error), and leads home sellers to over-rely on possibly erroneous algorithmic prices. As a consequence at the feedback loop equilibrium, sale prices can become entirely erratic (relative to true consumer preferences in absence of ML price interference). We then identify conditions (choice of ML models, seller characteristics and market characteristics) where the economic payoffs for home sellers at the feedback loop equilibrium is worse off than no machine learning. We also empirically validate primitive building blocks of our analytical model using housing market data from Zillow. % TODO: so what?
    % broad implications
    We conclude by prescribing algorithmic corrective strategies to mitigate the effects of machine learning feedback loops, discuss the incentives for platforms to adopt these strategies, and discuss the role of policymakers in regulating the same.
	}
	
	\maketitle    
	
	\vspace{-0.25 in}
	
	\smallskip

\clearpage
\section{Introduction}
    
Machine learning-based algorithmic pricing (``ML pricing'' henceforth) is increasingly used to facilitate transactions in markets for housing \citep{zillow2022zestimate,realtor2022realestimate}, property rentals \citep{redfin2022rent,airbnb2022smart}, peer-to-peer loans \citep{lendingclub2022modelrank}, and fine art \citep{liveart2022estimate}, among others \citep{pandey2021disparate}.
By providing consumers accurate and on-demand estimates of product values without a labor-intensive appraisal process, ML prices\footnote{ML pricing could dictate (such as for ride-shares on Uber, for example), suggest (such as for rents on Airbnb, for example), or simply display a price to consumers. On Zillow, for example, the machine learning-based ``Zestimate'' is simply displayed as an \textit{estimate} of the current \textit{value} of a home. We will formally define how value of a home relates to transaction or sale price if homeowner was in market to make a sale.} efficiently reduce pricing uncertainty and friction for all buyers and sellers \citep{forbes2021price} and democratize access to information for those who lack pricing experience \citep{huang2021seller, kehoe2018dynamic}. For example, a rideshare driver and rider avoid the friction of negotiating the price for every trip because the price is set by an algorithm. An investor with optimism about the art market but no artistic expertise can purchase art pieces for an ML price, benefiting both the investor and the artist (Bailey 2020).\\

\begin{wrapfigure}[11]{r}{0.3\linewidth}
    \vspace{-0.25in}
    \begin{center}
      \includegraphics[width=0.3\textwidth]{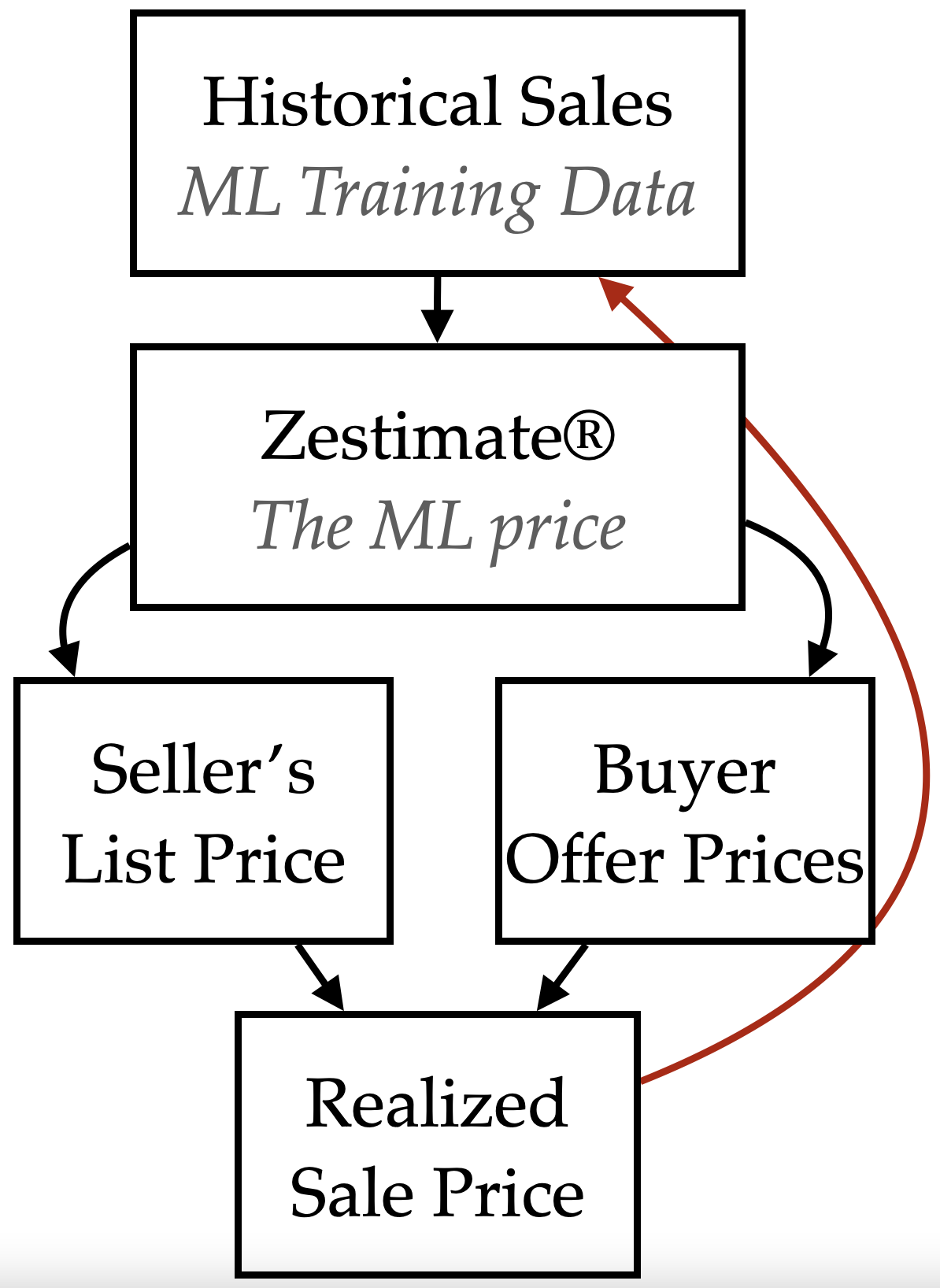}
    \end{center}
    % \vspace{-0.3in}
    % \caption{The machine learning feedback loop in the context of the housing market and Zillow.com.}
    \label{fig:intro}
\end{wrapfigure}

ML prices are typically generated by algorithms that capture high-dimensional product characteristics \citep{bertini2021pitfalls} and dynamically adapt to evolving market conditions \citep{brown2021competition}. These algorithms are trained (and periodically re-trained) to maximize the accuracy of estimated prices by uncovering patterns in historical realized prices in the market.

ML prices are widely believed to anchor realized sales prices \citep{era2019zestimate,oxford2021avm,agentfire2022zestimate}, which are in turn used to train future iterations of the ML pricing algorithms. This creates a feedback loop between the ML pricing algorithm and its own training data; the figure above illustrates a feedback loop in the context of the housing market and Zillow's ML price. Such feedback loops have been reported to limit the ability of machine learning algorithms to learn from their errors, among other undesirable outcomes \citep{chaney2018algorithmic,jiang2019degenerate}. However, the economic effects of ML pricing on consumers in the presence of such feedback loops remain unexplored.\\
% Given all these obvious advantages, ML pricing is being increasingly adopted in many markets (Pandey and Caliskan 2021). This paper questions if ML prices that are designed to maximize statistical accuracy automatically aid market objectives like lower pricing friction, greater and equitable surplus for heterogeneous participants. At the outset, ML prices merely crunch and summarize market information, which may otherwise be costly for participants to collect manually. But as an increasing proportion of buyers and sellers rely on ML pricing, it gains the power to shift market outcomes. In effect, ML pricing would be learning from and shifting the market outcomes simultaneously in a loop. There is a pressing need to understand ML prices embedded in such feedback loops and call attention to potential risks. This paper highlights detrimental implications to pricing and payoffs, grounded in the context of Zillow's ML pricing for residential homes. These implications will guide recommendations for intermediaries that are looking to offer ML pricing and policymakers who monitor and regulate those ML models.\\

% what we do, why is it hard
In this work, we analytically characterize how markets are affected by machine learning-based pricing algorithms when the algorithms influence and learn from consumer behaviors in a feedback loop. The key novelty is how we model the interplay between the algorithm's self-reported \textit{confidence} (displayed as a confidence interval, for example) and the consumers \textit{reliance} (sensitivity of consumers' beliefs or actions to the ML price) in the algorithm. We show that these reinforce each other other due to the feedback loop resulting in over-confidence and over-reliance. Extensive prior research has modeled the dynamics of consumer behavior (under a static algorithm) or of algorithms (under static consumer behavior) separately, we model and identify a joint equilibrium.\\

Our analytical model is grounded in institutional details of one specific context: the housing market and the ML prices (called Zestimates) on Zillow, which is currently the dominant platform for listing and discovering homes for sale. It consists of two components. In the first component, we model how sellers'\footnote{We explicitly model sellers' belief construction, choice and payoffs. We only model buyers as a crowd. We qualitatively argue that formal results for sellers' extend to buyers as well.} beliefs drive realized sales prices and their economic payoffs. In the second component, we model how the algorithm learns from historical sales. These components are fused with sellers' beliefs dependent on ML prices and algorithm trained on realized sale prices. It is important to note that any significant mistakes in the ML price of a home is propagated to sale prices but gets slowly attenuated over the feedback loop because seller also rely on external signals. Further, any temporary price mistakes are uncorrelated across homes, this there are no price bubbles. The innovation in this paper is to reveal the reinforcing feedback between aggregate quantities (instead of individual prices) - how seller \textit{reliance} on ML prices increases with algorithms' reported \textit{confidence} and the confidence calculation improves with reliance. At equilibrium --- attained after the algorithm and the market evolve simultaneously until convergence --- we characterize sellers' private valuations,  and economic payoffs, sales prices, the algorithm's accuracy and self-reported confidence, and sellers' reliance on the algorithm.\\

Our model reveals three key findings. First, presence of ML prices can increase the deviation (in both positive and negative directions) of realized sales prices from the ``true'' home value (grounded in true underlying consumer preferences for home features without interference from ML pricing). Second, we show conditions where this adversely effect the economic payoffs for sellers. If left unchecked, this deviation amplifies until realized sales prices and ML prices are entirely random (and uncorrelated with the ``true'' home value) at equilibrium. At this equilibrium : (i) the buyers and sellers fully rely on the ML price, (ii) the ML price are identical to the eventual realized sales price (akin to a self fulfilling prophecy), (iii) the algorithm's self-reported confidence is maximal, and (iv) sellers' may be worse off than in the absence of ML algorithm. These findings are counter to conventional wisdom that ML prices (by crunching large amounts of revealed preference data) are useful in inferring underlying preferences. Third, we identify seller characteristics (impatience or cost, risk aversion, ability to price home in absence of ML prices) where this equilibrium and its adverse economic implications are worse. We also discuss role of the exogenous factors such as level of ML adoption, the ML algorithm capacity\footnote{A high capacity algorithm has access to more training sample and trainable parameters to better fit data patterns. For example, a deep neural network with thousands of parameters and millions of training samples has a higher capacity than a degree-2 polynomial.} and other market characteristics.\\
%  \footnote{This paper does not model process of increasing adoption, instead just takes level of adoption as exogenous given. We examine outcomes if this exogenous level of adoption increases in future.}\\

In our model, buyers and sellers' do not correct \textit{over-reliance} and the platform does not correct algorithms' \textit{over-confidence}. Buyers and sellers do not correct because they trust the algorithms' confidence calculation\footnote{Platforms' like Zillow broadly publicize \cite{zillow2020zestimateaccuracy} that their algorithms' confidence calculation is tied to typical data science practice of measuring out of sample errors. Our analytical formulation is tied to this definition. In the absence of the feedback loop this would in fact be the correct way to measure confidence}. Our findings are moderated but not eliminated if buyers and sellers are fully rational about the presence of the feedback loop phenomenon and (correctly) calibrate their reliance of ML prices. The overconfidence of ML algorithm and increased deviation of realized sales prices from the ``true'' home value are not eliminated. We do not model platform as a strategic agent to correct the  \textit{over-confidence} or benefit from it. We identify various strategies that platforms could employ to correct the over-confidence. But all strategies effectively limit the visibility of ML prices to buyers and sellers. Our model enables analyzing the trade-off between direct positive impact of making ML prices visible to a single home and indirect negative impact of pervasive influence of ML prices (via adverse feedback loop). We qualitatively discuss why these corrective strategies may not be in line with platforms typical revenue streams from ad sales and iBuying.\\
% First, platforms could use low-capacity machine learning algorithms, which adapt less to feedback. Second, platforms could adjust the self-reported (over)-confidence of the pricing algorithm after accounting for the feedback loop (by temporarily hiding the ML price to construct an ``unconfounded'' evaluation dataset, for example). Third, platforms could eliminate the feedback loop altogether (by not re-training the algorithm with new sales, for example). 

Our model is built on two primitive assumptions: (i) that buyers and sellers indeed rely on the ML price, and (ii) that the machine learning algorithm is periodically re-trained with data from recent sales\footnote{The feedback loop would be absent if the pricing algorithm were driven instead by rules coded by domain experts. The feedback loop would be too slow to have practical implications if the ML price were reliant on older sales and thus relatively static.}. We provide empirical evidence to support these assumptions using real-world housing market transactions from Zillow. To support the first assumption, we collect data from Zillow (Appendix A.1) and use updates to the Zestimate algorithm by Zillow as an instrument to quantify the reliance on Zestimate (Appendix A.2). Reliance is defined as the sensitivity of the sale price (or sellers' list price) to change in Zestimate visible to buyers and sellers. We find an average reliance of 15\%; for example, if the Zestimate visible to a home was increased by \$10,000 the expected sale price would increase by \$1,500 (+15\% $\times$ \$10,000). We further find that the reliance on the Zestimate varies with the width of the displayed Zestimate range (a confidence interval which quantifies the algorithm's self-reported confidence). To support the second assumption, we measure empirical correlations between changes in the Zestimate for a home and new sales in that home's neighborhood (Appendix A.3). We find that if the sale price of a home were \$10,000 higher, the Zestimate of (approximately) 25 peer homes (that are similar in characteristics) increases by \$2,000 or more. Thus, empirical evidence supports the analytical model assumptions\footnote{It does not confirm the mechanism or findings of our model.}.\\

\begin{figure}[t]
    \centering
    \includegraphics[width=0.9\textwidth]{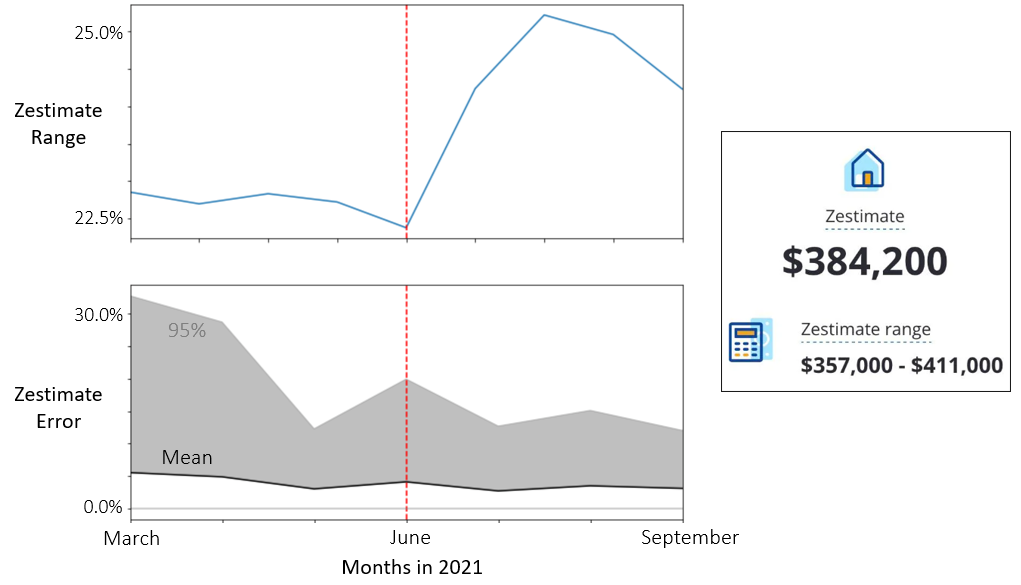}
    \caption{Zillow announced a major update to its Zestimate algorithm on June 15, 2021 (red line). (Top Left) Zillow's self-reported confidence in its Zestimate (as quantified by the width of the displayed Zestimate range) increased discontinuously from 22.69\% (3 months before June 2021) to 24.66\% (3 months after June 2021).
    % \emaad{@Nikhil: add a small screenshot of the Zestimate range.}
    (Bottom Left) In the same time period (3 months before and after June 2021), empirical average error in the Zestimate (relative to the realized sales price) decreased from 4.36\% to 3.41\%, and the 95\% error in the Zestimate (relative to the realized sales price) decreased from 23.41\% to 16.37\%. (Right) A snapshot of Zestimate range of 14\% (\$384,000 $\pm$ 7\%) for a home on Zillow.}
    \label{fig: MotivatingEvidence}
\end{figure}

A key intermediate finding of our model is the over-confidence of machine learning-based pricing algorithms in their ML prices. This over-confidence is also evident anecdotally\footnote{It does not confirm our model but simply says the algorithm over-confidence (a finding of our model) is empirically plausible.}, as illustrated in Figure \ref{fig: MotivatingEvidence}. Figure \ref{fig: MotivatingEvidence} (bottom) shows that the Zestimate is increasingly more accurate over time, likely driven by continuous improvements to the underlying algorithm by Zillow. However, the algorithm's confidence (as measured by the width of the Zestimate range or confidence interval) does not show a similar trend. In fact, the algorithm's confidence has a discontinuous reduction in June 2021, when Zillow announced a major update to the algorithm. A plausible explanation for this discontinuous reduction is that the reported Zestimates were overly-confident before being fixed in June 2021. Until the eventual fix by Zillow, the overly-confident algorithm was active in production, without any oversight to limit its potentially adverse effects.\\

ML prices that both influences and learn from consumer behaviors are increasingly being used to democratize access to information in a variety of markets \citep{zillow2022zestimate,realtor2022realestimate,forbes2021price,airbnb2022smart,lendingclub2022modelrank,liveart2022estimate}. To the best of our knowledge, our work is the first to jointly model the interdependent dynamics of machine learning based-algorithms and consumer behavior. As such, we contribute to the literature on machine learning feedback loops \citep{bottou2013counterfactual, perdomo2020performative, wager2014feedback, sinha2016deconvolving}, which does not consider consumer behavior and economic outcomes, and to the literature on dynamic consumer behavior, which does not consider the dynamics of machine learning algorithms. This paper is one of the first (i) to empirically measure the impact of ML pricing on home sale prices (in the context of Zillow and the Zestimate) and (ii) to begin to disentangle Zillow's proprietary Zestimate algorithm. These two smaller contributions should spark more academic research on and policymaker scrutiny of the dominant role of Zillow's ML pricing algorithm in the housing market.\\

More broadly, our work showcases the perils of deploying algorithms that are myopically optimized for statistical objectives such as accuracy, at the expense of long run economic objectives. Such algorithms, as a consequence of feedback loops, could reinforce non-diverse and self-fulfilling preferences (such as recommending fashion content to women and sports content to men, or recommending higher interest rates for loan applications from historically under-served groups due to higher predicted risks). If left unchecked, in the long run, these reinforced preferences could fully diverge from the ``ground truth'' i.e., preferences in absence of algorithmic interference. We unpack how this divergence from the ``ground truth'' can make consumers worse off. As such, our work suggests that policymakers monitor and regulate ML pricing. Our work also suggests that intermediary platforms consider the long-term implications of their deployed pricing algorithms when subject to feedback loops. The eventual divergence from ground-truth predicted by our analytical
model poses risks to platforms' reputation and brand perception, which partly depends on the accuracy of their deployed pricing algorithms.

\section{Related Literature}

A stream of literature has looked at pricing in the housing market before introduction of ML. \cite{linneman1986empirical} showed a large variance buyers and sellers home value estimates, which is unsurprising given that most buyers and sellers transact infrequently (say, once in 10 years), and, unlike some other assets, houses have a large diversity of features. Field surveys and experimental research tried to estimate valuation errors before the advent of pricing algorithms; error estimates include 14\% \citep{goodman1992accuracy}, 5.3\% \citep{kiel1999accuracy}, and 16\% \citep{ihlanfeldt1986alternative}. One would expect that agents, brokers, and other market experts could correct valuation errors \citep{han2015microstructure}, but sellers and expert agents' contract under information asymmetry. The seller's inability to observe his agent's efforts creates a moral hazard in the principal (seller) – agent contract \cite{anglin1991residential}. When searching (for buyers) is costly, the agent has an incentive to undervalue the home and save on search costs; when the market is competitive, the agent has an incentive to overvalue the home to outbid competing agents. Further, the adverse selection problem prevents the seller from accurately judging whether the agent is knowledgeable about the state of the market. The challenges faced by buyers and sellers in pricing homes is further supported by WakeField research survey (Melcher 2021) which finds that the average US home-buyer tours 15 homes and makes offers on 10 homes, spending a cumulative \$845 million of work time on home search and pricing. 85\% of first-time buyers say that it is challenging to make offers and stressful to be rejected or outbid.\\

In this context it is not surprising that ML prices have influence. Participants are likely to find ML pricing attractive because (i) it is free and thus highly accessible, unlike an appraiser; (ii) it seems impartial, unlike agents; and (iii) the platforms (e.g., Zillow and RedFin) report low error rates between ML price and eventual sale price to convey accuracy. Some experts have been highly critical of Zillow's Zestimate (or Redfins' Estimate) for being inaccurate, using outdated data, missing local non digitized information and even altogether unusable \citep{houwzer2020zillow}. Zillow's received customer ratings of 3.8, 2.8 and 1.6 on consumeraffairs.com, sitejabber.com and trustpilot.com. Given these concerns we empirically validate that buyers and sellers are in-fact sensitive to changes in Zestimate.\\

Another stream of literature has looked at feedback loops in a wide range of online learning settings where an ML algorithm learns by making mistakes \citep{barocas2017fairness}. Such feedback designs are innocuous in settings where the ML predictions do not contaminate the ground truth label, but elsewhere, the feedback design can slowly accrue a technical debt \citep{sculley2015hidden} that eventually has profound effects \citep{amodei2016concrete}. \cite{perdomo2020performative} identify conditions under which a feedback loop will converge to a stable point. Our analytical framework roughly concurs, but we are less concerned with statistical properties and more concerned with the payoffs at the equilibrium. We discern how much of the covariate distribution shift comes from the evolution of intrinsic housing preferences versus from the ML feedback itself. If the latter dominates, the social surplus may be lost even as the ML algorithm reaches optimal accuracy.\\

ML feedback loops have been documented in settings such as - ad placement \citep{bottou2013counterfactual}, search engine rankings \citep{wager2014feedback} and recommender systems \citep{chaney2018algorithmic,sinha2016deconvolving}. Our model of sellers has similarities with the model of \citep{schmit2018human}, who assume that users are naïve in believing that ML recommendations are unbiased, and users are myopic and honest about their current action without regard to its impact on future states via the feedback loop. Importantly, in all these examples, individuals interact with ML predictions without an outside option. For example, a user who is seeking relevant search results does not have an alternative mechanism besides the ML algorithm; she cannot realistically achieve the same outcome by interacting with a crowd of her peers. In the housing market, however, buyers and sellers can interact to determine prices in the absence of ML pricing. This interaction may happen indirectly, for example when a seller observes lack of visits or offers from buyers. Thus, the introduction of ML pricing to the housing market is unique in that it replaces the wisdom of the crowd with a single correlated signal. To our knowledge, we are one of the first to evaluate the impacts of ML feedback loops on a market as a whole.\\

The business literature has identified consequences of ML algorithmic pricing besides the feedback loop. For example, bias propagated by the algorithmic pricing of hotels, car insurance, loans \citep{israeli2020algorithmic}, and ride-hailing \citep{pandey2021disparate}. \citep{bertini2021pitfalls} find that customers misperceive the motives of firms that offer algorithmic pricing. \citep{assad2020algorithmic} and \citep{brown2021competition} study whether algorithms provide competitive advantages or lead to collusive outcomes. \citep{huang2021seller} identifies settings in which algorithmic pricing may increase market friction. \citep{yu2020algorithmic} argue that algorithmic price might mitigate racial disparities in the housing market. The present paper is unique in this literature because it does not model ML pricing or its statistical properties as static. Instead, we model ML algorithm's learning in conjunction with the evolving market as both move toward equilibrium.

\section{Model}

In this section, we model the interdependent evolution of consumer behavior and a machine learning-based pricing algorithm in the context of a housing market. We decompose our model into two components. In the first component, we model how a seller determines the price at which to list their home, buyer offers, the realized sale prices and we define true value of a home (Section 3.1). In the second component, we model how a machine learning-based pricing algorithm estimates home values (Section 3.2). In Section 4, we allow both model components to evolve simultaneously and interdependently, and characterize the equilibrium of the feedback loop between the market and the pricing algorithm.

\subsection{First Component: Market Participants}
\label{subsec:firstcomponent}
\textbf{Modeling home-buyers.} Consider a home for sale on the market in time period $\tau$. We focus on modeling an ``exemplar buyer'' of this home: the buyer who has the highest willingness to offer among all buyers interested in this home. Let the willingness to offer of the exemplar buyer be a random variable $y_{\tau}$ drawn from a distribution with mean $\mu$ (assume to be unique for each home) and variance $\sigma_b^2$ (assumed to be identical across homes), where the variance captures the heterogeneity in buyers' preferences. Since the seller (the current homeowner) only entertains the highest offer in each time period, it is sufficient to model the exemplar buyer instead of all prospective buyers.

Let the home be listed at price $l_{\tau}$. The exemplar buyer will then offer $\min(y_{\tau}, l_{\tau})$, without revealing their willingness to offer $y_{\tau}$\footnote{We assume that ``highest and best offer'' negotiations are absent.}. The home sells to the exemplar buyer at the end of time period $\tau$ for $\min(y_{\tau},l_{\tau}) = l_{\tau}$ if $y_{\tau} \geq l_{\tau}$, and remains unsold otherwise. We denote by $P(\textrm{sale}|l_{\tau}) = P(y_{\tau} \geq l_{\tau};\mu,\sigma_b^2)$ the probability that the home sells in time period $\tau$.
% Since, homes are unique with hundreds of differentiating features, typically it would be difficult for the homeowner to exactly know $\mu$ (and therefore $P(sale|l_{\tau})$) when listing the home on the market at $l_{\tau}$.
While individual buyers can update their valuations during their home search, their aggregate offer distribution $(\mu,\sigma_b^2)$ is assumed to be static\footnote{Modeling buyers' aggregate offer distribution as declining over time on market $\tau$ does not change any of our conclusions.}.\\

\begin{figure}
    \centering
    \includegraphics[width=0.8\textwidth]{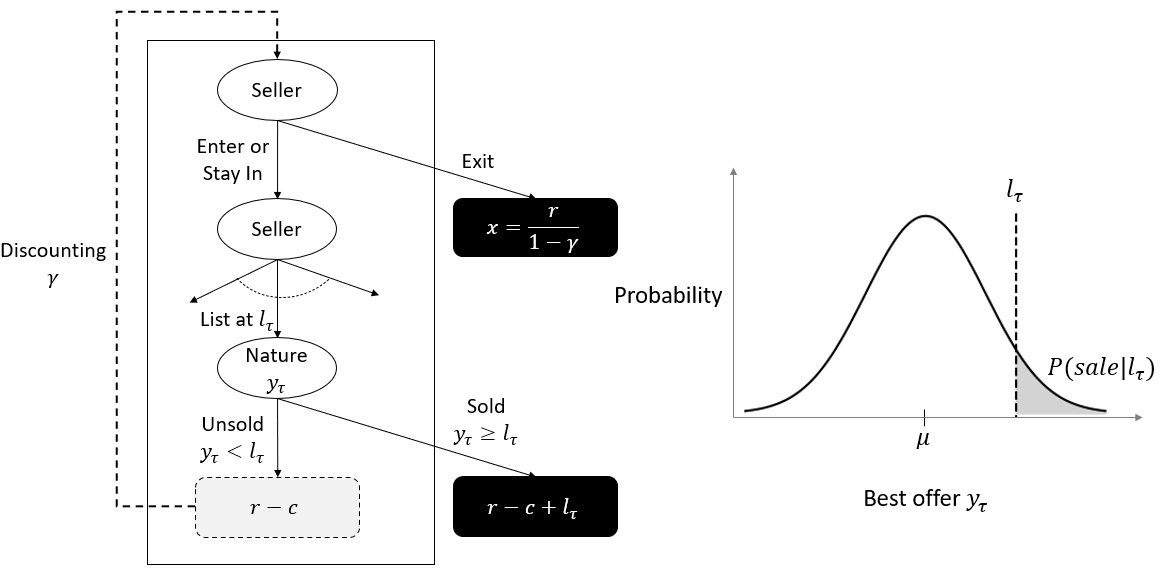}
    \caption{(Left) The home sale process as a single-period game with the seller as the focal agent. The game repeats until the home is sold or the seller decides to exit the market. (Right) A hypothetical distribution of the list price, best offer distribution, and probability of sale.}
    \label{fig: SellerGame}
\end{figure}

\textbf{Modeling home-sellers.}
Our model of a home-seller is illustrated in Figure \ref{fig: SellerGame}. At the start of each time period $\tau$, the seller decides whether to list her home for sale, and determines the list price $l_{\tau}$. At the end of the time period, the home (if listed) sells at the list price $l_{\tau}$ with probability $P(\textrm{sale}|l_{\tau})$, based on our model of home-buyers. If the home does not sell at the list price, the seller returns to the decision of whether to continue listing her home for sale\footnote{In Appendix \ref{app:bargaining}, we discuss why we do not model the seller's choice to accept an offer below the list price.}. In each time period, the seller earns a flow payoff (regardless of whether or not her home sells) from retaining ownership of the home (such as rental income $r$) less market participation costs $c$ (such as maintaining the home for open-houses). If the seller decides to exit the the market, she receives a lifetime payoff $x$ from retaining ownership of the home (such as via rental income $r$ at discount $\gamma$). We treat this as a terminal state: the seller cannot change her decision to exit the market. 

As an example, consider a seller who stays in the market for time periods $1,\dots,T$ and lists her home at a sequence of prices ${l_1, ..., l_{\tau}, ..., l_T}$, until her home sells in period $T$. Her payoff $\pi$ and the outside option value $x$ (had she not entered the market at all) are given by\footnote{Note that $\gamma \approx 1$ if each period is 1 month long; a 3\% yearly interest rate (or 0.97 yearly discounting) is equivalent to 0.997 monthly discounting. Since homes are typically listed for a few months before selling, $\gamma^T$ is also close to 1. We will use these approximations to simplify our calculations.},
\begin{align}
    \pi = \frac{1-\gamma^T}{1 - \gamma} \times (r - c) + \gamma^T \times l_T \text{ \quad ; \quad} x = \frac{r}{1 - \gamma}
\end{align}
The seller enters the market if her estimate of the expected payoff $\tilde{\mathbb{E}}[\pi]$ is greater then the outside option value $x$. The expectation $\tilde{\mathbb{E}}[\pi]$ is over all possible realizations of sale prices, on-market durations, and whether or not the home sells. The seller lists her home at the optimal price $l_{\tau}^*$ at the beginning of every period $\tau$ to maximize her expected payoff $\tilde{\mathbb{E}}[\pi | l_{\tau}]$ from period $\tau$ onwards.

We use the notation $\tilde{\mathbb{E}}[\cdot]$ to distinguish between the seller's estimate of the expected payoff and the actual expected payoff $\mathbb{E}[\cdot]$. The seller's estimate of the expected payoff depends on her estimate of the sale probability $\tilde{\mathbb{P}}(\textrm{sale}|l_{\tau})$, which in turn depends on her estimate of the offer distribution parameter $\tilde{\mu}_\tau$ in period $\tau$. The seller also has some uncertainty (variance) about her estimate, which we denote by $\sigma^2_{s,\tau}$. The seller's uncertainty and the estimation errors arising from $\tilde{\mu}_\tau \neq \mu$, $\tilde{\mathbb{P}}(\textrm{sale}|l_{\tau}) \neq \mathbb{P}(\textrm{sale}|l_{\tau})$, and $\tilde{E}[\pi|l_\tau] \neq E[\pi|l_\tau]$) will lower payoffs, which platforms like Zillow attempt to alleviate with their machine learning-based pricing algorithms.

We denote the seller's initial estimates as $\tilde{\mu} = \tilde{\mu}_{\tau = 0}$ and $\sigma_s^2 = \sigma^2_{s,\tau = 0}$. In each subsequent time period $\tau$, the seller learns and updates her initial estimates and uncertainty ($\tilde{\mu}_{\tau},\sigma^2_{s,\tau}$). We model these updates as a Bayesian learning process in which the seller combines her current estimate $(\tilde{\mu}_{\tau},\sigma^2_{s,\tau})$ with the observable market signal $(\mu,\sigma^2_{\textrm{signal}})$ to develop an updated estimate $(\tilde{\mu}_{\tau+1},\sigma^2_{s,\tau+1})$.  The market signal is observed when the seller interacts with prospective buyers, some of whom implicitly of explicitly reveal the price they are willing to offer for the seller's home. We assume that the market signal is noisy, but unbiased with respect to the buyer's offer distribution parameter $\mu$. With time on the market, we assume that the seller's estimate of $\mu$ improves and her uncertainty reduces: $\tilde{\mu}_{\tau} \rightarrow \mu, \sigma^2_{s,\tau} \rightarrow 0$ as $\tau \rightarrow \infty$. \\

\textbf{Home values.} Intuitively, a home's value $v$ is its market clearing price. This is easy to formalize for, say, a single IBM stock, with millions of identical units transacted among thousands of buyers and sellers every day. In contrast, a single home is bought and sold only a few times in many years. Hence, we consider a thought experiment where the same home is sold thousands of times. The value of the home $v$ in this thought experiment is its sale price $p$ (which depends on the buyer offer distribution and the seller's estimate of this parameter) averaged over all the times it is sold:
\begin{align}
    v = \mathbb{E}[\mathbb{E}[p|\mu, \tilde{\mu},\sigma_b,\sigma_s,\sigma_{signal}]]
    \label{eq:valueDef}
\end{align}
Next we will consider machine learning based estimation of this home value. Note that the machine learning algorithm does not model the market structure (offers $(\mu,\sigma_b)$, beliefs $(\tilde{\mu},\sigma_s$, and signals $\sigma_{signal}$).  

\subsection{Second Component: Machine Learning-Based Pricing Algorithm}
% Things the model needs to have:
%
% - Home feature vector
% - Outcome: predicted sale price (value) of a home k at time t
% - Value of the home evolves
%
% - Note that prices on Zillow are rounded to 10s
% - Coarsening in practice is very common

% Motivation for this section
We assume that the home value $v_{k,t} = G_t(\mathbf{X}_k)$ of a home $k$ at time $t$ is a time-evolving function of the home's characteristics $\mathbf{X}_k$. This can be interpreted as home buyers' preferences for the home's characteristics evolving over time, while the home's characteristics remain unchanged. We further assume that the resulting home value evolves as a random walk:
\begin{align}
    v_{k,t} = v_{k,t-1} + e^{\textrm{rw}}_{k,t}, \qquad v_{k,t} = G_t(\mathbf{X}_k), ~ e^{\textrm{rw}}_{k,t} \sim N(0, \sigma^2_{\textrm{rw}})
    \label{eq:randomwalk}
\end{align}
In this section, we propose a general machine learning framework for home value estimation. For analytical simplification we assume that the machine learning model parameters $\theta_t$ are retrained every period $t$ to approximate the underlying true home preferences $G_t(\mathbf{X}_k)$\footnote{This simplifies the analytical expressions without changing the results.}. Finally we derive analytical expressions for the machine learning-based prices as a function of observed sale prices.\\

\textbf{General machine learning framework.} Our framework assumes that the machine learning model of home values is trained to (i) accurately estimate the value of each home in the training data, and (ii) estimate similar home values for homes with similar features. Satisfying objective (i) increases in-sample accuracy, but decreases out-of-sample accuracy due to over-fitting. Essentially, the inclusion of objective (ii) regularizes the model to increase its generalization power. Our framework is a generalization of the network lasso framework \citep{hallac2015network} to non-convex optimization objectives.

Formally, let $\mathcal{H}_t$ be a set of $N$ homes used to train the machine learning model at time $t$\footnote{The machine learning training period duration $t$ (say 1 year) is typically different than the home listing period duration $\tau$ (say 1 week or 1 month). We also skip the details of hold out sample for validation within the training data.}. For each home $k \in \mathcal{H}_t$, let $v_{k,t}$ be the true value of the home at time $t$, and $\mathbf{X}_k \in \mathbb{R}^{\bar{Q}}$ be a $\bar{Q}$-dimensional vector of the home's characteristics (such as its number of bedrooms and year of construction). We assume that the machine learning model parameters $\theta_t$ are given by:
\begin{align}
    \theta_t =
    \textrm{argmin}_{\theta}
    ~\left[~
    \underbrace{\sum_{k \in \mathcal{H}_t} f_k(\mathbf{X}_k | \theta )}_{\textrm{objective (i)}}
        \quad+\quad
    \underbrace{\sum_{m,n \in \mathcal{H}_t} g_{mn}(\mathbf{X}_m, \mathbf{X}_n  | \theta)}_{\textrm{objective (ii)}}
    ~\right]~
    \label{eq:networklasso}
\end{align}
Here, $f_k(\mathbf{X_k}|\theta)$ is a loss function capturing the first objective of accurate in-sample estimation for each home $k \in \mathcal{H}_t$ in the training data, and $g_{mn}(\mathbf{X_m}, \mathbf{X_n}|\theta)$ is a loss function capturing the second objective of making similar estimations for similar homes. Note that we make no assumptions about the functional form of either loss function, or about the architecture of the machine learning model. As such, our framework generalizes a variety of common machine learning models, including deep neural networks. \\

\textbf{Example.} As a specific example of our framework, let $\hat{v}_k = F(\mathbf{X}_k, \Theta_k)$ be the estimated price of each home $k$ derived using a neural network $F(\cdot)$ with weights $\Theta_k$. While the weights are different for each home $k$, a single neural network $F(\cdot)$ is trained to learn the weights $\Theta_1, \dots, \Theta_N$ of all $N$ homes in the training data $\mathcal{H}_\tau$; these weights can be viewed as homes' embeddings. To maximize in-sample accuracy, we minimize the mean squared error by setting $f_k(\mathbf{X_k}|\theta) = [\hat{v}_k - v_{k,t}]^2 = [F(\mathbf{X}_k,\Theta_k) - v_{k,t}]^2$. To improve out-of-sample accuracy, we  penalize differences between the learned weights for pairs of similar homes. Specifically, we set $g_{mn}(\mathbf{X}_m, \mathbf{X}_n  | \theta) = \lambda \|\mathbf{X}_m - \mathbf{X}_n\|_2 \|\Theta_m - \Theta_n\|_2$, where $\|\cdot\|$ is the Euclidean distance or $L_2$ norm, and $\lambda$ is a regularization hyperparameter. Note that the parameters $\theta = \{F(\cdot), \lambda\} \cup \{ \Theta_k\}_{k \in \mathcal{H}_t}$ include the neural network architecture $F(\cdot)$, the regularization hyperparameter $\lambda$, and the weights $\Theta_k$ for each home $k \in \mathcal{H}_t$.\\

\textbf{Regularization and home clustering.} Solving for $\theta_\tau$ in Eq. \ref{eq:networklasso} essentially performs joint home value estimation and home clustering, where a pair of homes $i, j$ with similar learned weights $\Theta_i, \Theta_j$ can be assigned to the same cluster by discretizing the weight values (by rounding them to the nearest integer, for example). After discretization, $Q$ homes in a cluster $\mathcal{C}_i$ for $i=1,\dots,Q$ have the same learned weights $\Theta_k$ and similar characteristics $\mathbf{X}_k$ (and hence, similar estimated home values). The regularization parameter $\lambda$ controls the clustering granularity: $\lambda = 0$ permits each home to form its own cluster, while $\lambda \rightarrow \infty$ leads to a single cluster containing all homes in the training data. We denote by $Q$ the number of clusters, and assume (for analytical simplification) that each cluster contains exactly $N/Q$ homes.\\

\textbf{Generating algorithmic prices.} The clustering of homes facilitates making out-of-sample predictions. Let $\bar{\mathbf{X}}_\mathcal{C}$ denote the centroid of cluster $\mathcal{C}$ (the average of $\mathbf{X}_k$ over all homes $k \in \mathcal{C}$), and let $\bar{v}_{\mathcal{C},t}$ be the average value of all homes $k \in \mathcal{C}$ at time $t$. We assign each out-of-sample home $q$ with characteristics $\mathbf{X}_{q}$ to its nearest cluster $\mathcal{C}(q) \in \{\mathcal{C}_1, \dots, \mathcal{C}_J\}$ with the smallest Euclidean distance $\|\mathbf{X}_{q} - \bar{\mathbf{X}}_\mathcal{C}\|_2$, and could subsequently use $\bar{v}_{\mathcal{C}(q),t}$ as its predicted home value. However, true home values are unobservable in practice. Hence, a common proxy for home values is their sales price, which are observable in the past. Using this proxy plays an important role in the machine learning errors that we discuss later in this section. Let $p_{k,t}$ be the sale price of home $k$ at time $t$, and let $\bar{p}_{\mathcal{C}, t}$ be the average sale price of all homes in cluster $\mathcal{C}$. Then the predicted price of an out-of-sample home $q$ assigned to cluster $\mathcal{C}(q)$ (as described above) at time $t+1$ is given by:
\begin{align}
    z_{q,t+1} = \bar{p}_{\mathcal{C}(q),t} = \frac{Q}{N} \times \sum_{k \in \mathcal{C}(q)} p_{k,t}
    \label{eq:3}
\end{align}

\begin{table}[]
\caption{Assumptions about the distributions in our simple and full models.}
\begin{tabular}{l|l|l|}
\cline{2-3}
\textbf{}                                      & \multicolumn{1}{c|}{\textbf{\begin{tabular}[c]{@{}c@{}}Simple Model\end{tabular}}} & \multicolumn{1}{c|}{\textbf{\begin{tabular}[c]{@{}c@{}}Full Model \end{tabular}}} \\ \hline
\multicolumn{1}{|l|}{Buyer Offer Distribution $P(y_\tau)$} & $U[\mu - \sigma,\mu + \sigma]$                                                                                                         & $N(\mu,\sigma_b^2)$                                                                                                       \\ \hline
\multicolumn{1}{|l|}{Seller Estimate $P(\tilde{\mu})$}          & $P(\tilde{\mu}=\mu-2\sigma)=P(\tilde{\mu}=\mu)=P(\tilde{\mu}=\mu+2\sigma)=1/3 $                                                                                                         & $N(\mu,\sigma_s^2)$                                                                                                       \\ \hline
\multicolumn{1}{|l|}{Seller Learning $\sigma_{\textrm{signal}}^2$}          & 0                                                                                                        & $\kappa \sigma_b^2$                                                                                                   \\ \hline
\end{tabular}
\label{table:SimpleAndFull}
\end{table}

\begin{table}[]
\centering
\caption{Table of notation and definitions}

\textbf{Exogenous Variables}\\
\hfill

\begin{tabular}{|l|l|}
\hline
\textbf{Notation} & \textbf{Description}                 \\ \hline
$v_t$         & True value (market clearing price) of home at time $t$   \\ \hline
$\sigma_{rw}^2 = Var_t[v_{t+1} - v_t]$  & Variance of true value changes of a home over time.   \\ \hline
$\sigma_v^2 = Var_k[v_t]$   & Variance of true value across all homes in one period. \\ \hline
$\sigma_e^2 = Var_i[\tilde{v}_{i,t} - v_t]$      & Variance of participant $i$ valuations (before introducing ML) \\ \hline
$N$         & Total number of homes sold in one period.      \\ \hline
\end{tabular}

\hfill
\center{\textbf{Key Endogenous Variables}}\\
\hfill

\begin{tabular}{|l|l|}
\hline
\textbf{Notation} & \textbf{Description}                 \\ \hline
${l_{\tau}}_t$         & Sequence of list prices set by seller for home at time $t$        \\ \hline
$p_t$         & Realized sale price        \\ \hline
$\tilde{v}_{i,t}$         & Participant $i$ valuation of home at time $t$        \\ \hline
$z_t$         & ML (Machine Learning) price of home at time $t$   \\ \hline
$Q$          & ML model hyperparameter controlling number of clusters   \\ \hline
\end{tabular}

\hfill
\center{\textbf{Outcomes Variables (as function of reliance $\alpha$)}}\\
\hfill

\begin{tabular}{|l|l|}
\hline
\textbf{Notation} & \textbf{Description}                 \\ \hline
$\alpha(\hat{\sigma}^2_z)$  & Participant reliance on ML price   \\ \hline
$\sigma^2_z(\alpha)$  & True ML price error ($\sigma^2_z(\alpha = 0)$ denoted by $\sigma^2_z$)  \\ \hline
$\hat{\sigma}^2_z(\alpha)$  & Estimated ML price error ($\hat{\sigma}^2_z(\alpha = 0)$ denoted by $\hat{\sigma}^2_z$)  \\ \hline
$\pi, \Pi(\alpha)$  & Realized payoff and expected risk averse payoff  \\ \hline
\end{tabular}
\label{table:notation}
\end{table}

\section{Results}
\label{sec:results}

In Section \ref{subsec:firstcomponent}, we described our model of buyer-seller interactions grounded in the structure of the housing market --- the offer distribution parameters ($\mu, \sigma_b^2$), the seller's estimate of the offer distribution mean $\tilde{\mu}$, her uncertainty $\sigma^2_{s}$, the seller's learning aided by a market signal $(\mu, \sigma^2_{\textrm{signal}})$, and the seller's outside option value $x$ and market participation costs $c$. We further formalized the sellers' optimal listing price choice $l_{\tau}$ in terms of these parameters. However, a signal such as the ML price $z$ does not convey information about the offer distribution $\mu$ or the list price $l_\tau$, nor is it tailored to an individual sellers characteristics. Instead, the ML price $z$ is an estimate of a home's value $v$ (equation \ref{eq:valueDef}) \textit{summarized} over these market structure details and individual heterogeneity. 

In section 4.1, we derive the home value $v$ and distribution of sale prices ($E[p],Var[p]$). We consider two sets of assumptions (listed in Table \ref{table:SimpleAndFull}) about the distributions of offers $y_\tau$, the seller's estimate $\tilde{\mu}_{\tau}$, and the market signal $\sigma_{\textrm{signal}}^2$. In our results described in Section \ref{sec:results}, we use the ``simple" model for analytical closed-form solutions. In Appendix B.2, we employ numerical simulations under the ``full'' model to verify that our findings are consistent with the ``simple'' model. In section 4.2, we will derive machine learning pricing errors (true $\sigma_z^2 = Var[v - z]$ and estimate $\hat{\sigma}_z^2  = Var[p - z]$). In section 4.3, we will identify equilibrium of the feedback loop between the machine learning pricing error estimate $\hat{\sigma}_z^2(\alpha)$ and reliance on machine learning prices $\alpha(\hat{\sigma}_z^2)$ which depend on each other. Finally, in section 4.4, we will use these equilibrium expressions to formulate payoffs for sellers at the equilibrium. Table \ref{table:notation} and \ref{table:LemmaPropMap} summarize the notation and key results.

\subsection{Home Value and Sale Prices}

Consider an oracle who knows a home's true offer distribution ($\mu, \sigma_b^2$), and knows that the sellers have potentially erroneous estimates. The oracle could calculate the home's value $v(\mu)$ by integrating over the sellers' estimates of the buyer offer distribution $\tilde{\mu}$, and over the stochasticity in buyers' offers (embedded in $\mathbb{E}[p|\mu, \tilde{\mu}]$), as follows:
\begin{align}
v(\mu) = \mathbb{E}[\mathbb{E}[p|\mu, \tilde{\mu}]] = \int_{\tilde{\mu}} P(\tilde{\mu} | \mu, \sigma) \times \mathbb{E}[p|\mu, \tilde{\mu}]
\end{align}

Similarly, we can formalize the seller's estimate $\tilde{v}$ of her home's value in time period $\tau$ (where we drop the subscript $\tau$). The seller does not know her home's true offer distribution ($\mu, \sigma_b^2$), However, the seller knows that her estimate $\tilde{\mu}$ is drawn from a distribution with mean $\mu$ and variance $\sigma^2_s$. The seller's valuation of her home $\tilde{v}$ is then given by $\tilde{v} = \tilde{\mathbb{E}}[v(\mu)]$, where the expectation is over all possible estimates of $\mu$ by the seller. Now consider an external signal from an ML price $z$ that also claims to estimate the home's value: $z = \hat{E}[p] = \hat{v}$, where we use $\hat{E}[\cdot]$ to differentiate between the algorithm's estimate and the seller's estimate (denoted by $\tilde{\mathbb{E}}[\cdot]$). After observing this signal, the seller updates their valuation $\tilde{v}$ by combining their prior valuation (before observing and independent of the ML price) and the ML price $z$ as follows:
\begin{align}
\tilde{v} = (1-\alpha) \times \tilde{\mathbb{E}}[v(\mu)] + \alpha \times z \quad ; \quad \textrm{Var}[\tilde{v}] := (1 - \alpha)^2 \textrm{Var}[\tilde{v}]
\label{eq: 1}
\end{align}
This formulation for impact of ML price $z$ condenses complex details on how sellers absorb the ML price $z$. The seller (jointly with their agent) receives informative signals from a lot of sources (agent, appraisers, neighbors and market experts) all embedded into their private valuation $\tilde{\mathbb{E}}[v]$. The ML price $z$ is yet another informative signal, specially treated in our model because we want to isolate its impact, to construct a final valuation $\tilde{v}$. The seller uses the valuation ($\tilde{v}$ e.g., \$520k) to infer likely offers ($\tilde{\mu} \pm \sigma_b$ e.g., \$500k to \$550k) and a corresponding good list price ($l^*$ e.g., \$540k). A similar influence occurs for buyers. Individual buyers (jointly with any buyer agent) incorporate ML price $z$ as yet another informative signal into constructing a valuation $\tilde{v}$. This in turn updates individual buyers' willingness to offer and consequently the offer distribution from the exemplar buyer every period $\mu$. We do not explicitly model the estimation and willingness to offer for individual buyer. But implicitly the ML price $z$ impacts valuations of both buyers and sellers, and thereby impacts the realized sale price $p$. Going forward we will use the phrase buyer-seller when discussing impact of ML price. 

Using the definitions above, we can derive expressions for home value $v$, buyer-seller valuation $\tilde{v}$ and the sale price $p$ \footnote{Proofs in Appendix \ref{app: proofs}.}. 

\lemma{The true home value $v$ is given by,
\begin{align}
v = (\mu + 2\sigma/3 - \sqrt{2c \sigma})
\end{align}
Each seller's valuation (and the variance of this valuation) is given by,
\begin{align}
\tilde{v}(\alpha = 0) = \tilde{\mathbb{E}}[p|z] = \tilde{\mu} + 2\sigma/3 - \sqrt{2c\sigma} \text{ ; \quad } \sigma_e^2(\alpha = 0) = \sigma_e^2 = Var[\tilde{v}] = 8 \sigma^2/3 \nonumber \\
\tilde{v}(\alpha) = (1-\alpha) \times \Big( \tilde{\mu} + 2\sigma/3 - \sqrt{2c \sigma}\Big) + \alpha \times z \text{ ; \quad } \sigma_e^2(\alpha) = (1-\alpha)^2 \sigma_e^2
\end{align}
The distribution of sale price is given by,
\begin{align}
\mathbb{E}[p|z] = (1-\alpha) \times \Big( \mu + 2\sigma/3 - \sqrt{2c\sigma}\Big) + \alpha \times z \text{ ; \quad } \sigma_{\epsilon}^2(\alpha) = Var[p] = \delta \sigma_e^2(\alpha) \text{ ; \quad } \delta = (1/6)
\end{align}
Reliance on ML price ($\alpha > 0$) adds bias in valuations and sale prices i.e., $v \neq \mathbb{E}[p] = \mathbb{E}[\tilde{v}]$ if $z \neq p$. It also reduces variance in valuations and sale prices i.e., $Var[p]$ and $Var[\tilde{v}]$ are decreasing in $\alpha$.
\label{lemma:1}} \normalfont\\ 

% \emaad{Add some notation to define p}

Let the subscript $i$ index each seller. We can express each seller's valuation as $\tilde{v} = \mathbb{E}[\tilde{v}_i] + e_i$ where $e_i = (1-\alpha)(\tilde{\mu} - \mu)$ is the noise in valuation across individual sellers with $E[e_i] = 0$. We can also express the realized sale price as $p = \mathbb{E}[p] + \epsilon$ where $\epsilon$ is the noise in realized sale prices (across multiple hypothetical sale instances) with $E[\epsilon] = 0$. There are two intermediate results worth highlighting here. First, the true expected sale price is equal to the expected seller valuation: $\mathbb{E}[p] = \mathbb{E}[\tilde{v}_i]$. At $\alpha = 0$ we have $v = \mathbb{E}[p] = \mathbb{E}[\tilde{v}_i]$ i.e., prior valuations (before observing ML price) are unbiased with respect to the true value $v$. At $\alpha > 0$, the valuation are not unbiased anymore i.e., $\mathbb{E}[\tilde{v}_i] \neq v$ if $z \neq p$. Second, the variance in sale prices $Var[p]$ is proportional to variance (disagreement) of seller's valuation $Var[\tilde{v}]$\footnote{Variance (disagreement) in individual seller's valuation is a theoretical measure in the housing market setting because only one homeowner can be the seller for a unique home.}. The constant $\delta$ can be interpreted as the degree to which disagreement in private valuations is reduced by participating in the market (buyers and sellers learning and attaining consensus). The variance in valuations $Var[\tilde{v}]$, and consequently variance in sale prices $Var[p]$ are diminished by a fraction $(1-\alpha)^2$. So overall, ML price is adding some bias but removing some variance from valuations and sale prices. The payoff implications of this will be expressed in Lemma 2 and Lemma 3 in section 4.4.

\subsection{Machine Learning Pricing Errors}
\begin{figure}
    \centering
    \includegraphics[width=\textwidth]{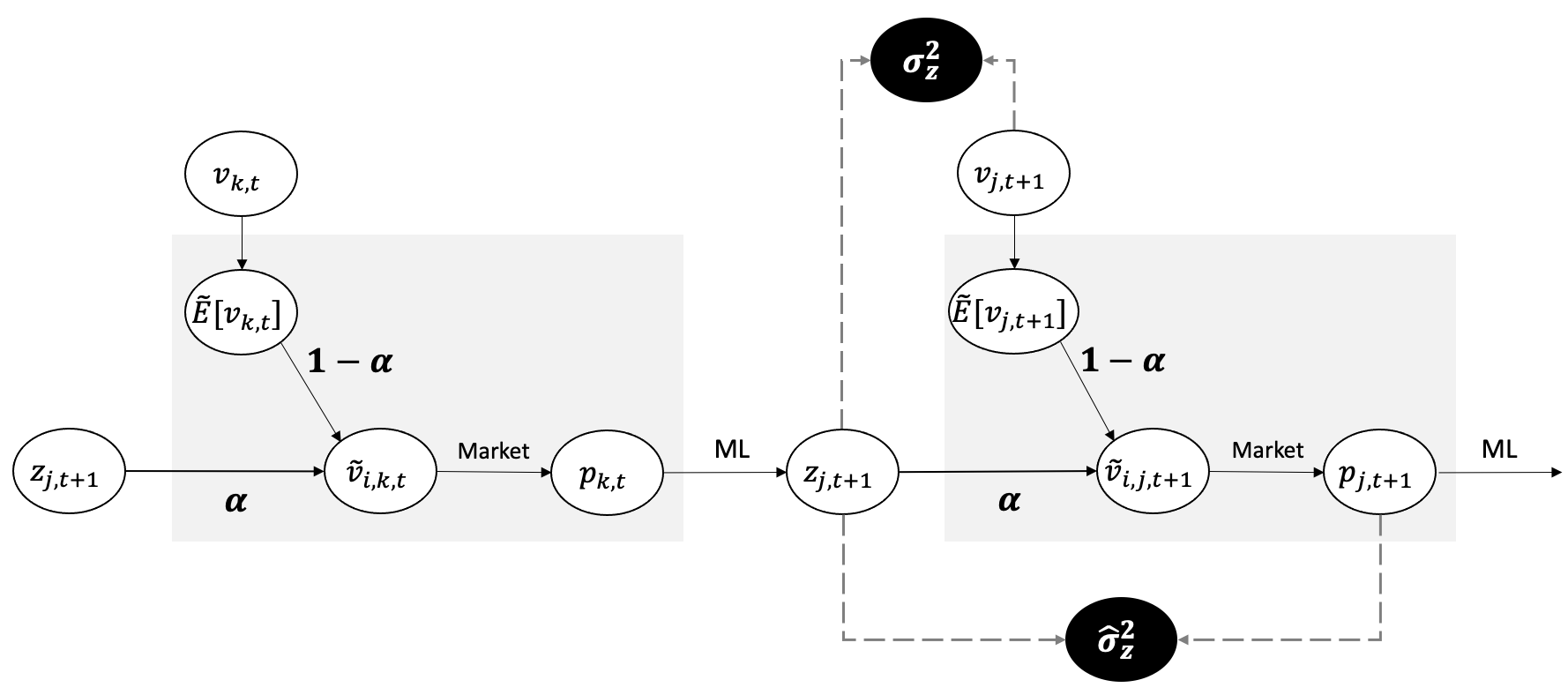}
    \caption{Buyer-seller $i$ of home $k$ construct home valuation $\tilde{v}_{i,k,t}$ using ML price $z_{k,t}$ and private estimation with weights $\alpha$ and $1-\alpha$ respectively. The market interaction among buyer-sellers with similarly constructed valuations produces a sale prices $p_{k,t}$. $N$ home sales similar to home $k$ form training sample for Machine Learning model. In the next period, a home $j$ is influenced by ML price $z_{j,t+1}$. The machine learning error can be estimated $\hat{\sigma}_z^2$ using ML price $z_{j,t+1}$ and realized sale price $p_{j,t+1}$. This is an approximation since the true home value $v_{j,t+1}$ is not observed.}
    \label{fig: MLFeedbackModel}
\end{figure}

We can use equations \ref{eq:3} (ML price $z$ as function of sale prices $p$), and Lemma \ref{lemma:1} (sale prices $p$ as function of ML price $z$) to formulate the feedback loop. Figure \ref{fig: MLFeedbackModel} visually illustrates this loop. In this section, we will start by calculate the true Machine Learning pricing errors $\sigma_z^2 = Var[v - z]$ and its empirical estimate $\hat{\sigma}_z^2  = Var[p - z]$ (because true value $v$ is not observed, only sale prices $p$ are observed).\\

\begin{table}[]
\centering
\caption{At a glance view of Lemma and Propositions}
\begin{tabular}{|l|l|}
\hline
Lemma 1        & Expression for home value and sale price distribution \\ \hline
Lemma 2       & Expression for expected payoff                 \\ \hline
Lemma 3       & Expression for variance of payoffs   \\ \hline
Proposition 1 & Confounded ML error is inflated i.e., $\sigma^2_z(\alpha) \geq \sigma^2_z(\alpha = 0)$ \\ \hline
Proposition 2 & Confounded ML error estimate is deflated i.e., $\hat{\sigma}^2_z(\alpha) \leq \sigma^2_z(\alpha)$ if $\alpha$ large enough \\ \hline
Proposition 3 & Confounded ML error estimate $\hat{\sigma}^2_z(\alpha)$ is decreasing in $\alpha$ if $\alpha$ large enough \\ \hline
Proposition 4 & Proposition 2 and 3 always true at equilibrium $\alpha^*$ \\ \hline
Proposition 5 & $\alpha^* = 1$ always an equilibrium and only equilibrium if $\sigma_z^2/\sigma_e^2$ is small enough\\ \hline
Proposition 6 & Variance in payoff $Var[\pi]$ increasing in $\alpha$ if $\alpha$ is large enough \\ \hline
Proposition 7 & Risk averse payoff $\Pi(\alpha^* = 1) < \Pi(\alpha = 0)$ if $\sigma_z^2/\sigma_e^2$ is large enough \\ \hline
\end{tabular}
\label{table:LemmaPropMap}
\end{table}

The ML price $z_{k,t+1}$ for a focal home $k$ is given by $z_{k,t+1} = \bar{p}_{K,t}$ from equation \ref{eq:3}, where $K=\mathcal{C}(k)$ is the focal home's peer cluster at time $t$. We now focus on the focal home $k$ and drop the subscript $k$. The error in ML price is the difference between the actual realized home value $v_{t+1}$ and ML price $z_t$. This error can be decomposed into three components as,
\begin{align}
    v_{t+1} - z_{t+1} = \underbrace{v_{t+1} - v_t}_{\text{Random Walk}} + \underbrace{v_t - v_{K,t}}_{\text{Unpriced Features}} + \underbrace{v_{K,t} - \bar{p}_{K,t}}_{\text{Finite Sample Error}}
\end{align}
Home values in the current period $v_t$ can not forecast the random walk of preferences and values into the next period $v_{t+1}$. We denote variance of \textbf{random walk} error as an exogenous and constant quantity $Var[v_{t+1}-v_t] = \sigma^2_{rw}$.

We can interpret the clustering of homes into $Q$ clusters (described in Section 3.2) in terms of matching on homes' features (home characteristics such as its age and size). Specifically, given a total number of features $\bar{Q}$, we can view the homes within a cluster as being identical or matching on $Q$ features. In using the peer cluster's mean sale price, $\bar{Q} - Q$ unique features of the focal home are left unpriced i.e., an error $v_t - v_{K,t}$. Intuitively, the variance of \textbf{unpriced features} should depend on variance of all features and the number of priced features $Q$. The variance of all features, also the heterogeneity in housing stock, $\sigma^2_v$ is treated as an exogenous constant. The choice of priced features $Q$ explains increasingly greater proportion of the total variance. This is captured by monotonically decreasing function $h(Q)$ (possibly with positive second derivative because of diminishing returns). Thus we have $Var[v_t-v_{K,t}] = h(Q)\times\sigma^2_v$. Intuitively, as the number of ``clustering features'' (features used to place homes in the same cluster) increases, clusters will have fewer, very similar homes. Hence, the variance of home values in a cluster will be low. Similarly, as the number of ``clustering features'' decreases, clusters will have more, dissimilar homes. Hence, the variance of home values in a cluster will be high.\\

The ML price is effectively the sample mean of cluster sale prices. The \textbf{finite sample error} is the difference between the mean of cluster sale prices and the true cluster value i.e., $v_{K,t} - \bar{p}_{K,t}$. Since there are $N/Q$ home sales in the cluster, we can express $\bar{p}_{K,t}$ as,
\begin{align}
    \bar{p}_{K,t} = (Q/N) \times \sum_{k} \Big[ p_{k,t} \Big] \nonumber \\
     = (Q/N) \times \sum_{k} \Big[ \mathbb{E}[v_{i,k,t}] + \epsilon_{k,t} \Big] \nonumber \\
     = (Q/N) \times \sum_{k} \Big[ \mathbb{E}[(1-\alpha) \times \tilde{v}_{i,k,t} + \alpha \times z_{k,t}] + \epsilon_t \Big] \nonumber \\
     = (Q/N) \times \sum_{k} \Big[ (1-\alpha)\times v_{k,t} + \alpha \times (v_{k,t} + e^z_{k,t}) + \epsilon_{k,t} \Big] \nonumber \\
     = (Q/N) \times \sum_{k} \Big[ v_{k,t} + \alpha e^z_{k,t} + \epsilon_{k,t} \Big] \nonumber \\     
     = v_{K,k,t} + \alpha (Q/N) \times \sum_{k} e^z_{k,t} + (Q/N) \times \sum_{k} \epsilon_{k,t} 
\end{align}
Note that under $\alpha = 0$, the second components disappears. Under $\alpha > 0$, this additional component captures the \textit{confounding} of the sale price from the ML price.

We can now express the variance of the finite sample error as,
\begin{align}
    Var[v_{K,t} - \bar{p}_{K,t}] = Var \Big[(Q/N) \times \sum_{k} (\alpha \mathbb{E}[ e^z_{k,t}] + \epsilon_{k,t}) \Big] \nonumber \\
    = \alpha^2 Var[(Q/N) \times \sum_{k} e^z_{k,t}] + Var[(Q/N) \sum_{k} \epsilon_{k,t}] \nonumber \\
    = \alpha^2 Q \sigma_z^2 / N + \delta Q (1 -  \alpha)^2 \sigma^2_e / N
\end{align}

We can now write the full ML price error variance as,
\begin{align}
    \sigma_z^2 (\alpha) = Var \Big[ (v_{t+1} - v_t) + (v_t - v_{K,t}) + (v_{K,t} - \bar{p}_{K,t}) \Big] \nonumber\\
    = \sigma^2_{rw} + h(Q)\times\sigma^2_v + \alpha^2 Q \sigma_z^2 (\alpha)/N + (1 -\alpha)^2 \delta Q \sigma^2_e/N \nonumber\\
    = \Big( \frac{1}{1 - Q\alpha^2/N} \Big) \times (\sigma^2_{rw} + h(Q)\times\sigma^2_v + (1 -\alpha)^2 \delta Q \sigma^2_e/N) \nonumber\\
    = \sigma^2_{rw} + h(Q)\times\sigma^2_v + (1 -\alpha)^2 \delta Q \sigma^2_e/N
    \label{eq:sigma_z}
\end{align}
The denominator $(1 - Q\alpha^2/N)$ is very close to 1 since $Q << N << \infty$ and set to 1 going forward as a conservative assumption for analytical simplification. The ML error is increasing in random walk of home preferences $\sigma^2_{rw}$, heterogeneity in housing stock $\sigma^2_v$ and error in private valuations $\sigma^2_e$. The ML error sensitivity to $Q$ is mixed - the unpriced feature error component $h(Q)\times\sigma^2_v$ is decreasing in $Q$ while the finite sample error component $(1 -\alpha)^2 \delta Q \sigma^2_e/N$ is increasing. We will discuss endogenization of $Q$ later in this section. The ML error is increasing in $\alpha$. In fact, we can express the ML error in terms of the \textit{un-confounded} ML error $\sigma_z^2 = \sigma_z^2 (\alpha = 0)$ as,
\begin{align}
    \sigma_z^2 (\alpha) = \sigma_z^2 +  \alpha(2 - \alpha) \delta Q \sigma^2_e / N  \nonumber \\
    \text{where \quad} \sigma_z^2 = \sigma_z^2(\alpha = 0) = \sigma^2_{rw} + h(Q)\times\sigma^2_v + \delta Q \sigma^2_e / N
    \label{eq: result1a}
\end{align}
The un-confounded ML error $\sigma_z^2$ are valid in a limited setting where the platform does not reveal the ML price or the buyers-sellers do not use the ML price at all. In the rest of the paper, we will continue to compare confounded results at $\alpha > 0$ with unconfounded results at $\alpha = 0$ to highlight impact of the confounding over the feedback loop. In comparing expressions for confounded ($\sigma_z^2 (\alpha > 0)$) and unconfounded ($\sigma_z^2 = \sigma_z^2(\alpha = 0)$) ML errors, the additive term $\alpha(2-\alpha)\delta Q \sigma^2_e / N$ captures the amplification in ML error because of the confounding.

\proposition{The ML error ($\alpha > 0$) is strictly greater than un-confounded ML error ($\alpha = 0$) i.e., $\sigma_z^2(\alpha) > \sigma_z^2 = \sigma_z^2(\alpha = 0)$.} \normalfont \\

If this ML error $\sigma_z^2(\alpha)$ were directly observable, a data scientist could set $Q$ to minimize $\sigma_z^2(\alpha)$. Unfortunately, the ML price error $e^z_t$ was defined as $z_{t+1} - v_{t+1}$, where the true home value $v_t$ is not observed by the platform and their data scientist. In practice, the Machine Learning platform evaluates its algorithm by comparing its ML price with the eventual sale price i.e., $\hat{e}^z_{t+1} = z_{t+1} - p_{t+1}$. The estimated ML price error can be expressed as,
\begin{align}
    \hat{\sigma}_z^2 (\alpha) = Var [z_{t+1} - p_{t+1}]
    = Var [(z_{t+1} - v_{t+1}) + (v_{t+1} - p_{t+1})] \nonumber\\
    = Var [e^z_{t+1} + (v_{t+1} - (v_{t+1} +  \alpha e^z_{t+1} + \epsilon_{t+1}))] \nonumber\\
    = Var [(1-\alpha) e^z_{t+1} - \epsilon_{t+1}] \nonumber\\
    = (1 - \alpha)^2 \times \sigma_z^2 (\alpha) + (1 - \alpha)^2 \times \delta \sigma_e^2
    \label{eq:sigma_hat}
\end{align}

Under $\alpha = 0$, $\hat{\sigma}_z^2 > \sigma_z^2$ because $\delta \sigma_e^2 > 0$. This means that the platform is under reporting its accuracy\footnote{This is result at $\alpha = 0$ is contrary to the final result in this paper at $\alpha > 0$ which concludes that the platform over-reports its accuracy and confidence.}. There is no cause for alarm since the ML platform is acting in a conservative fashion. But, for $\alpha > 0$, when comparing true ($\sigma_z^2 (\alpha)$) and estimated ML errors ($\hat{\sigma}_z(\alpha)^2$), the conclusion is not trivial. The additive term $\delta \sigma_e^2$ is same as in the unconfounded setting. But, the fraction $(1-\alpha^2)$ captures the secondary effect due to the ``self fulfilling prophecy" over the feedback loop. This secondary effect may dominate when $\alpha$ is large enough. The platform will be presenting an overly optimistic claim of accuracy to the buyers-sellers.

\proposition{The ML error estimate is less than the ML error i.e., $\hat{\sigma}_z^2 (\alpha) < \sigma_z^2 (\alpha)$ when $\alpha > \alpha_1$ where $\alpha_1$ is unique solution (guaranteed to be in $(0,1)$) to,
\begin{align}
    \frac{\sigma_e^2}{\sigma_z^2 (\alpha)} = \frac{1}{\delta} \times \Big(\frac{1}{(1-\alpha)^2} - 1 \Big)
    \label{eq: result1b}
\end{align}
At $\alpha = 0$ (un-confounded setting) ML error estimate is strictly greater than the ML error i.e., $\hat{\sigma}_z^2(\alpha = 0) > \sigma_z^2(\alpha = 0)$. At $\alpha = 1$, $\hat{\sigma}_z^2 (\alpha) = 0$}  \normalfont \\

Let us now consider the choice of priced features $Q$. First, consider the un-confounded setting $\alpha = 0$. $Q$ would be set to minimize $\hat{\sigma}_z^2$. The unpriced feature error $h(Q)\sigma_v^2$ is decreasing in $Q$ while the finite sample error $\delta Q \sigma_e^2 /N$ is increasing in $Q$. Let $Q=Q^*_{\alpha = 0}=Q^*$ minimizes the error $\hat{\sigma}^2_z$, effectively trading-off these two components. At $Q=Q^*=$ we have, 
\begin{align}
    - \frac{\partial }{\partial Q} (h(Q) \sigma_v^2) |_{Q=Q^*} = \frac{\partial }{\partial Q} (\delta Q \sigma_e^2 / N) |_{Q=Q^*}
    \label{eq:derivativesQ}
\end{align}
Any $Q<Q^*_{\alpha = 0}$ would increase the sample size $N/Q$ within a cluster and reduce the finite sample error, but the larger cluster size comes with more unpriced home features and a wider range of heterogeneous homes within the cluster. On the other hand, any $Q>Q^*_{\alpha = 0}$0 would better distinguish unique homes at the cost of an erratic estimate of the mean cluster price (due to the smaller sample). Since $\sigma^2_e$ is independent of $Q$, the choice of priced features $Q$ simultaneously maximizes both the true ML price error $\sigma_z^2$ and its estimate $\hat{\sigma}_z^2$. Both these points confirm that there is no cause for alarm and the formulation captures conventional wisdom.\\

Now consider the confounded setting ($\alpha > 0$), the unpriced feature error remains constant while the finite sample error variance reduces. At the same $Q=Q^*$ the comparison of derivatives (similar to equation \ref{eq:derivativesQ}) now favors the unpriced feature error i.e., this component diminishes faster with increasing $Q$. As a result, the the unpriced feature error component will dominate more than before in determining the new $Q=Q^*_{\alpha}$ that minimizes the error $\hat{\sigma}^2_z(Q,\alpha)$. Thus, endogenously setting $Q$ would result in more clusters ($Q^*_{\alpha} > Q^*$), smaller cluster size, fewer home sales in every cluster ($N/Q$) and more adverse confounding from feedback. For simplicity, we take the conservative assumption that $Q$ is held constant under the feedback loop at $Q=Q^*$ i.e., to minimize $\hat{\sigma}^2$.\\

Thus far we have summarized ML price error (true vs. estimated, confounded vs. un-confounded) for an exogenous level of reliance on ML price $\alpha$. We conclude by examining how the estimated ML price error changes with $\alpha$.

\proposition{The ML price error estimate $\hat{\sigma}_z^2(\alpha)$ is decreasing in $\alpha$ when $\alpha > \alpha_2$ where,
\begin{align}
    \alpha_2 = 1 - \sqrt{\frac{1}{2} + \frac{N}{2Q} + \frac{N}{2 Q \delta}\frac{\sigma_z^2}{\sigma_e^2}}
\end{align}
} \normalfont

Intuitively, we expect that buyers-sellers give more weight to the ML price (larger $\alpha$) if the ML price is presented with lower error estimate ($\hat{\sigma}_z^2 (\alpha)$). When ($\alpha > \alpha_2$) error estimate ($\hat{\sigma}_z^2 (\alpha)$) is decreasing in $\alpha$, we may have an alarming re-inforcing ``self fulfilling prophecy". In order to formulate this, we will next formally endogenize $\alpha$ and formulate the equilibrium ML price reliance $\alpha^*$ and resulting ML price error $\sigma_z^2(\alpha),\hat{\sigma}_z^2(\alpha)$. 

\subsection{Feedback Loop Equilibrium}

To endogenize ML reliance $\alpha$, note that individuals have knowledge of the error in their own private signal $\sigma^2_e$ and the ML platform provides estimated ML price error $\hat{\sigma}_z^2$. Individual can weigh the two signals based on relative noisiness. For example, if ML price is accurate (small $\hat{\sigma}_z^2$) the individual can rely less on their private valuation. Thus, the reliance $\alpha$ can be endogenized as,
\begin{align}
    \alpha = \frac{\sigma_e^2}{\sigma_e^2 + \hat{\sigma}_z^2}
    \label{eq:alpha}
\end{align}
Now consider the feedback loop between $\hat{\sigma}_z^2$ and $\alpha$. A low estimated ML price error $\hat{\sigma}_z^2$ increases weight on ML price $\alpha$. This in turn shifts the $\hat{\sigma}_z^2$. Proposition 3 captures the critical range of $\alpha$ above which $\hat{\sigma}_z^2$ is decreasing in $\alpha$ and therefore the feedback loop is re-inforcing. Now we can substitute $\alpha$.

\proposition{At all feedback loop equilibria $\alpha^*$, the estimated ML price error is less than true ML price error $\hat{\sigma}_z^2 (\alpha) < \sigma_z^2 (\alpha)$ i.e., $\alpha^* > \alpha_1$ (see Proposition 2). Also, at all equilibria $\alpha^*$, the estimated ML price error is decreasing in $\alpha$ i.e., $\alpha^* > \alpha_2$ (see Proposition 3). \footnote{Detailed proof in Appendix \ref{app: proofs} by substituting $\alpha$ from equation \ref{eq:alpha} into Proposition 2 and 3 respectively.}
}  \normalfont \\

We can also formulate all equilibria $(\hat{\sigma}_z^2(\alpha)^*,\alpha(\hat{\sigma}_z^2)^*)$ using equations $\ref{eq:sigma_hat}$ and $\ref{eq:alpha}$.

\proposition{The feedback loop equilibria are given by solutions to\footnote{Detailed proof in Appendix \ref{app: proofs}.},
\begin{align}
\frac{\sigma_z^2}{\sigma_e^2} + \delta +  \frac{Q\alpha(2 -\alpha)}{N} = \frac{1}{\alpha (1 - \alpha)}
\end{align}
Full reliance on ML price ($\alpha^* = 1$) and the estimated ML price error collapsed to zero ($\hat{\sigma}_z^* = 0$) is always a solution to this and an equilibrium of the feedback loop. Full reliance on ML price ($\alpha^* = 1, \hat{\sigma}_z^* = 0$) is also the only equilibrium if ,
\begin{align}
    \frac{\sigma_z^2}{\sigma_e^2} < 4 - \delta - \frac{\delta Q}{N}
\end{align}
} \normalfont

\textbf{Hypothetical Example}: Let us consider a hypothetical parameter setting to understand these results better. Say a new Machine Learning algorithm is introduced in the market. Before revealing the ML prices to the buyer-sellers, the platform evaluates the performance and estimates the ML error to be $\hat{\sigma}_z = \$20,000$ ($\hat{\sigma}_z^2 = 4\times10^8$) on \$1Mn homes. While the platform can not measure the true error, let us assume that the true error is $\sigma_z^2 = 3.5\times10^8$. In comparison, say individual buyer-sellers make an error $\sigma_e = \$10,000$ ($\sigma_e^2 = 1\times10^8$) when privately estimating the home value i.e., $\hat{\sigma}_z^2 = 4 \sigma_e^2$. The platform now starts revealing these ML prices on its website. Substituting $\hat{\sigma}_z^2$ in equation \ref{eq:alpha}, the reliance of buyer-sellers $\alpha$ on the ML price rises from 0 to 0.2. This is relatively low because the ML price error is much larger than participant's private valuation errors. Substituting $\alpha$ in equation \ref{eq:sigma_hat}, $\hat{\sigma}_z^2$ shrinks by a factor of 0.64 i.e., from $4 \sigma_e^2$ to $2.56 \sigma_e^2$. Again substituting $\hat{\sigma}_z^2$ in equation \ref{eq:alpha}, this underestimation of the ML error increases individuals' reliance on the ML price from 0.2 to 0.28. The feedback loop repeats until reliance and the ML error are in equilibrium at $\alpha^*=0.5$ and $\hat{\sigma}_z^{2^*} = \sigma_e^2$. At this equilibrium the estimated ML error is at least 3.5 times smaller than the true ML error. While this is alarming, this feedback cycle did not collapse all the way to $(\hat{\sigma}_z^2,\alpha)^* = (0,1)$ because the original ML error (unconfounded or before introduction) was large. The feedback cycle is more acute if the original ML error is low to begin with say $\hat{\sigma}_z^2 = 2 \sigma_e^2$. In this case at equilibrium individuals eventually rely entirely on the ML price $\alpha^*=0 \rightarrow ... \rightarrow 1$ and estimated ML error collapses $\hat{\sigma}_z^2 = 2 \sigma_e^2 \rightarrow ... \rightarrow 0$.\\

\textbf{Intuition using Toy Scenario}: Consider first a baseline in absence of ML pricing. Individuals construct private valuations for a product (e.g., a $v=\$100$ painting may have private valuations $\tilde{v}_i$ distributed in $\$100 \pm 20$). Individuals then enter a room and spend time to learn from each other in the crowd to resolve disagreement in their private valuations. While time consuming, correction is possible since private valuations are unbiased ($v=\mathbb{E}[\tilde{v}_i]$) and uncorrelated ($\mathbb{E}[(\tilde{v}_i-v)*(\tilde{v}_j-v)] = 0$) across individuals. The learning from crowd mitigates error in valuations. Subsequently the product seller contracts with a willing buyer (both of whom have improved their valuation in this process) on a sale price ($p$ realized in $\$100 \pm 4$) that deviates a lot less than its true worth ($E[|p-v|] < E[|\tilde{v}_i - v|]$).\\ 

Introduction of an accurate ML price ($z=\$100$) provides a common signal to all. Now individuals can construct their valuation by placing some reliance or weight ($\alpha = 0.5$) on the ML price ($\tilde{v}_i := (1-\alpha)\tilde{v}_i + \alpha z$). The new valuations are more narrowly distributed in $\$100 \pm 10$. This alleviates some of the costly valuation disagreement among individuals. However, if the ML price has error ($z=\$110$) this is universally propagated to all individuals ($\tilde{v}_i$ in $\$105 \pm 10$). Going forward we treat error in valuations ($\tilde{v}_i-v$) as split into – (i) component capturing disagreement from the crowd ($\tilde{v}_i-\mathbb{E}[\tilde{v}_i$]) called variance and (ii) component common across the crowd ($\mathbb{E}[\tilde{v}_i]-v$) called bias\footnote{One could argue that the true product value moves up from $\mathbb{E}[v_i] \rightarrow \mathbb{E}[\tilde{v}_i]$ if an error systematically moves everyone in the market to value the product more. So, bias would always be zero by definition. But such contamination by systematic error will not sustain indefinitely. Thus, our bias measurement captures a short-run systematic error in valuations for one product. This bias (positive or negative) is not in the same direction across all products in the market. So, it does not represent inflation or deflation of (housing) market as a whole. The ML model bias-variance and product valuation bias-variance measure different quantities, but they are intricately related in our model. The ML model bias-variance will drive valuation bias-variance.}. In the example above, introduction of ML pricing reduced variance but added bias. Participants can correct variance (disagreement or random error) in private valuation via learning from the crowd, but they cannot correct bias (common or systematic error). Thus, any valuation bias added by ML price gets propagated to sale prices ($\tilde{p}_i$ realized in $\$105 \pm 2$). Whether the sale price errors\footnote{The economic significance of statistical error in valuations and sale prices become clear once we elaborate payoffs.} ($|\tilde{p}_i-v|$) is alleviated or amplified depends on the size of ML error ($|e_z|=|z-v|=10$) and reliance on ML price ($\alpha=0.5$). ML price alleviates error if reliance is optimal given the ML error (e.g., low reliance under large error or high reliance under small error) but it amplifies error if reliance is inflated (e.g., high reliance under large error). In fact, this impact is true of any signal, say real estate expert opinion, that is widely influential in the market.\\ 

The uniqueness of the ML price signal becomes apparent once we endogenize reliance or weight on the ML price $\alpha$. The individuals determine reliance $\alpha$ by observing ML error estimate $\hat{\sigma}_z$ presented alongside the ML price. The ML error $\hat{\sigma}_z$ is endogenously estimated in the ML framework by comparing ML prices with sale prices. When buyer-sellers rely on the ML price to any extent ($\alpha>0$), the sale prices settle closer to the ML price than they would if the ML price were hidden ($\alpha=0$). As a result, ML error $\hat{\sigma}_z$ is underestimated. The underestimation of the ML error $\hat{\sigma}_z$ inflates reliance $\alpha$. In turn, an increase in reliance $\alpha$ further underestimates the ML error $\hat{\sigma}_z$ due to the self-fulfilling nature of the ML price, which leads to worsening over reliance, and so on as the feedback loop iterates\footnote{Hypothetically if a real estate expert, say Warren Buffet, becomes increasingly boisterous and influential as their prophecies are fulfilled, it would lead to the same feedback loop mechanism.}. At equilibrium with deceptively large ML error and over-reliance $\alpha^*$, the ML price can overall amplify sale price errors. This reinforcing loop depends on – (i) limited rationality of buyer-sellers who do not realize the underestimation in $\hat{\sigma}_z$ and (ii) passive behavior of platform that presents the ML price without correcting the underestimation in $\hat{\sigma}_z$. In Section 6, we will qualitatively discuss when these two assumptions stop holding true and how they moderate (but don't eliminate) the problem.

\subsection{Payoffs}
Until now we have examined the statistical properties of the ML price at the feedback loop equilibrium. In this section, we discuss the implications for seller's payoffs. First, we will examine the payoff for an exogenous level of reliance on ML price $\alpha$. This will shed light on how the payoff varies as the feedback loop strength increases. Finally, we will consider the payoff at the extreme feedback loop equilibrium $\alpha^* = 1$.\\

In our model \textit{price noise} in the market $Var[\tilde{v}] = \sigma^2_e(\alpha)$ (which depends on ML reliance $\alpha$) impacts the seller through two channels. First channel is the heterogeneity in buyer offers. The seller can benefit from this \textit{price noise} if they can wait long enough for a high draw from the offer distribution. Second channel is noise in seller's guess. The seller has disutility from this because they may not enter the market altogether when it may have been profitable to do so or list the home at too low a price. In our simple model, the seller spends cost $c$ in the first period and resolve this noise in pricing entirely. If the cost $c$ is small enough, the seller benefit from the first channel dominates over disutility from the second channel. We can model the cost $c$ as linearly growing with seller's price uncertainty $\sigma_e(\alpha)$ as $c = \kappa \sigma_e(\alpha)$.

\lemma{The expected seller payoff is given by,
\begin{align}
\mathbb{E}[\pi(\mu)] = \mu - \Gamma(\kappa) \times \sigma_e(\alpha) \text{\quad where \quad}  \nonumber \\
\text{where \quad} \Gamma(\kappa) = \kappa + (5/3)*(3/2)^{1/4} \sqrt{\kappa} - (1/3)*(3/2)^{1/2}
\end{align}
The expected seller payoff is decreasing in $\sigma_e$ if (proof in Appendix \ref{app: proofs}),
\begin{align}
    \frac{\partial \mathbb{E}[\pi]}{\partial \sigma_e(\alpha)} < 0 \implies \Gamma(\kappa) > 0
\end{align}
\label{lemma:2}
} \normalfont
Thus, expected seller payoff is decreasing in $\sigma_e$ when market participation cost $c$ (and therefore $\kappa$) is high\footnote{Numerically $\Gamma(\kappa) > 0$ solves to $\kappa > 0.04$ which is a fairly relaxed requirement.}. We consciously choose to only examine this high cost parameter range such that the seller has an overall disutility from \textit{price noise}\footnote{We believe this better reflects individual sellers. It is a conservative assumption with respect to our claim about negative implications of the feedback loop.}. From Lemma \ref{lemma:1}, \textit{price noise} $\sigma_e(\alpha)$ is monotonically decreasing in level of reliance $\alpha > 0$ on the Machine Learning price as\footnote{We follow the convention that $\sigma_e^2(\alpha=0)$ is denoted simply by $\sigma_e^2$.}, 
\begin{align}
    \sigma_e^2(\alpha) = Var[\tilde{v}] = (1-\alpha)^2 \sigma_e^2
\end{align} 
In fact at $\alpha^* = 1$ we have $\sigma_e(\alpha^*=1) = 0$ and consequently the expected seller payoff is maximized at $E[\pi] = \mu$. Surprisingly, this suggest that the Machine Learning feedback loop and the resulting over-reliance on ML prices $\alpha$ appears to have no negative consequence on seller's payoff. In order to fully understand the impact of the ML feedback loop, let us examine the resulting error in valuations and sale prices. The error in valuations can be formulated as,
\begin{align}
    E[(\tilde{v}_i - v)^2] = \underbrace{E[(\tilde{v}_i - E[\tilde{v}_i])^2]}_{\text{Price Noise}} + \underbrace{E[(E[\tilde{v}_i] - v)^2]}_{\text{Price Bias}}
\end{align}
While reliance on Machine Learning $\alpha$ is reducing \textit{price noise}, it is increasing \textit{price bias} because $E[(E[\tilde{v}_i] - v)^2] \neq 0$. From Lemma 1, we have $Var[p] = \delta \times Var[\tilde{v}_i]$. Using this, we can write error in sale price as,
\begin{align}
    E[(p - v)^2] = \delta \times \underbrace{E[(\tilde{v}_i - E[\tilde{v}_i])^2]}_{\text{Price Noise}} + \underbrace{E[(E[\tilde{v}_i] - v)^2]}_{\text{Price Bias}} 
\end{align}
The \textit{price noise} noise is reduced ($\delta < 1$) as buyers and sellers learn and converge towards a consensus. Note that these sale price errors $(p-v)$ are not correlated across homes i.e., any random sample of homes will not have systematic upward or downward error. In other words, Machine Learning is not causing systematic ``price bubbles" in the market. While ML is not causing price bubble, it may be increasing randomness to the sale prices and therefore the payoffs.

\lemma{The error in valuations, sale prices and variance in payoffs are given by,
\begin{align}
E[(\tilde{v}_i - v)^2] = (1-\alpha)^2 \sigma_e^2 + \alpha^2 \sigma_z^2(\alpha) \nonumber \\
E[(p - v)^2] = \delta \times (1-\alpha)^2 \sigma_e^2 + \alpha^2 \sigma_z^2(\alpha) \nonumber \\
Var[\pi] = \Omega(\kappa;\beta) (1-\alpha)^2 \sigma_e^2 + \alpha^2 \sigma_z^2(\alpha) \nonumber \\
\small{\text{ where \quad } \delta = (1/6) \text{ \quad and \quad } \Omega(\kappa;\beta) = \beta_0 + \kappa^{1/2} \beta_1 + \kappa \beta_2 + \kappa^{3/2} \beta_3}
\end{align}
} \normalfont 

\proposition{The error in valuations $E[(\tilde{v}_i - v)^2]$, error in sale prices $E[(p - v)^2]$ and variance in payoff $Var[\pi]$ is increasing with reliance $\alpha$ when,
\begin{align}
 \alpha > \frac{\sigma_e^2}{\sigma_e^2 + \sigma_z^2} \quad , \quad \alpha > \frac{\sigma_e^2}{\sigma_e^2 + \sigma_z^2/\delta} \quad , \quad \alpha > \frac{\sigma_e^2}{\sigma_e^2 + \sigma_z^2/\Omega}   
\end{align}
respectively. Equivalently we can express as condition on confounded ML price error estimate $\hat{\sigma}_z^2(\alpha)$ dropping below $\sigma_z^2,  \sigma_z^2 / \delta $ and $\sigma_z^2 / \Omega$ respectively over the feedback loop.
} \normalfont

More randomness in the seller’s payoff $Var[\pi]$ means that the seller may get very lucky or unlucky. This is akin to the seller gambling on a coin toss instead of a deterministic payoff. Randomness is not an issue for risk-neutral sellers but is undesirable to risk-averse sellers. Consider a constant absolute risk-averse (CARA) seller with a concave utility $u(\pi)=1-e^{-a\pi}$ corresponding to a constant risk-aversion coefficient $a$. The expected utility $\Pi = E[u(\pi)]$ for this CARA seller is linearly decreasing in $Var[\pi]$ as,
\begin{align}
\Pi(a) = \mathbb{E}[\pi] - 0.5 a \times \sqrt{Var[\pi]} \nonumber \\
\Pi(a) = \mu - \Gamma(\kappa) \times \sigma_e(\alpha) - 0.5 a \times \sqrt{\Omega(\kappa) \sigma_e^2(\alpha) + \alpha^2 \sigma_z^2(\alpha)} \nonumber \\
\Pi(a) = \mu - \sigma_e(\alpha) \times \Big[ \Gamma(\kappa) + 
\frac{a}{2} \sqrt{\Omega(\kappa) + \alpha^2 \frac{\sigma_z^2(\alpha)}{\sigma_e^2(\alpha)}} \Big]
\label{eq: riskAversePayoff}
\end{align}
We can break down payoffs $\Pi$ implications into two forces – (a) ML price reduces \textit{price noise} $\sigma_e(\alpha)$. This improves $\mathbb{E}[\pi]$ by minimizing need for a slow and costly learning among the crowd to resolve disagreements. Further, the realized sale prices (and therefore payoffs) have lower variance because buyer-sellers inherently have lower disagreement at the beginning. This is the positive force. (b) But ML price, under strong feedback loop, has a deceptively large error ($\sigma_z^2(\alpha)/\sigma_e^2$ inflates) and therefore a large valuation bias ($E[\tilde{v}] - v$) which makes the market resemble a coin toss or lottery (increased $Var[\pi]$) where the seller may arbitrarily get a lucky or unlucky draw. This is the negative force. In essence ML price replaces the ``slow crowd learning” nature of the market with one that resembles a ``quick lottery”. Note that the exogenous parameters $a,\Gamma(\kappa),\Omega(\kappa)$ are all positive. We can compare payoff with ML (at $ \alpha^*=1$ equilibrium) and without ML as,
\begin{align}
\Pi(\alpha = 0) > \Pi(\alpha = 1) \nonumber \\
\mu - \sigma_e(\alpha = 0) \times \Big[ \Gamma(\kappa) + 
(a/2) \sqrt{\Omega(\kappa)} \Big] > \mu - (a/2) \alpha \sigma_z(\alpha = 1)
\end{align}

\proposition{The risk neutral payoff $\Pi(a=0)$ is always increasing in reliance $\alpha$ (and severity of feedback loop). The risk averse payoff $\Pi(a > 0)$ at feedback loop equilibrium ($\alpha^* = 1$) is worse off than risk averse payoff under no ML if,
\begin{align}
    \frac{\sigma^2_z}{\sigma^2_e} > \Big( \sqrt{\Omega(\kappa)} + \frac{2\Gamma(\kappa)}{a} \Big)^2 - \frac{\delta Q}{N}
    \label{eq: prop7eq2}
\end{align}
} \normalfont

The condition is relaxed (i.e., payoff under ML more worse off), under - higher risk aversion parameter $a$ and lower $\kappa$ (smaller $\Gamma(\kappa),\Omega(\kappa)$). The role of remaining exogenous parameters ($\sigma_z,\sigma_e,\delta,Q$ and $N$) is mixed. They determine whether $\alpha^*=1$ is the only equilibrium (Proposition 5) and favor-ability of risk averse payoff at this extreme equilibrium (Proposition 7). The implications are mixed because these exogenous parameters may make the $\alpha^* = 1$ equilibrium condition relaxed (tighter) but makes the unfavorable payoff condition at the $\alpha^* = 1$ equilibrium tighter (relaxed). The conditions in Proposition 5 and Proposition 7 are simultaneously satisfied if,
\begin{align}
   \Big( \sqrt{\Omega(\kappa)} + \frac{2\Gamma(\kappa)}{a} \Big)^2 - \frac{\delta Q}{N} < \frac{\sigma^2_z}{\sigma^2_e} < 4 - \delta - \frac{\delta Q}{N}
\end{align}
This is viable if,
\begin{align}
   \Big( \sqrt{\Omega(\kappa)} + \frac{2\Gamma(\kappa)}{a} \Big)^2 - \frac{\delta Q}{N} < 4 - \delta - \frac{\delta Q}{N} \nonumber \\
   \Big( \sqrt{\Omega(\kappa)} + \frac{2\Gamma(\kappa)}{a} \Big)^2 + \delta < 4
\end{align}
Exogenous parameter settings satisfying the condition above suggest that $\alpha^*=1$ is the only equilibrium and seller payoff is worse than no ML at this equilibrium. In the next section, we elaborate on the exogenous factors - choice of Machine Learning model (which depends on $N$ and $Q$, and drives $\sigma_z^2$), market characteristic ($\delta$), and seller characteristics (risk aversion $a$, cost $(c,\kappa)$ and noise in private valuations $\sigma_e$). 

\section{Exogenous Factors}
Conventional wisdom would suggest that ML price as a source of information has larger benefits when - ML model has access to more training data (large $N$ and $Q$) and has high capacity (small $\sigma_z$), and sellers' have large errors in their private valuations (large $\sigma_e$). We find that all of these circumstances lead to a ``stronger" feedback loop potentially collapsing to full reliance equilibrium ($\alpha^* = 1$, Proposition 5). This kicks in the adverse payoff implications of the feedback loop. The adverse implications are minimal for sellers' that are impatient (large $\kappa$) and risk-neutral ($a=0$).\\

\textbf{Training Data and ML Model} $(N,Q)$: Consider introduction of ML price in a hypothetical where the confounding feedback is absent. As the platform gets access to more data $N$ and more features describing each home, it can increase the number of features priced $Q$ by deploying a higher capacity ML model e.g., neural network with millions of tunable parameters instead of linear model with a dozen parameters. This reduces the un-confounded true ML price error $\sigma^2_z$. Consistent with conventional wisdom, this improves payoffs in our model. Next, consider ML price in presence of confounding feedback. Reduced $\sigma^2_z$ has the obvious positive effect of providing more accurate information, but it has a potentially negative effect arising from the feedback loop. If $\sigma_z$ is small enough, it satisfies the condition in Proposition 5 where the feedback loop equilibrium collapses to full reliance ($\alpha^* = 1,\hat{\sigma}_z^* = 0$). Conventional wisdom suggests that a more powerful Machine Learning model (more data, more training features and better un-confounded accuracy) is more beneficial to the market. Counter to this intuition, a higher capacity ML model results in stronger feedback loop that may collapse to full reliance (Proposition 5) and amplify pricing errors (large $\sigma_z^2(\alpha)$). \\

\textbf{Market Characteristics} $\delta$: The exogenous parameter $\delta$ is defined as the ratio $Var[p]/Var[\tilde{v}]$\footnote{Lemma 1 calculates a numerical value of $\delta = 1/6$ using the ``simple" distributional assumptions}. It captures the degree to which the noise or disagreement in valuations $Var[\tilde{v}]$ are resolved in the market. If $\delta = 0$, all disagreement is resolved and the home always sells at the consensus price i.e.,  $p=E[\tilde{v}]$ and $Var[p] = 0$. If $\delta$ is large, the interaction among buyers and sellers does not lead to any learning or consensus among the crowd. Intuitively, any error in ML price is propagated to all buyer-sellers. A high degree of consensus among the crowd (low $\delta$) eliminates noise (as well as useful signal) in the private valuations. But, this consensus leaves the error in ML price unmitigated because the error is a common signal across the market. This strengthen the feedback loop. Consistent with this intuition, a small $\delta$ improves the payoffs relative to no Machine Learning (Proposition 7) but it results in stronger feedback loop potentially collapsing to full reliance (Proposition 5).\\

\textbf{Homeowner Characteristics} ($a,c,\sigma_e$): In equation \ref{eq: riskAversePayoff} we broke down the impact of ML price into positive forces (a) that alleviates ``slow crowd learning” and negative force (b) that molds the market to resembles a ``quick lottery”. First consider seller risk aversion $a$. If sellers' are risk averse (high $a$) the negative force (b) dominates because randomness of payoff (large $Var[\pi]$) hurts the seller. Consistent with this intuition, Proposition 7 condition is relaxed under - higher risk aversion parameter $a$. A relaxed condition suggest that payoff under ML is worse off than no ML under a wider (relaxed) range of remaining exogenous parameters. Second consider seller market participation cost or impatience\footnote{Cost can come from better outside option e.g., other sources of income to replace the slow and costly market participation.}. If a sellers is impatient the positive force (a) dominates. We model cost on staying on the market as $c = \kappa \sigma_e$. Consistent with this intuition, Proposition 7 condition is relaxed under - higher cost parameter $\kappa$ (thus higher $\Gamma(\kappa),\Omega(\kappa)$).\\ 

Finally, consider noise in seller's private valuation $\sigma_e = \sigma_e(\alpha = 0)$. A small noise in private valuations means that the seller has high reliance on private valuations when ML price is introduced. This results in $\alpha^*=1$ not being the only feedback loop equilibrium i.e., condition in proposition 5 is tighter. Next, consider the two forces (a) and (b) driving risk averse payoff in equation \ref{eq: riskAversePayoff}. The positive force (a) does not add a lot of value because $\sigma_e(\alpha = 0)$ is small to begin with and the marginal value of reducing it further ($\sigma_e(\alpha) - \sigma_e(\alpha = 0)$) may not be large enough. The feedback loop can only have negative implications via force (b). Consistent with this intuition, Proposition 7 condition is relaxed (payoff under ML is more likely to be worse off than no ML).\\

The equilibrium results have been derived assuming that sellers are homogeneous. We can consider a thought experiment where an individual seller differs from the remaining homogeneous population on some characteristics. This thought experiment allows us to (at least directionally) characterize implications for home sellers with heterogeneous characteristics. Under a strong feedback (inflated ML price error and over-reliance on ML price) a seller that is more impatient (high $\kappa$), risk-neutral ($a=0$) and high ability to price\footnote{An individual with strong ``ability to price" is one who can accurately guess offers a home would receive and likely sale price. Such an individual does not rely a great deal on the ML price and unaffected by errors in ML price.} (low $\sigma_e$) has more to gain from ML prices. Notably, sophisticated investors (impatient, risk-neutral and high ability to price traditionally stayed away from the housing market. The introduction of ML pricing may favor and encourage entry of such sophisticated investors. This is loosely consistent with the entry of iBuyers (1\% of all US home purchases in 2019) and large real estate investors (18\% of all US home purchases in Q3 2021) in residential housing market \citep{katz2021real}.

\section{Discussion}
\textbf{Feedback Correction}: A platform that offers ML pricing has a few options to mitigate or correct the ML price error underestimation. One option is to measure the reliance $\alpha$ on ML prices and use it to correct the estimated ML error $\sigma_z^2$. To measure $\alpha$, the platform could run a randomized experiment in which the ML prices for some homes are hidden while others remain available, but buyer-sellers may not perceive the hidden information as a random occurrence, confounding the experimental results. Alternatively, the platform could add random, small, positive, or negative errors to some ML prices. In fact, in Appendix A.2, we measure $\alpha$ with a similar natural experiment (i.e., we leverage unintentional random errors in ML prices). But it may be more challenging for the platform to intentionally and regularly add errors for the sole purpose of experimentation. Such experiments are common on websites, e-commerce platforms, and search engines, but the scrutiny in the housing market may be prohibitive. A second and more conservative option is to calculate the ML price for a home using only a sample of historical sale prices for which the ML price was hidden from the market. Even with this conservative approach, some feedback may seep in indirectly. For example, when an ML price is not available, buyer-sellers may look up the ML prices of similar neighboring homes. Further, this strategy would require the platform to leave a fraction of homes without an ML price for un-confounded training samples. In summary there is a trade-off between un-confoundedness of the ML prices and number of homes where ML prices are visible.\\

\textbf{Platform Incentives}: We do not model the platform as an agent in the analytical model, therefore we are limited in discussing platforms incentives to correct. We provide here some plausible discussion on how correction strategies may impact platforms revenue sources. The platforms' major revenue source is to sell ads on the website to local brokers, agents, and other real estate services. ML price presented alongside a small error (further underestimated due to the feedback loop) likely increases perceived informativeness of the ML price, site visits and therefore ad revenue. As a result, the platform may not have incentive to limit visibility of ML prices in pursuit of making them un-confounded. Further, the platform may participate as an iBuyer: an entity that purchases, upgrades, and flips (re-sells) homes at scale. The iBuyer is – (i) risk-neutral because random gains or losses over hundreds of home transactions average out, (ii) has a strong ability to price thanks to the access to proprietary data and ML model, and (iii) impatient because they prefer to flip homes quickly instead of holding large inventory. As discussed in the last section, this trio of characteristics positions the iBuyer to gain from ML pricing and an uncorrected feedback loop. In short, the platform's two sources of revenue (ads and iBuyer) do not appear to incentivize the platform to correct the feedback and error underestimation. The platform may have other incentives to correct (e.g., the risk of long-term reputation damage, threat of regulations, or pure ethics), but we leave a more thorough examination as open questions for future research.\\

\begin{figure}
    \centering
    \includegraphics[width=0.7\textwidth]{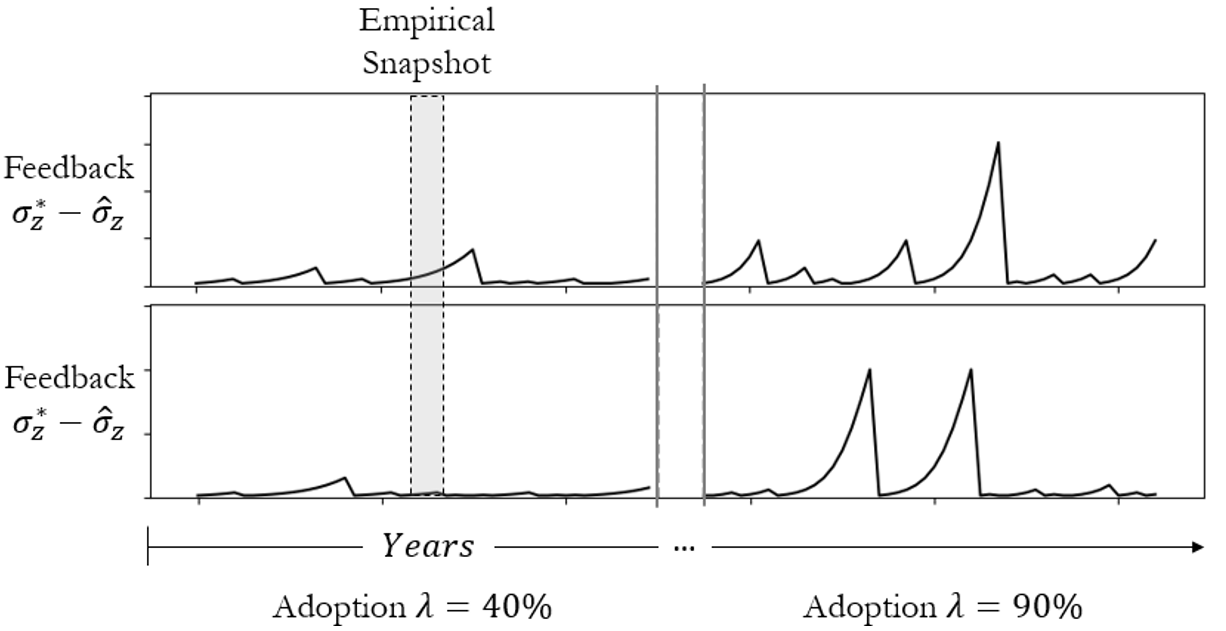}
    \caption{The simulated, cyclical formation and correction of feedback overconfidence in two submarkets with low (40\%) and high (90\%) rates of ML adoption. An empirical snapshot is likely too narrow to capture the cycles.}
    \label{fig: PriceBubbleSim}
\end{figure}

\textbf{Price Bubbles} are defined as systematically higher transaction price $p$ relative to an underlying true value $v$ of an asset (home). We have $E[p_k - v_k] = 0$ across the market at the feedback loop equilibrium. While there is no systematic ML price bubble across the entire market (or any large number of homes), the ML price may be above (or equally likely below) the true value for small sub-markets (e.g., 1500–2000 sq. ft. homes in one Austin neighborhood due to a dozen idiosyncratic inflated sales). Conventionally we expect the ML price in this sub-market to correct back down. But, the feedback loop may be gradually increasing degree of ML error underestimation ($\sigma_z - \hat{\sigma}_z >> 0$). If the feedback loop evolves toward ``self-fulfilling" ML prices ($\hat{\sigma}_z \rightarrow 0$) faster than ML prices can correct, it may result in a persistent over-pricing (or equally likely under-pricing) in the small sub-market. If this persistent over-pricing is very high (low), more and more sellers (buyers) will hesitate to deviate from the ML price until it becomes too difficult to make sales (purchases) at the ML price. Eventually this will result in a correction due to loss in trust in ML prices $\alpha = 0$. Modeling such correction is outside the scope of our paper. The analytical model is not informative about the time required to arrive at this ``self-fulfilling" equilibrium. Figure \ref{fig: PriceBubbleSim} depicts the simulated formation and correction of overconfidence ($\sigma_z - \hat{\sigma}_z$). Such formation and correction of overconfidence are uncorrelated across sub-markets. We may not observe the overconfidence throughout the housing market simultaneously.

We also expect expert opinions or model updates by the ML platform to correct or reset the ``self-fulfilling" feedback. Pricing bias in a sub-market (e.g., 5\% overpricing of 1500–2000 sq. ft. homes in one Austin neighborhood) persisting for 12–36 months should be identifiable by observing summary-level demand-supply data in comparable neighborhoods or by using knowledge of ML feedback (as described in this paper). However, market experts are not necessarily able to identify such bubbles. In fact, \citep{cheng2014wall} show that securitized home loan managers were unaware of the growing housing bubble in 2004–06, preceding the 2007–08 collapse. The ML feedback loop is a statistically complex phenomenon, like the risk pricing of securitized home loan assets, so it would not be surprising if pricing bias from the ML feedback loop remains opaque to experts for years.\\

\textbf{ML Price Trust and Adoption}: In equation \ref{eq:alpha} we modeled the seller's reliance on ML prices. We can express the reliance moderated by an exogenous lever $\lambda$ as,
\begin{align}
    \alpha = \lambda \times \frac{\sigma_e^2}{\sigma_e^2 + \hat{\sigma}_z^2}
\end{align}
At $\lambda = 0$, the buyer-sellers do not use the ML price at all even if its available. At $\lambda = 1$, the model represents a future state where ML price is available and used by all buyer-sellers in the rational fashion discussed above\footnote{The rationality is still bounded since the participant uses ML platform's estimate $\hat{\sigma}_z^2$ without adjusting for the platform's estimation limitation or presence of feedback confounding)}. All intermediate values of $\lambda$ can be roughly interpreted as different degree of trust and adoption of ML prices. This is more representative of the current state where some sellers' have not adopted (or simply do not observe the ML price at all) or simply underweight the the ML price. A lower level of trust and adoption moderates all implications of the ML Feedback Loop. The ML price error is underestimated ($\hat{\sigma}_z < \sigma_z$) but may not collapse entirely to $\hat{\sigma}_z = 0$. The over-reliance on ML prices $\alpha$ and increase in payoff variance $Var[\pi]$ will be moderated. The feedback overconfidence (Figure \ref{fig: PriceBubbleSim}) is relatively mild (severe) when the level of ML price adoption is low (high).\\

\textbf{Homeowner Rationality}: Our analytical model (Equation \ref{eq:alpha}) assumes that seller reliance on ML price $\alpha$ depends on the errors in private valuations $\sigma_e$ and estimated ML price error (presented by the platforms' like Zillow alongside the ML price). In doing so, the seller is not modeled as informed (or rational) about the ML feedback loop and the risk of under-estimated ML error and inflated true error i.e., $\hat{\sigma}_z(\alpha) < \sigma_z(\alpha)$. Let us consider a hypothetical where the seller with knowledge of the feedback loop (and all exogenous factors in our model), can determine the true ML error $\sigma_z(\alpha)$. This moderates sellers' reliance on ML price as 
\begin{align}
    \alpha = \frac{\sigma_e^2}{\sigma_e^2 + \sigma_z^2(\alpha)}
\end{align}
The feedback loop is not reinforcing anymore. While increasing $\alpha$ increases true ML error $\sigma_z^2(\alpha)$, increased $\sigma_z^2(\alpha)$ reduces $\alpha$. While this moderates but it does not eliminate the feedback loop. Following Proposition 1, the true ML error still remains inflated i.e., $\sigma_z^2(\alpha) > \sigma_z^2(\alpha = 0)$. Following Proposition 2, the estimated ML error can still be under-estimated i.e., $\hat{\sigma}_z^2(\alpha) < \sigma_z^2(\alpha)$. Homeowners' being fully rational about the ML Feedback Loop moderate their reliance on ML prices, but they still (knowingly) consume inflated errors (relative to un-confounded) in the ML price. Individual seller would prefer that ML prices were un-confounded i.e., ML prices are hidden for all other homes. But, no seller benefits from having ML price of their homes being hidden. The ``self fulfilling" nature of the feedback loop is moderated but not eliminated.\\ 

\textbf{Drivers}: There are three key ingredients for the findings. Our findings should be broadly applicable to any market with these ingredients. First, preferences for the product evolve smoothly over time and across a high dimensional product feature space (such that similar products have similar value). This motivates the use of an ML model to estimate prices as a function of home features, and it also necessitates \textbf{perpetual feedback in the training algorithm}. Second, most buyers and sellers participate rarely in the market (e.g., a homeowner sells once in a decade; an entrepreneur raises funds infrequently). This ensures that \textbf{buyer-sellers have a limited understanding of the ML model, feedback, and potential risks} \citep{schmit2018human}. Many markets meet these first two requirements and have known ML feedback loops. For example, the traffic routing model in Google Maps \citep{lau2020google} informs changes in driver behavior, which then are observed by Google Maps and used to update its ML routing model. In fact, feedback labels are useful in a wide range of online learning settings where an ML algorithm learns by making mistakes. In the ML routing model, a mistake is soon corrected because the resulting traffic congestion sends negative feedback to the ML model. Unfortunately, in the housing market, the feedback label is not visible. Third, individuals needs to guess how other individuals value a product. In housing, sellers' needs to guess how buyers will value their home. Individual buyers can determine their private valuation but need to guess how other buyers may value the product in case the current buyer needs to resell in the future. Such markets \textbf{lack ground truth} prices, the ML price influences both sides of the market, thus contaminating the resulting sale prices. As more buyer-sellers rely on ML pricing, the impartial ground truth is further obscured. This is true with products that have complex preferences e.g., housing, crowdfunding, peer-to-peer lending, art auction markets among others.

\section{Conclusion}
ML pricing is increasingly pervasive and purports high accuracy that can overcome pricing uncertainty and associated frictions in the market. But this argument presumes that the ML model was trained on large, independent ``ground truth" samples. In practice, ML training samples can be confounded by its own predictions, resulting in self-fulfilling feedback. We have shown that algorithm's self-reported confidence and the consumers reliance reinforce each other other due to the feedback loop resulting in over-confidence and over-reliance. We find that ML prices can increase the deviation of realized sales prices from the ``true'' home value, until realized sales prices (and ML prices) are entirely random. Second, we show conditions where this adversely effect the economic payoffs for sellers. Third, we identify seller characteristics where this equilibrium and its adverse economic implications are worse. Overall, our model suggests that introducing ML pricing into a market could adversely affect the economic outcomes.\\

The self-fulfilling ML feedback loop has similarities with the phenomena of “echo chambers” and “filter bubbles” in the personalized social media context \citep{pariser2011filter}, where ML models continuously learn from user behavior while also influencing those behaviors. Such ML models tend to reinforce selective preferences, attaining good prediction accuracy (likelihood that user clicks on recommended content) but potentially losing sight of long run outcomes. We expect that regulators eventually may have to enforce ML model controls, similar to asset pricing models in financial markets. We hope future research will investigate policies to incentivize the correction of ML feedback loops. Also, our limited data set enabled us only to verify primitives for the analytical model, but long-term field research is critical for uncovering how the introduction of ML pricing affects housing market characteristics.

\bibliographystyle{emaad}
\bibliography{Zillow_Feedback_Loop.bib}

\newpage

\begin{appendices}
\begin{table}[]
\centering
\caption{Table of Content for Appendix}
\begin{tabular}{|l|l|l|}
\hline
\textbf{Section}   & \textbf{Subsection} & \textbf{Title}                           \\ \hline
\multirow{4}{*}{A} &                     & Empirical Evidence to support analytical model primitives                      \\ \cline{2-3} 
                   & A.1                 & Data Description                         \\ \cline{2-3} 
                   & A.2                 & Forward Loop: Zestimate Impact on List and Sale Prices \\ \cline{2-3} 
                   & A.3                 & Backward Loop: Zestimate Calculation from Sale Prices   \\ \hline
\multirow{4}{*}{B} &                     & Support for analytical model simplifications                            \\ \cline{2-3} 
                   & B.1                 & Buyer-Seller Bargaining                  \\ \cline{2-3} 
                   & B.2                 & Full Model                             \\ \hline
\multirow{4}{*}{C} &                     & Proofs   \\ \cline{2-3} 
                   & C.1                 & Proofs for Lemma \\ \cline{2-3} 
                   & C.2                 & Proofs for Propositions  \\ \hline
\end{tabular}
\end{table}

\section{Empirical Evidence}  
\subsection{Data Description}

We use housing market data from Zillow\comment{\footnote{Data collected by Runshan Fu, NYU Stern School of Business.}}, which is an online real estate database company. Zillow provides information about home features (e.g., floor size, year built), location (e.g., county, zip code, street address), historical and current listing information (e.g., list price, sale price), and a price estimate called the “Zestimate”. Zillow describes the Zestimate as an “estimate of a home's market value”. From this and other publicly available posts \citep{zillow2020zestimateaccuracy}, we infer that the Zestimate is an ML-based estimation of the sale price as a function of home features, location, and the economic environment. We have access to data from over 750,000 homes in Austin (Travis County, Texas), Boston (Suffolk County, Massachusetts), and Pittsburgh (Allegheny County, Pennsylvania). Table \ref{table:datadesc} provides sample values of the features available for every home.

\begin{table}[]
\centering
\caption{Sample values of the house features, location, listing information, and Zestimate}
\begin{tabular}{|l|l|l|}
\hline
\textbf{Category}          & \textbf{Variable} & \textbf{Sample Value}                         \\ \hline
Location  & Latitude          & 29.7–42.3 degrees North                       \\ \cline{2-3} 
                           & Longitude         & 71.0–95.3 degrees West                        \\ \cline{2-3} 
                           & Neighborhood      & South Boston, Carrick, Brighton Heights, etc. \\ \cline{2-3} 
                           & Zip Code          & 15210, 15212, 15232, etc.                     \\ \cline{2-3} 
                           & County            & Suffolk, Allegheny, Travis                    \\ \hline
Features & Floor Size        & 100–10,000 sq. ft.                            \\ \cline{2-3} 
                           & Year Built        & 1799–2019                                     \\ \cline{2-3} 
                           & Last Remodel Year & 1799–2019                                     \\ \cline{2-3} 
                           & Bathrooms         & 0–15                                          \\ \cline{2-3} 
                           & Bedrooms          & 0–15                                          \\ \cline{2-3} 
                           & Parking           & 0–1000 sq. ft.                                \\ \cline{2-3} 
                           & Lot               & 100–10,000 sq. ft.                            \\ \cline{2-3} 
                           & Stories           & 0–50                                          \\ \cline{2-3} 
                           & Solar Potential   & 0–100                                         \\ \cline{2-3} 
                           & Type              & Single Family, Multi-Family, Condo, etc.      \\ \cline{2-3} 
                           & Structure Type    & Colonial, Victorian, Modern, etc.             \\ \cline{2-3} 
                           & Roof Type         & Composition, Shingle, Asphalt, etc.           \\ \cline{2-3} 
                           & Flooring          & Hardwood, Carpeted, Tile, etc.                \\ \cline{2-3} 
                           & Patio             & Porch, Deck, None, etc.                       \\ \cline{2-3} 
                           & Ex-Material       & Brick, Wood, Cement, etc.                     \\ \hline
Listing   & List Price        & $10,000–$10,000,000                           \\ \cline{2-3} 
                           & Days Listed       & 1–365 days                                    \\ \cline{2-3} 
                           & Sale Price        & $10,000–$10,000,000                           \\ \hline
ML Price                   & Z (Zestimate)     & $10,000–$10,000,000                           \\ \hline
\end{tabular}
\label{table:datadesc}
\end{table}

While home features and location are largely static, the Zestimate of a home can change as economic environment evolves. Zestimate may also change when Zillow (relatively infrequently) makes upgrades to the Zestimate algorithm. Consider as example an actual home in Boston. On January 1, 2020, Zillow presented a historical Zestimate trend from \$100,000 on January 1, 2010, to \$122,000 on January 1, 2020 (an annual increase of 2\%). Then, following an algorithm update on February 1, 2020, the historical Zestimate trend was updated as \$100,000 on January 1, 2010, to \$148,000 on January 1, 2020 (an annual increase of 4\%). Thus, the pre-update and post-update versions of the Zestimate assigned two different values (\$122,000 and \$148,000) to the same home on the same date (January 1, 2020). To track algorithm updates, we have access to 25 snapshots of Zillow information approximately every two weeks between February 2019 and March 2020. This data on algorithm updates becomes crucial for identification strategy described in the next section.

\subsection{Zestimate Impact on List and Sale Prices} 
\begin{figure}
    \centering
    \includegraphics[width=\textwidth]{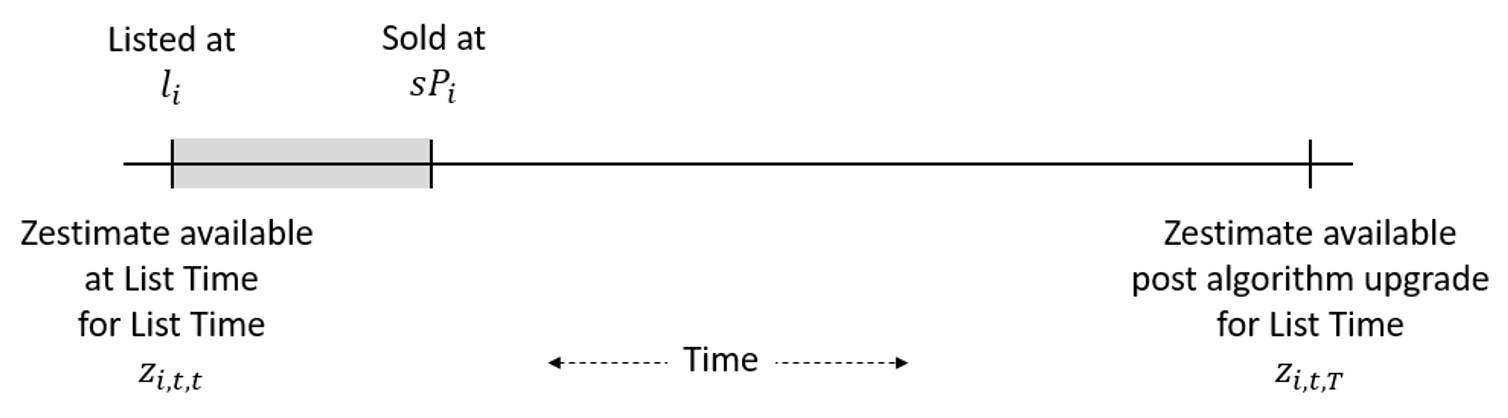}
    \caption{A timeline of the relationship between the list price $l_i$, the pre-update Zestimate $z_i$ ($z_{i,t,t}$) for home $i$ available at list time t, and the post-update Zestimate $\bar{z}$ ($z_{i,t,T}$) at time $T$ for home $i$ available at list time $t$.}
    \label{fig: Fig12}
\end{figure}

The first necessary primitive for our analytical model is that Zestimate $z_i$ has significant impact on the home sale price $p_i$. Both the Zestimate $z_i$ and buyers-sellers in the market may use local information unobserved to us, thus Zestimate $z_i$ is endogenous. To estimate the impact, we need to compare two groups of homes: those that received an erratic Zestimate over the true value, and those that received an erratic Zestimate under the true value. We cannot experimentally manipulate and add errors in Zestimate or simulate the home sale process in lab. Instead, we take advantage of the frequent upgrades to the Zestimate algorithm, from which we can infer historical instances in which the Zestimate was temporarily erratic (Figure \ref{fig: Fig12}). Then, we calculate the difference in the average sale price between the two groups to identify the impact of the Zestimate. 
\begin{align}
    z_i = v_i + e_i \quad ; \quad \bar{z}_i = v_i + \bar{e}_i \quad ; \quad z_i^e = \frac{z_i - \bar{z}_i}{\bar{z}_i}
\end{align}

Here $z_i$ represent the Zestimate for home $i$ when it was listed on the market at time $t$. This Zestimate contains some unobserved error $e_i=z_i-v_i$ relative to true value $v_i$. Let $\bar{z}_i$ represent the Zestimate for the same time snapshot $t$ after the Zestimate upgrade at $T$ ($>>t$). This upgraded Zestimate also contains some unobserved error $\bar{e}_i=\bar{z}_i-v_i$. The change in Zestimate ($\bar{z}_i - z_i = \bar{e}_i - e_i$) is correlated with the unobserved error $e_i$ presented on the platform when the home was on the market at time $t$.

Using this \textbf{naturally randomized treatment} $z_i^e$, we construct two groups of homes: the positive-error group is defined by $z_i^e > +1\%$, while the negative-error group  is defined by $z_i^e< -1\%$. These comparison of sale prices across these two groups can provide the treatment effect assuming $z_i^e$ is truly exogenous. For additional rigor, we also conduct propensity score matching (\textbf{PSM}) between the two groups i.e., propensity of $z_i^e > +1\%$ relative to $z_i^e < -1\%$. It is plausible that algorithm upgrade (and therefore treatment $z_i^e$) corrects pricing for home features that were originally under or over-priced. So, following equation 21 we conduct PSM (nearest neighbor matching) using the post-update Zestimate $\bar{z}_i$ (presumably less erratic than the pre-update Zestimate) and expansive set of home features $\vec{X}_i$ (such as floor area, number of bedrooms, year of construction, and many more). The validity of this pseudo randomization depends on two assumptions – (i) algorithm does not model home features that are unobserved or hidden on the platform and (ii) algorithm does not have data leakage. Zillow does not disclose the exact algorithm or its upgrade, but these assumptions are plausible given Zillow’s qualitative discussion of the algorithm \citep{zillow2020zestimateaccuracy}. In summary, the identification of Zestimate impact on sale prices relies on $z_i^e$ being exogenous to local information that is observed by buyers-sellers on the ground but unobserved to us (researchers).
\begin{align}
    pScore_i(\bm{\phi}) = \frac{1}{1 + exp(-[\bar{z}_i,\vec{X}_i] \times \bm{\phi})}
\end{align}
After conducting PSM, the \textit{standardized percentage  bias} (averaged across all observables) between the two groups is 0.93\%. The \textit{Rubin's R} and \textit{Rubin's B} are 1.01 and 0.30\% respectively\footnote{Rubin's R in range of $[0.5,2.0]$ and Rubin's B less than 25\% are considered a good match.}.

\begin{table}[]
\centering
\caption{Descriptive Statistics for key house features (subset of $\vec{X}$), Zestimate, List and Sale Prices for model of Zestimate impact on List and Sale Prices}
\begin{tabular}{|l|r|r|r|r|}
\hline
\textbf{Variable}         & \multicolumn{1}{l|}{\textbf{Mean}} & \multicolumn{1}{l|}{\textbf{Std. Dev}} & \multicolumn{1}{l|}{\textbf{Min}} & \multicolumn{1}{l|}{\textbf{Max}} \\ \hline
Z (pre algorithm update)  & 433737.6                           & 393379.4                               & 25065                             & 8244491                           \\ \hline
Z (post algorithm update) & 433321.7                           & 382211.1                               & 21226                             & 7911868                           \\ \hline
Z Error \%                & 0.109                              & 0.09                                   & -20                               & 20                                \\ \hline
Z Confidence Interval \%  & 17.3                               & 10.34                                  & 10                                & 130                               \\ \hline
Sale Price                & 387602.5                           & 376852.3                               & 13000                             & 4950000                           \\ \hline
List Price                & 438729.8                           & 390021                                 & 4800                              & 6675000                           \\ \hline
Markup \%                 & 1.08                               & 0.12                                   & -9.55                             & 22.5                              \\ \hline
Time to Sale              & 27.8                               & 47.94                                  & 1                                 & 350                               \\ \hline
Floor Size                & 1768.6                             & 831.3                                  & 293                               & 6000                              \\ \hline
Year Built                & 1961                               & 37.56                                  & 1799                              & 2019                              \\ \hline
Bathrooms                 & 2.3                                & 0.96                                   & 1                                 & 14                                \\ \hline
Bedrooms                  & 3.1                                & 1.2                                    & 1                                 & 12                                \\ \hline
Parking                   & 156.6                              & 220.55                                 & 0                                 & 995                               \\ \hline
Lot                       & 4396.4                             & 5113.88                                & 293                               & 9000                              \\ \hline
Stories                   & 1.9                                & 2.39                                   & 0                                 & 25                                \\ \hline
Last Remodel Year         & 1973.2                             & 39.91                                  & 0                                 & 2019                              \\ \hline
Solar Potential           & 73                                 & 25.47                                  & 0                                 & 95.66                             \\ \hline
\end{tabular}
\label{Table: PostPSMDescriptive}
\end{table}

\begin{figure}
    \centering
    \includegraphics[width=\textwidth]{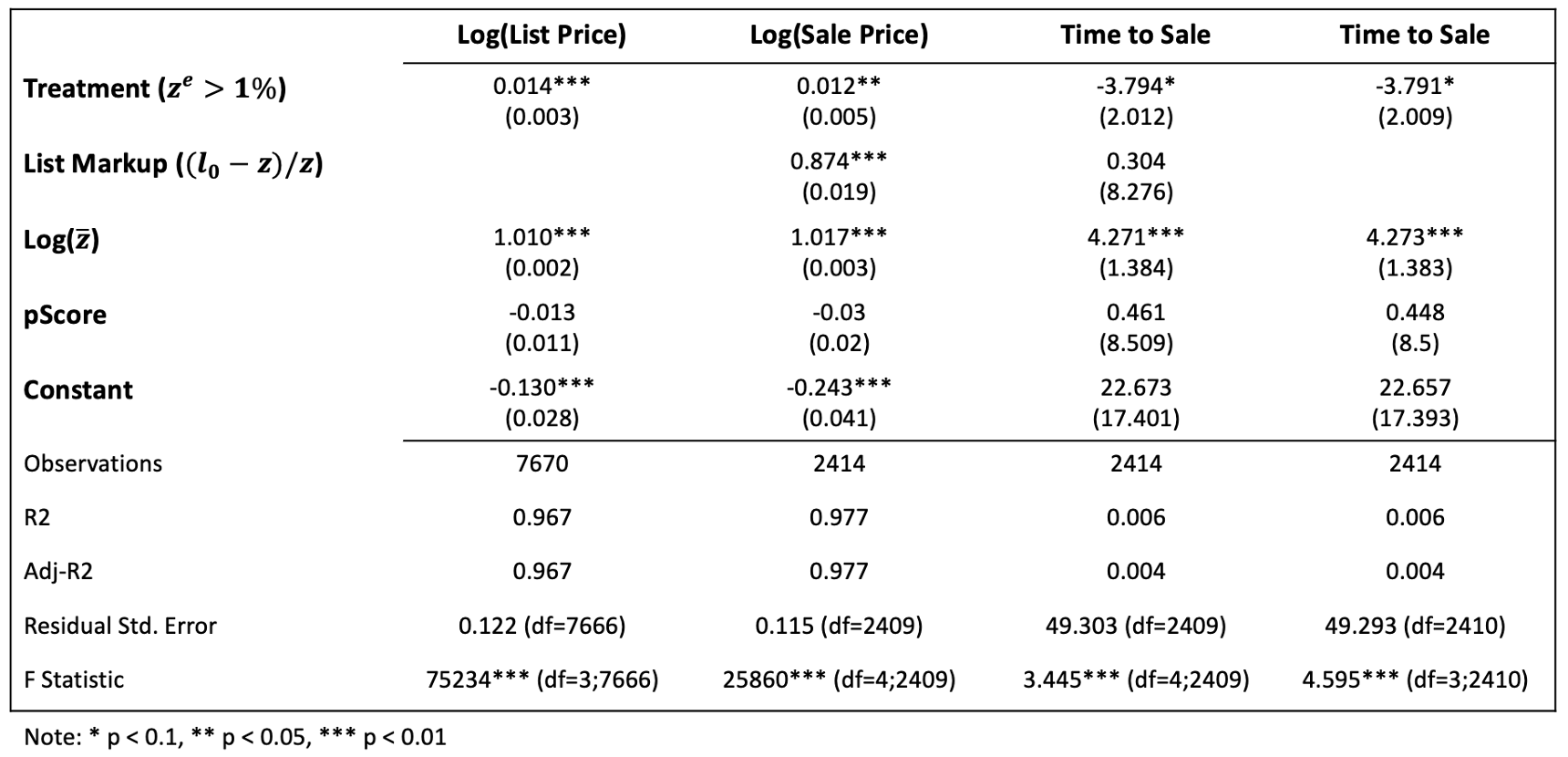}
    \caption{(Model 1) Primary specification for the impact of the Zestimate error on the sale price (equation 22). (Model 2) Additional interaction of the Zestimate error with the home size. (Model 3) Additional interaction of the Zestimate error with the Zestimate. (Model 4) Additional regression controls for home features (more than 50 estimates skipped here).}
    \label{fig: RegressionTable5}
\end{figure}

\begin{figure}
    \centering
    \includegraphics[width=\textwidth]{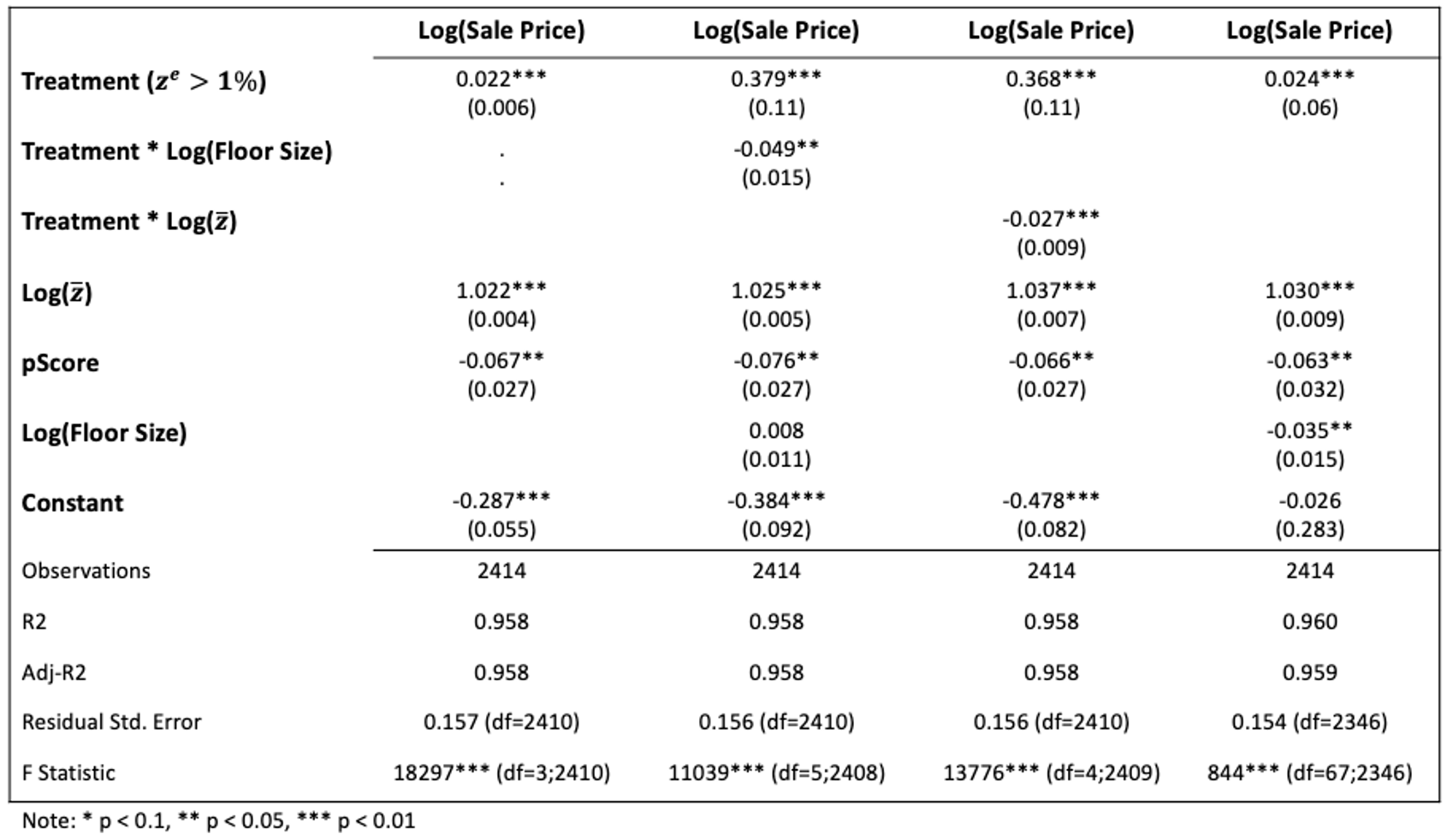}
    \caption{(Model 1) Impact of the Zestimate error on the list price (equation 23). (Model 2) Impact of the Zestimate error on the sale price, controlling for the list price (equation 24). (Models 3 and 4) Impact of the Zestimate error on the time to sale (in days)}
    \label{fig: RegressionTable6}
\end{figure}

After matching, the positive-error group  had an average Zestimate error of $z_i^e= 7.1\%$ while the negative-error group had an average Zestimate error of $z_i^e= -7.1\%$. Table \ref{Table: PostPSMDescriptive} provides descriptive statistics for dependent and independent regression variables after PSM. Figure \ref{fig: RegressionTable5}, Model 1 following specification\footnote{Including or excluding the propensity score in the regressions specification does not change the results.} in equation 22 and matched sample reports a sale price difference of 2.22\% between the two groups. The difference in sale prices has a sensitivity of roughly 15\% (15.5 = 2.22/14.2) to the Zestimate errors because buyers and sellers rely only partially on the Zestimate to determine their list prices and offers. In Model 3, the difference in sale prices is as high as 7.5\% for small homes (average approximately \$50,000), which tend to be more standardized and may involve buyers and sellers who rely more heavily on the Zestimate. The difference is only 1.4\% for large homes (average approximately \$500,000). Thus, we infer that the degree of reliance on the Zestimate ranges from 0.1 (14.2\% $\rightarrow$ 1.4\%) to 0.5 (14.2\% $\rightarrow$ 7.5\%). Going forward, we use the range 0.1 - 0.5 as a loose range of reliance on the Zestimate across housing submarkets.
\begin{align}
    log(p_i) = \beta_0 + \beta_1 * log(\bar{z}_i) + \beta_2 * \underbrace{(z_i^e > 1\%)}_{\text{Treatment}} + \beta_3 * pScore_i + \xi_{1,i}
\end{align}
Next, we use the observed list price and time to sale to uncover the impacts of the Zestimate on buyers and sellers, independently. For sellers, we examine the list prices, which are set by sellers alone (without explicit influence from buyers). As expected, Figure \ref{fig: RegressionTable6}, Model 1, reports a difference of 1.4\% in the initial list price between the two groups, confirming that the Zestimate impacts sellers. For buyers, we reason that if the Zestimate had no effect, then homes in the positive-error group should not be any harder to sell than homes in the negative-error group (as we conducted PSM to create groups with similar true prices). If, however, the Zestimate does affect buyers, then the sale price and time to sale should differ between the groups. Indeed, Model 2 (controlling for the initial list price) reports a difference of 1.2\% in the sale price, and Model 3 reports a shorter time to sale by 4 days. This serves as a motivating evidence for the assumption in the analytical model - Zestimate influences buyers-sellers and as a result the sale price of homes.
\begin{align}
    log(l_i) = \beta_0 + \beta_1 * log(\bar{z}_i) + \beta_2 * \underbrace{(z_i^e > 1\%)}_{\text{Treatment}} + \beta_3 * pScore_i + \xi_{2,i} \\
    log(p_i) = \beta_0 + \beta_1 * log(\bar{z}_i) + \beta_2 * \underbrace{(z_i^e > 1\%)}_{\text{Treatment}} + \beta_3 * pScore_i + \beta_4 \times \Big(\frac{l_i - \bar{z}_i}{\bar{z}_i}\Big) + \xi_{3,i}
\end{align}

\subsection{Zestimate Calculation from Sale Prices}
The Zestimate algorithm is proprietary to Zillow and thus opaque to us. Zillow describes the Zestimate as composed of both expert-driven economic modeling and data-driven predictive ML. The feedback loop phenomenon is relevant only if the data-driven predictive ML component is a significant driver. We look for some descriptive evidence that data-driven predictive ML methodology is a significant driver of Zestimate. After careful exploration of the Zillow pages for each home, we note that Zillow reports the sales of 4 or 5 ``peer homes" that are in the geographic vicinity or have similar features as the focal home. We first attempt to reverse engineer the choice of these peer sales. Then we test a hypothesis that Zestimate is driven primarily by a simple variable: the average sale price of peer homes. If so, it would greatly simplify the data-driven methodology behind the Zestimate calculation.

\begin{figure}
    \centering
    \includegraphics[width=\textwidth]{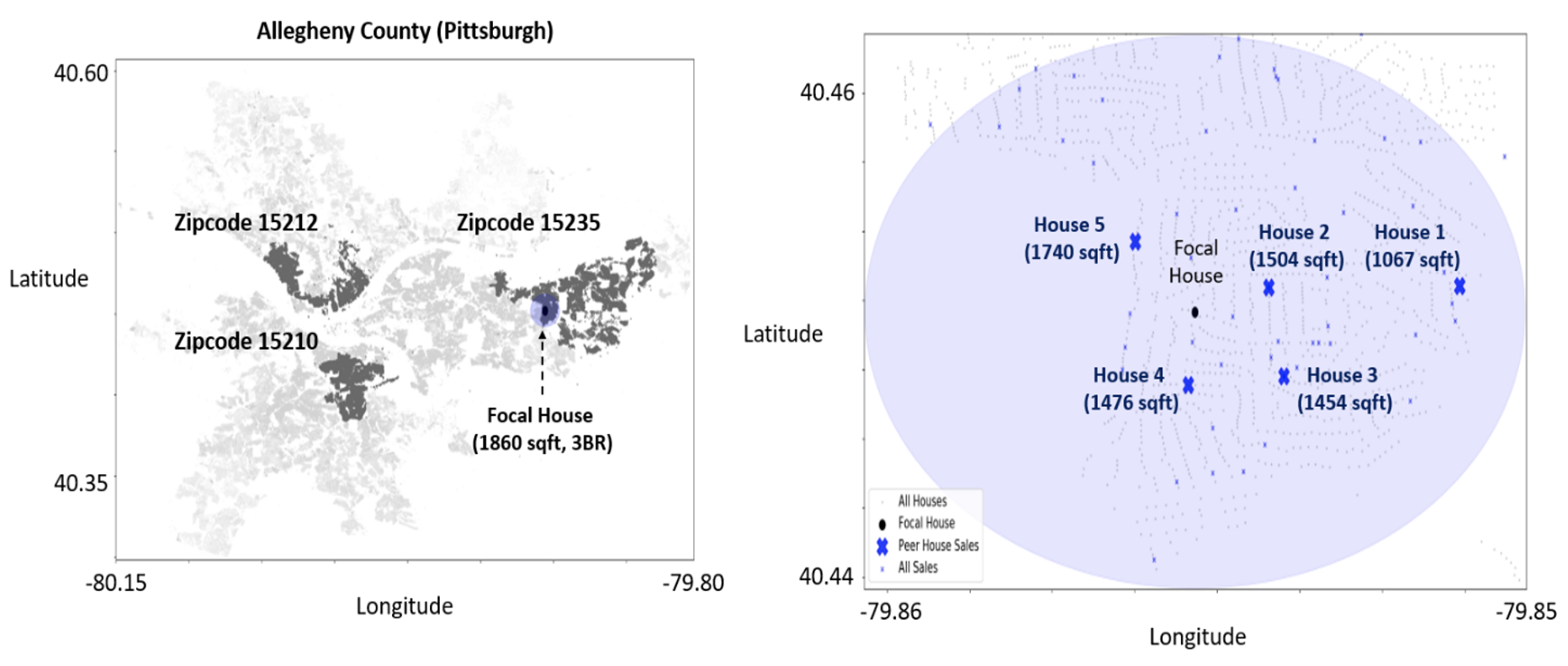}
    \caption{(Left) All homes in Allegheny County (small grey dots); our dataset includes homes from three zip codes (small dark grey dots). Consider an example of a focal home: a 3-bedroom, 1660 square-foot home in zip code 15235 (large black dot). (Right) The focal home’s five peers are located within 2 km of the focal home (the shaded circular region), sold in the past 12 months (small blue x), and have a similar floor size (large blue X).}
    \label{fig: FigA9}
\end{figure}

\begin{figure}
    \centering
    \includegraphics[width=\textwidth]{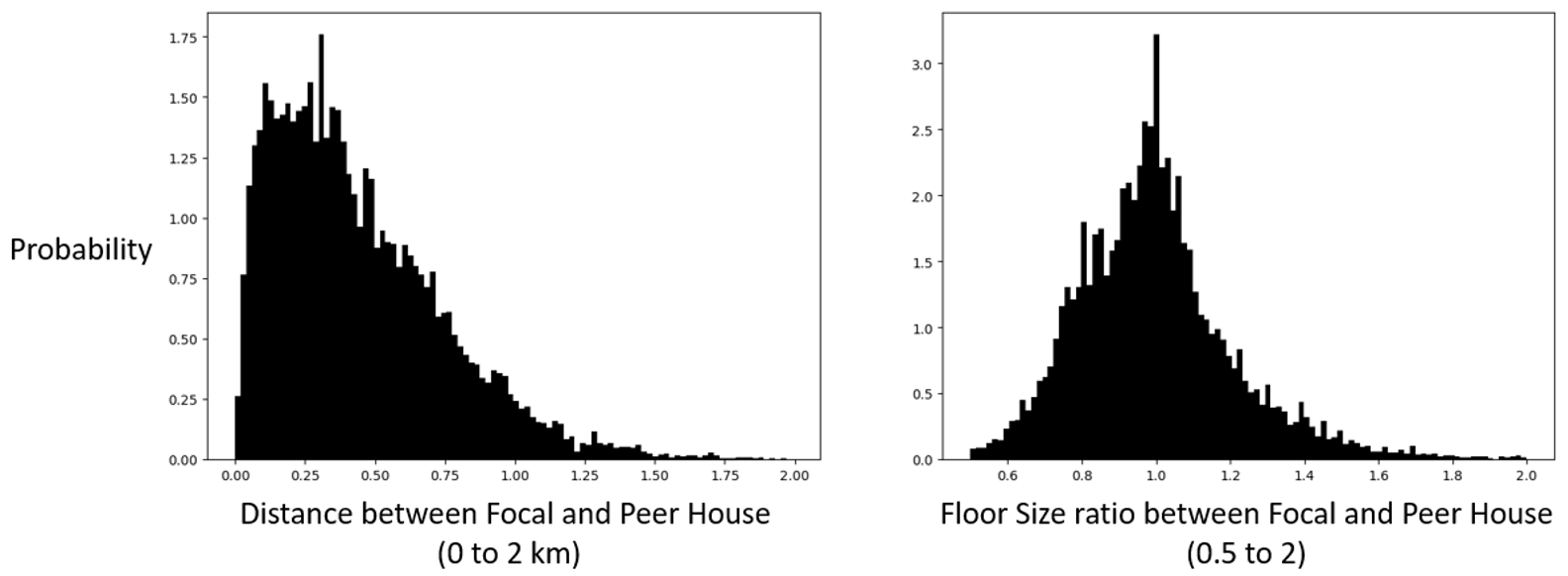}
    \caption{Probability distribution of geographical distance (Left) and floor size ratio (Right) between focal and peer sales. The distributions are empirically calculated using Zillow’s reported peer sales for houses in Zip Codes (15210, 15212, 15235) between March and August 2019.}
    \label{fig: FigA10}
\end{figure}

\textbf{Model of Peer Sales}: We analyze 1346 houses ($i \in \{1,...,1346\}$) in three zipcodes (15212, 15210 and 15235) of Allegheny County. Each home is observed over seven (approximately biweekly) time snapshots $t \in \{1,...,7\}$. In Allegheny County, Zillow reported 1408 home sales ($j \in \{1,...1400\}$) at the same time going up to 12 months back. Among the 1408 candidates only 4-5 houses are selected as peer sales for each house. All these recent sales $j$ are candidates to be in the ``peer homes" set for every house $i$.  We use $isPeer_{i,j,t}={0,1}$ to represent if house sale $j$ is tagged as a peer sale of house $i$ at time snapshot t. Given the characteristics of house $i$ and $j$, we want to predict if the pair would be tagged as peers. Figure \ref{fig: FigA9} illustrate an example home and its five peers.
\begin{align}
    P\Big( isPeer_{i,j,t} = 1 \Big) = (t - saleTime_j < 12) \times (Dist_{i,j} < 2) \nonumber \\
    \times (0.5 < \frac{floorSize_j}{floorSize_i} < 2) \times \frac{1}{1+exp(-y_{i,j,t})} \nonumber \\
    y_{i,j,t} = \beta_0 + \beta_1 * Dist_{i,j} + \beta_2 * (ZipCode_j = ZipCode_i) + \beta_3 * abs \Big( log \Big( \frac{floorSize_j}{floorSize_i} \Big) \Big) \nonumber \\
    + \beta_4 * (bedrooms_j - bedrooms_i) + \beta_5 * abs(bathrooms_j - bathrooms_i)
\end{align}
We infer from simple descriptive analysis (Figure \ref{fig: FigA10}) that a house has a very high likelihood of being tagged as a peer ($isPeer_{i,j,t}= 1$) if – (i) it is sold within past 12 months, (ii) it is within 2 km of the focal house and (iii) it has a floor size that is no less than half and no more than double of the focal house. To further predict $isPeer_{i,j,t}$ among houses that satisfy the three criteria above, we use a simple logistic regression model using – distance between house $i$ and $j$, binary indicator whether the two houses are in the same zipcode, ratio of floor sizes, absolute difference of number of bedrooms and bathrooms.

Figure \ref{Table: TableA4} reports that homes in close geographical vicinity (distance and zip code) and similar house features (size, bedrooms, bathrooms) are more likely to be peers. An accuracy of 98.7\% and an F1 score of 0.32 (compared with F1 score of 0.02 for a random model) suggest that our simple predictive model performs well in picking out peers of a house (Table \ref{Table: confusionmatrix} and \ref{Table: logitPerformance}).

\begin{table}[]
\centering
\scriptsize
\caption{Model of whether a house, which satisfies sale time within 12 months, distance within 2 km and floor area between 0.5 to 2 times of focal house, is a peer to the focal house.}
\begin{tabular}{lcc}
\cline{2-3}
                                  & \multicolumn{2}{c}{\textbf{Peer Match}}                                                                                       \\
                                  & \textbf{\begin{tabular}[c]{@{}c@{}}OLS\\ (1)\end{tabular}}  & \textbf{\begin{tabular}[c]{@{}c@{}}Logistic\\ (2)\end{tabular}} \\ \hline
\textbf{Peer Distance}            & \begin{tabular}[c]{@{}c@{}}-0.090***\\ (0.003)\end{tabular} & \begin{tabular}[c]{@{}c@{}}-3.716***\\ (0.016)\end{tabular}     \\
\textbf{Peer Floor Ratio}         & \begin{tabular}[c]{@{}c@{}}-0.069***\\ (0.001)\end{tabular} & \begin{tabular}[c]{@{}c@{}}-3.049***\\ (0.038)\end{tabular}     \\
\textbf{Peer Bedroom Difference}  & \begin{tabular}[c]{@{}c@{}}-0.010***\\ (0.002)\end{tabular} & \begin{tabular}[c]{@{}c@{}}-0.425***\\ (0.010)\end{tabular}     \\
\textbf{Peer Bathroom Difference} & \begin{tabular}[c]{@{}c@{}}-0.017***\\ (0.000)\end{tabular} & \begin{tabular}[c]{@{}c@{}}-0.822***\\ (0.010)\end{tabular}     \\
\textbf{Peer Zipcode Match}       & \begin{tabular}[c]{@{}c@{}}0.008***\\ (0.00)\end{tabular}   & \begin{tabular}[c]{@{}c@{}}0.013\\ (0.019)\end{tabular}         \\
\textbf{Constant}                 & \begin{tabular}[c]{@{}c@{}}0.170***\\ (0.001)\end{tabular}  & \begin{tabular}[c]{@{}c@{}}0.756***\\ (0.022)\end{tabular}      \\ \hline
Observations                      & 1,383,795                                                   & 1,383,795                                                       \\
R2                                & 0.086                                                       & 0.086                                                           \\
Adjusted R2                       & 0.086                                                       & 0.086                                                           \\
Log Likelihood                    & \multicolumn{1}{l}{}                                        & \multicolumn{1}{l}{-136,567.200}                                \\
Akaike Inf. Crit.                 & \multicolumn{1}{l}{}                                        & \multicolumn{1}{l}{273,146.500}                                 \\
Residual Std. Err.                & 0.172                                                       &                                                            \\ \hline
\end{tabular}
\label{Table: TableA4}
\end{table}

\begin{table}[]
\centering
\caption{Confusion Matrix and classification evaluation metrics (TPR, FPR, Accuracy, F1 Score) for the peer prediction model.}
\begin{tabular}{|cc|cc|}
\hline
\multicolumn{2}{|c|}{\multirow{2}{*}{Confusion Matrix}}       & \multicolumn{2}{c|}{Actual}            \\ \cline{3-4} 
\multicolumn{2}{|c|}{}                                        & \multicolumn{1}{c|}{Peer} & Not a Peer \\ \hline
\multicolumn{1}{|c|}{\multirow{2}{*}{Predicted}} & Peer       & \multicolumn{1}{c|}{4}    & 16         \\ \cline{2-4} 
\multicolumn{1}{|c|}{}                           & Not a Peer & \multicolumn{1}{c|}{1}    & 1379       \\ \hline
\end{tabular}
\label{Table: confusionmatrix}
\end{table}

\begin{table}[]
\centering
\caption{Performance of logistic regression components $y_{i,j,t}$ within the overall predictive model}
\begin{tabular}{|l|l|}
\hline
True Positive Rate  & 80\%   \\ \hline
False Positive Rate & 1.14\% \\ \hline
Precision           & 20\%   \\ \hline
Accuracy            & 98.7\% \\ \hline
F1 Score            & 0.32   \\ \hline
\end{tabular}
\label{Table: logitPerformance}
\end{table}

\textbf{Model of Zestimate from Peer Sales}: Next, we hypothesize that Zestimate $z_i$ is driven primarily by a simple variable: the average sale price of ``peer homes" set $J_i$ i.e. $z_i \approx \bar{p}_{J_i}$. To establish the role of $\bar{p}_{J_i}$, we could consider a pair of adjacent similar homes A and B. At time $t_1$, both homes have same Zestimate and the same peer sets ($J_A = J_B$), so they have the same $\bar{p}_{J_A} = \bar{p}_{J_B}$. At time $t_2$, the peer set for one home stays the same while the peer set for the other home changes. Given the geographical proximity and similarity in features, any difference in the Zestimate ($z_{B,t_2} - z_{A,t_2}$) at time $t_2$ must arise from the change in the peer set ($J_B=\{1,2,3,4\} \rightarrow \{1,3,4,5\}$). A plausible reason why peer set for only home B changes is that a new sale $j = 5$ is within the ``peer boundary" (e.g., 2 km distance) of B, but just outside A. The sharp boundaries that define ``peer homes" create this natural experiment\footnote{It is possible that such situations never occur i.e., similar and proximal homes always have the same peer set. In this case, our approach would have failed.}.
\begin{align}
    pScore_i(\bm{\phi}) = \frac{1}{1 + exp\Big( \Big[ (\bar{p}_{J_{i,t}}),z_{i,t_1},\theta_{i,t_1},\Delta\theta_i,\vec{X}_i \Big] * \bm{\phi} \Big)} \quad ; \quad (\bar{p}_{J_i,t}) = \Big( \sum_{j \in \text{peers of $i$ at $t$}} p_j \Big)
\end{align}

\begin{figure}
    \centering
    \includegraphics[width=\textwidth]{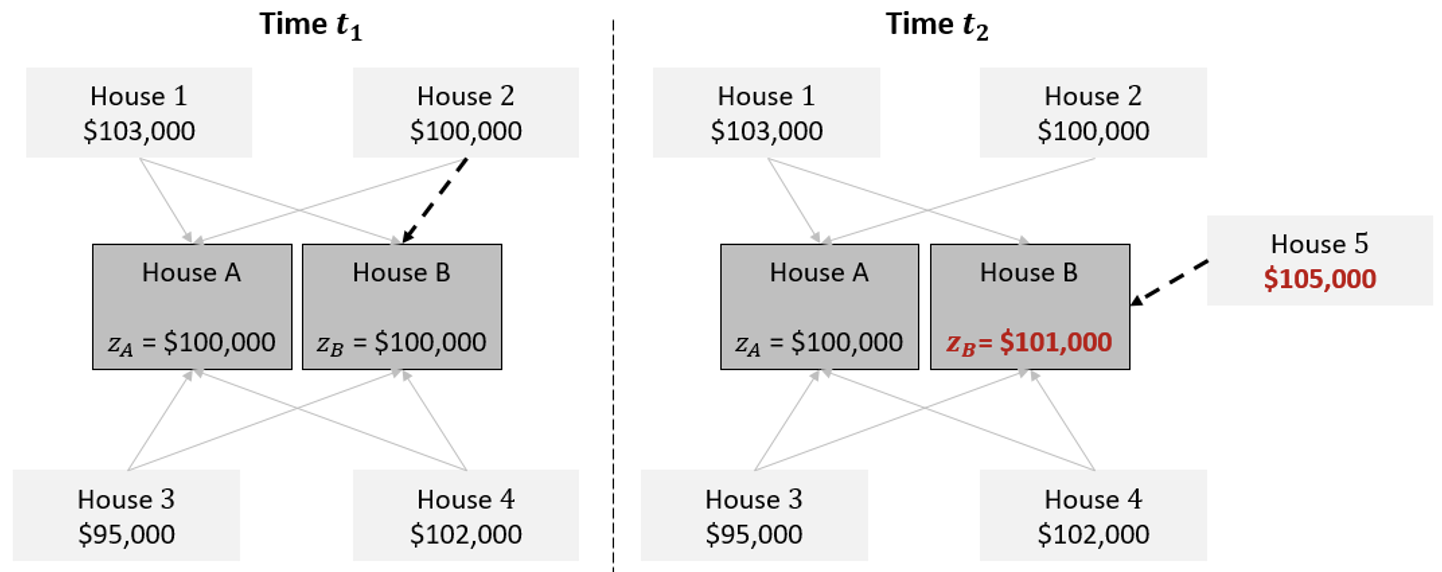}
    \caption{A hypothetical example. At time $t_1$, Homes A and B have the same Zestimate, the same set of peer homes ${1,2,3,4}$, and the same average peer sale price ($\bar{p}_{J_A} = \bar{p}_{J_B}$). At time $t_2$, House 5 replaces House 2 in the peer set of Home B, causing the $\bar{p}_{J_B}$ for Home B to increase by \$1,250 and its Zestimate to increase by \$1,000.}
    \label{fig: FigA11}
\end{figure}
We create a control group (no change in the peer set) and a treatment group (a change in the peer set). We restrict to small positive change in peer set sale price average i.e.
$\Delta \bar{p}_{J_i}$ is 0-5\%. The average $\Delta \bar{p}_{J_i}$ is 2.3\%. We use Propensity Score Matching (PSM) to create a sample that matched treat and control units on the peer set at $t_1$, Zestimate at $t_1$, home features $X_i$, and other potential drivers\footnote{For example tax rate changes between $t_1$ and $t_2$. Even though such change should impact both the treatment and control units.} of a change $\theta$ in the Zestimate. Now we can measure the treatment effect of $\Delta \bar{p}_{J_i}$ as $\beta_2$ in equation below. 
\begin{align}
    \Delta z_i = (z_{i,t_2} - z_{i,t_1}) = \beta_0 + \beta_1 * \Delta \theta_i + \beta_2 * \underbrace{(\Delta \bar{p}_{J_i})}_{\text{Treatment}} + \beta_3 * pScore_i
\end{align}

\begin{table}[]
\centering
\scriptsize
\caption{Model of Zestimate using average peer sale price, weighted peer sale prices, average peer floor size, average peer list price and county tax estimate.}
\begin{tabular}{lcccl}
\cline{2-5}
                                   & \multicolumn{4}{c}{\textbf{Zestimate}}                                                                                                                                                                                                                                   \\
                                   & (1)                                                             & (2)                                                               & (3)                                                               & (4)                                                            \\ \hline
\textbf{Avg. Peer Sale Price}      & \begin{tabular}[c]{@{}c@{}}1.050***\\ (0.002)\end{tabular}      & \begin{tabular}[c]{@{}c@{}}1.049***\\ (0.002)\end{tabular}        & \begin{tabular}[c]{@{}c@{}}1.042***\\ (0.002)\end{tabular}        & \begin{tabular}[c]{@{}l@{}}0.077***\\ (0.007)\end{tabular}     \\
\textbf{Time Since Sale Weight}    &                                                                 & \begin{tabular}[c]{@{}c@{}}0.934***\\ (0.142)\end{tabular}        & \begin{tabular}[c]{@{}c@{}}1.015***\\ (0.136)\end{tabular}        &                                                                \\
\textbf{Peer Distance Weight}      &                                                                 & \begin{tabular}[c]{@{}c@{}}-1.395***\\ (0.137)\end{tabular}       & \begin{tabular}[c]{@{}c@{}}-1.186***\\ (0.132)\end{tabular}       &                                                                \\
\textbf{Peer Floor Ratio Weight}   &                                                                 & \begin{tabular}[c]{@{}c@{}}-0.900***\\ (0.095)\end{tabular}       & \begin{tabular}[c]{@{}c@{}}-1.065***\\ (0.091)\end{tabular}       &                                                                \\
\textbf{Avg. Peer Time Since Sale} &                                                                 &                                                                   & \begin{tabular}[c]{@{}c@{}}-323.955***\\ (64.591)\end{tabular}    &                                                                \\
\textbf{Avg. Peer Floor Size}      &                                                                 &                                                                   & \begin{tabular}[c]{@{}c@{}}12.474***\\ (0.543)\end{tabular}       &                                                                \\
\textbf{Avg. Peer List Price}      & \multicolumn{1}{l}{}                                            & \multicolumn{1}{l}{}                                              & \multicolumn{1}{l}{}                                              & \begin{tabular}[c]{@{}l@{}}-0.003\\ (0.004)\end{tabular}       \\
\textbf{Tax Estimate}              &                                                                 &                                                                   &                                                                   & \begin{tabular}[c]{@{}l@{}}0.268***\\ (0.005)\end{tabular}     \\
\textbf{Constant}                  & \begin{tabular}[c]{@{}c@{}}5,916.929***\\ (372.91)\end{tabular} & \begin{tabular}[c]{@{}c@{}}-5,751.926***\\ (368.627)\end{tabular} & \begin{tabular}[c]{@{}c@{}}-3,907.004***\\ (422.500)\end{tabular} & \begin{tabular}[c]{@{}l@{}}-3,517.140***\\ (307.487)\end{tabular} \\ \hline
Observations                       & 9,383                                                           & 9,383                                                             & 9,366                                                             & 9,203                                                          \\
R2                                 & 0.958                                                           & 0.959                                                             & 0.963                                                             & 0.973                                                          \\
Adjusted R2                        & 0.958                                                           & 0.959                                                             & 0.963                                                             & 0.973                                                          \\
Residual Std. Err.                 & 12,730.440                                                      & 12,578.420                                                        & 12,016.530                                                        & 10,240.530                                                     \\ \hline
\end{tabular}
\label{Table: TableA7}
\end{table}

The average increase of 2.3\% in $\bar{p}_{J_i}$ corresponds to a 1.9\% increase in the Zestimate $z_i$. We repeat the same steps with two alternative treatment groups: 5–10\% increase in $\bar{p}_{J_i}$ and 10–15\% increase in $\bar{p}_{J_i}$. Table \ref{Table: A6} reports the results from all three treatment specifications. Overall, a 1\% increase in $\bar{p}_{J_i}$ corresponds to an increase of 0.66–0.83\% in the Zestimate $z_i$.

\begin{table}[]
\centering
\caption{Treatment effects using three different treatment groups with different ranges of the change in the average peer sale price $\bar{p}_{J_i}$.}
\begin{tabular}{|l|l|l|l|l|}
\hline
\textbf{Treatment Group} & \textbf{Range of Change in $\bar{p}_{J_i}$} & \textbf{Mean Change in $\bar{p}_{J_i}$} & \textbf{\begin{tabular}[c]{@{}l@{}}Treatment \\ Effect ($\beta_2$)\end{tabular}} & \textbf{$\beta_2/\bar{p}_{J_i}$} \\ \hline
I                        & 0–5 \%                      & 2.3\%                   & 1.9\%                                                                    & 0.83       \\ \hline
II                       & 5–10 \%                     & 6.7\%                   & 4.4\%                                                                    & 0.66       \\ \hline
III                      & 10–15\%                     & 11.5\%                  & 8.6\%                                                                    & 0.75       \\ \hline
\end{tabular}
\label{Table: A6}
\end{table}

\begin{table}[]
\centering
\scriptsize
\caption{Model of Zestimate using different weighted peer sale prices.}
\begin{tabular}{lccc}
\cline{2-4}
                                                 & \multicolumn{3}{c}{\textbf{Zestimate}}                                                                                                                                                                     \\
                                                 & (1)                                                               & (2)                                                               & (3)                                                                \\ \hline
\textbf{Avg. Peer Sale Price}                    & \begin{tabular}[c]{@{}c@{}}1.051***\\ (0.001)\end{tabular}        & \begin{tabular}[c]{@{}c@{}}1.050***\\ (0.001)\end{tabular}        & \begin{tabular}[c]{@{}c@{}}1.030***\\ (0.002)\end{tabular}         \\
\textbf{Peer Sale Price Deviation}               & \begin{tabular}[c]{@{}c@{}}-0.000\\ (0.004)\end{tabular}          & \begin{tabular}[c]{@{}c@{}}-0.004\\ (0.008)\end{tabular}          & \begin{tabular}[c]{@{}c@{}}-0.022\\ (0.172)\end{tabular}           \\
\textbf{Peer Price Deviation * Peer Distance}    &                                                                   & \begin{tabular}[c]{@{}c@{}}0.033***\\ (0.004)\end{tabular}        &                                                                    \\
\textbf{Peer Price Deviation * Peer Floor Ratio} &                                                                   & \begin{tabular}[c]{@{}c@{}}0.119***\\ (0.0027)\end{tabular}       &                                                                    \\
\textbf{Peer Price Deviation * Time Since Sale}  &                                                                   & \begin{tabular}[c]{@{}c@{}}-0.009***\\ (0.001)\end{tabular}       &                                                                    \\
\textbf{Peer Price Deviation * Ind w1}           &                                                                   &                                                                   & \begin{tabular}[c]{@{}c@{}}0.018*\\ (0.011)\end{tabular}           \\
\textbf{Peer Price Deviation * Ind w2}           &                                                                   &                                                                   & \begin{tabular}[c]{@{}c@{}}-0.008\\ (0.172)\end{tabular}           \\
\textbf{Peer Price Deviation * Ind w3}           &                                                                   &                                                                   & \begin{tabular}[c]{@{}c@{}}0.067\\ (0.173)\end{tabular}            \\
\textbf{Peer Price Deviation * Ind w4}           &                                                                   &                                                                   & \begin{tabular}[c]{@{}c@{}}0.133\\ (0.124)\end{tabular}            \\
\textbf{Constant}                                & \begin{tabular}[c]{@{}c@{}}-6,152.341***\\ (160.929)\end{tabular} & \begin{tabular}[c]{@{}c@{}}-6,692.196***\\ (185.685)\end{tabular} & \begin{tabular}[c]{@{}c@{}}-4,586.921***\\ (1705.038)\end{tabular} \\ \hline
Observations                                     & 46,735                                                            & 46,735                                                            & 10,609                                                             \\
R2                                               & 0.961                                                             & 0.961                                                             & 0.961                                                              \\
Adjusted R2                                      & 0.961                                                             & 0.961                                                             & 0.961                                                              \\
Residual Std. Err.                               & 12,233.42 (df = 46732)                                            & 12,184.070 (df = 46726)                                           & 11,954.88 (df = 10598)                                             \\ \hline
\end{tabular}\label{Table: TableA8}
\end{table}

To further substantiate the role of $\bar{p}_{J_i}$ in driving Zestimate $z_i$, we test the out-of-sample explanatory power, and we find that $\bar{p}_{J_i}$ alone explains almost 96\% of all variation in the Zestimate. Although the actual Zestimate model may be significantly more sophisticated than our simple approximation, the extremely high out-of-sample explanatory power ($R^2 \approx 0.96$) suggests that significant fraction of full model’s output is driven by information contained in peer sales. We iterate over various alternative predictors – (i) weighted (instead of unweighted) average of peer sale prices, (ii) House Features $X_i$, (iii) Location (Neighborhood, ZipCode, County) fixed effects, (iv) Time fixed effects, (v) Tax Estimate. We find that additional features (House Features, Location fixed effects, Time fixed effects, and Tax Estimate do not contain significant Zestimate explanatory power on their own. Table \ref{Table: A9} shows that all these features and weighted peer sale price add very little to the explanatory power when used alongside the simple average peer sale price $\bar{p}_{J_i}$ metric.

\begin{table}
\centering
\caption{Alternative Models and features to predict Zestimate}
\begin{tabular}{|l|ll|}
\hline
\multirow{2}{*}{Features}    & \multicolumn{2}{c|}{Out of Sample}                         \\ \cline{2-3} 
                             & \multicolumn{1}{l|}{Linear Model}   & Support Vector Model \\ \hline
Average Peer Sale Price only & \multicolumn{1}{l|}{\textbf{96.15}} & 95.97                \\ \hline
+ Kernel Weights             & \multicolumn{1}{l|}{96.59}          & 96.61                \\ \hline
+ All other covariates       & \multicolumn{1}{l|}{96.79}          & 97.34                \\ \hline
\end{tabular}
\label{Table: A9}
\end{table}

\section{Model Choices}

\subsection{Buyer-Seller Bargaining} \label{app:bargaining}
We model seller's choice of list price such that - if an offer meets the list price the home sells otherwise the home does not sell. We make a three assumptions here – (a) The seller does not receive offers above the list price, (b) The seller cannot reject an offer at (or above their list price) and (c) The seller cannot accept an offer below the list price. \\

\begin{figure}
    \centering
    \includegraphics[width=0.7\textwidth]{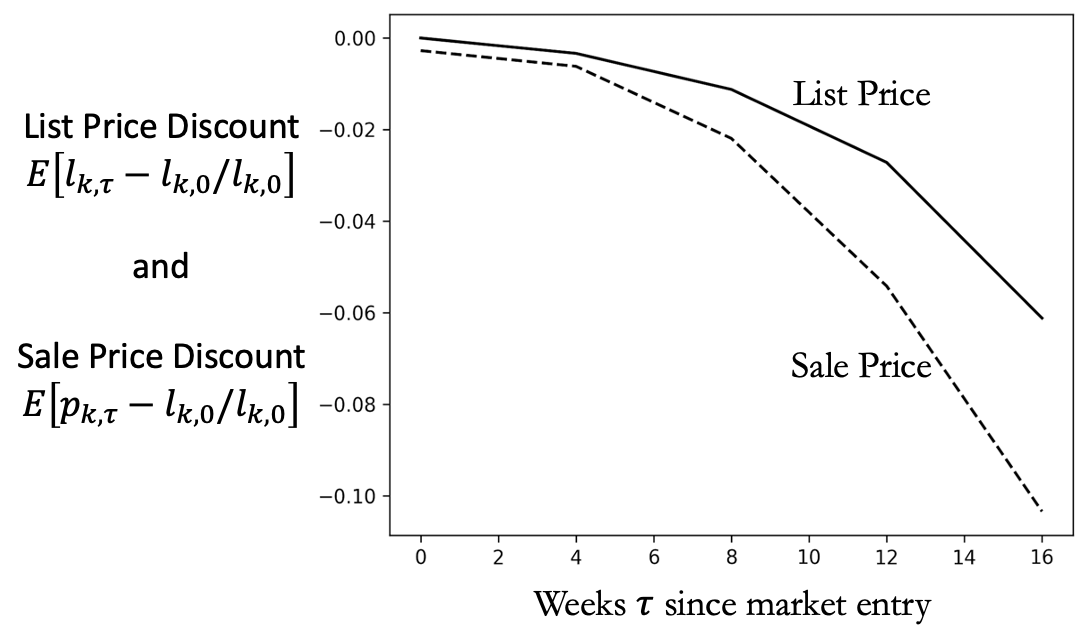}
    \caption{Average list price discount $\mathbb{E}[(l_{k,\tau}-l_{k,\tau=0})/l_{k,\tau=0}]$ and average sale price discount $\mathbb{E}[(p_{k,\tau}-l_{k,\tau=0})/l_{k,\tau=0}]$ relative to the first list price.}
    \label{fig: ListPriceDiscounting}
\end{figure}

Let's consider the assumption (a). Since, the seller's list price is visible to the buyer, even if the buyer has a higher willingness to pay, they have no reason to make an offer above the list price. A buyer may make an offer above the list price if they face competition from other buyers bidding for the same home. Empirically we observe that 81\% of homes sell below their list price. On an average homes seller 0.5\% below their list price (Figure \ref{fig: ListPriceDiscounting}). Given these observations we choose to not model buyer competition and resulting offers above list price. Regarding (b), if the buyer makes an offer at the list price, we assume that seller cannot reject the offer i.e., ``enforced full price offer contract" \citep{guerra2018real}. In practice, contracts and regulations do not allow enforcing a seller to accept offers at list price. However, brokers may demand commission if the seller chooses to reject an offer at list price. This indirectly discourages the seller from listing lower than their reservation price and subsequently studying offers and making choices to accept or reject. We do not model broker commission, instead we directly assume that seller does not have a choice to reject a full price offer and therefore does not list below their reservation price.

Let's consider the assumption (c). The seller could choose to list at a very high price, well above their reservation price. Once they receive the offers, they could choose to accept if the offer is above the reservation price. In doing so, the seller doesn't reveal their reservation price to the buyer. In such a model, the buyer is forced to reveal their full willingness to pay or engage in bargaining as both the buyer and seller choose to not reveal their reservation price. In practice, listing very high has some negative implications – (i) Buyer's may not want to enter lengthy negotiations, (ii) Broker may not be willing to spend effort in advertising a home where an offer meeting the list price is unlikely, (iii) The home may be left out of buyer's consideration, since similar homes priced more competitively substitute out the home with the high list price. We do not model seller's competition against similar homes, instead we directly assume that seller does not list significantly above their reservation price. The Figure \ref{fig: ListPriceDiscounting} below shows the empirically observed evolution of list price and corresponding sale prices. Empirically we observe that 55\% of homes sell at or withing 1\% of their list price. The evidence reinforces the assumption that seller does not receive and accept offer significantly below their list price. However, this assumption is not always true. We do observe that 10.7\% of homes sell more than 5\% below their list price. Our analytical model is limited in representing these outcomes.

\subsection{Full Model}
\begin{figure}
    \centering
    \includegraphics[width=0.9\textwidth]{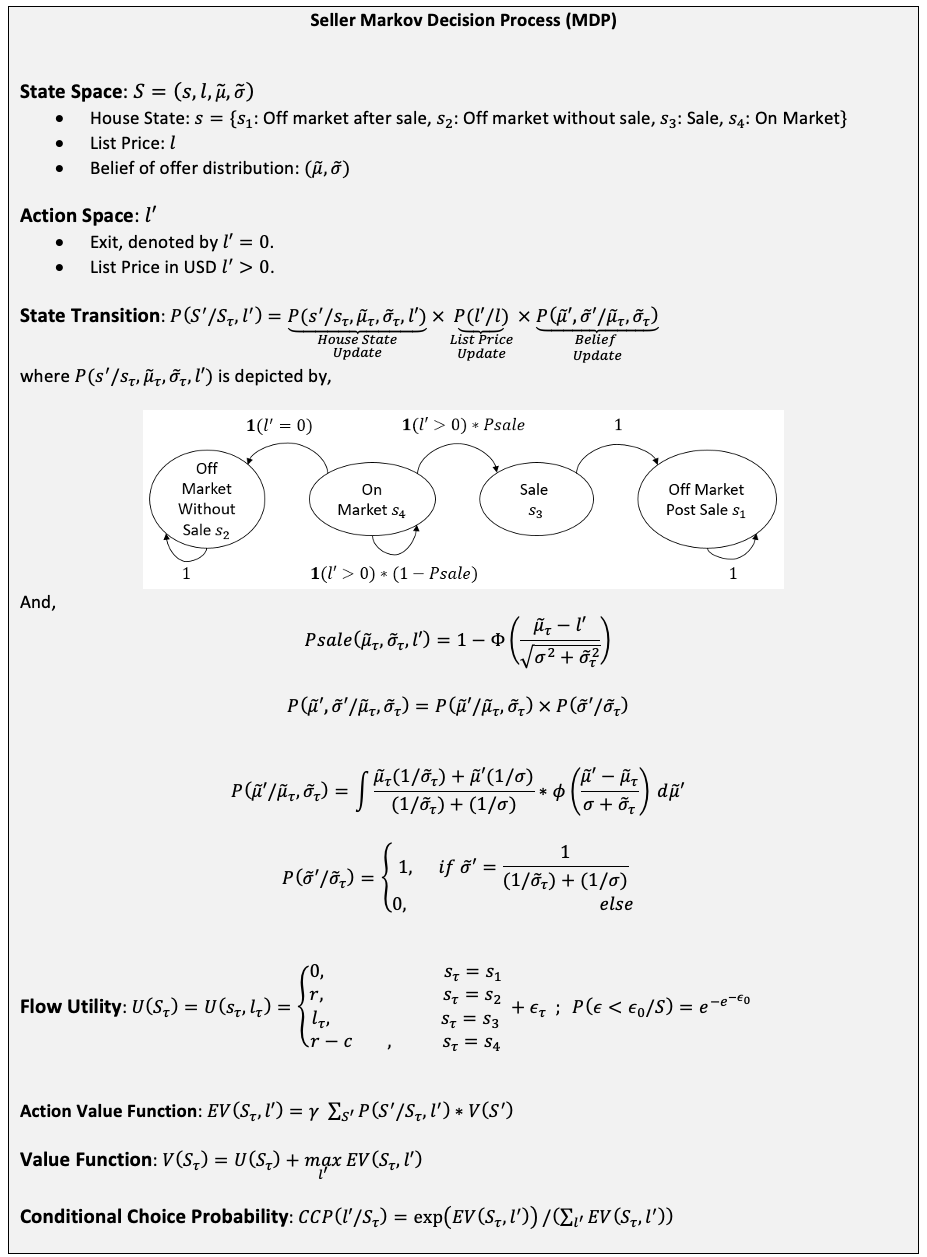}
    \caption{Seller Markov Decision Process (MDP) under the full model.}
    \label{fig: FullModelMDP}
\end{figure}

In section 3.1, we described a model of home sale and then we use \textbf{simple} distributional assumptions to derive closed form expressions summarized in Lemma 1 and Lemma 2. Lemma 1 establishes relationships (between seller beliefs, valuations, sale prices and home value) that are a building block for examining pricing implications of the feedback loop. Lemma 2 establishes relationships (between seller beliefs and payoffs) that are building block for examining implications of the feedback loop for seller payoffs. In this section, we use more comprehensive distributional assumptions (Table \ref{table:SimpleAndFull}). The purpose is to at least qualitatively support Lemma 1 and Lemma 2. This would provide some confidence that results are not driven by simplified modeling assumptions. In this full model environment, the seller follows an infinite horizon discounted Markov Decision Process (MDP) summarized in Figure \ref{fig: FullModelMDP}.

The seller \textbf{state} $s_{\tau}$ is fully characterized by – whether their house is on or off the market ($s_{\tau} \in \{s_1,s_2,s_3,s_4\}$), current list price ($l_{\tau}$), and their \textbf{belief} about best buyer offers ($N(\tilde{\mu}_{\tau},\tilde{\sigma}_{\tau})$). If the house is on the market $s_{\tau}=s_4$, the seller has a choice of \textbf{actions} – they can exit the market or continue in the market at a new list price. If the house is not on the market $s_{\tau} \neq s_4$, the seller actions are irrelevant. The belief update follows Bayesian learning as the seller combines their current beliefs $N(\tilde{\mu}_{\tau},\tilde{\sigma}_{\tau})$ with noisy but unbiased signal of offer distribution from the market $N(\mu,\sigma)$. The more time they spend in the market the closer their belief gets to actual average offer $\tilde{\mu} \rightarrow \mu$ and $\tilde{\sigma} \rightarrow \sigma$. This equilibrium for the full model is only calculated numerically.

\begin{figure}
    \centering
    \includegraphics[width=0.9\textwidth]{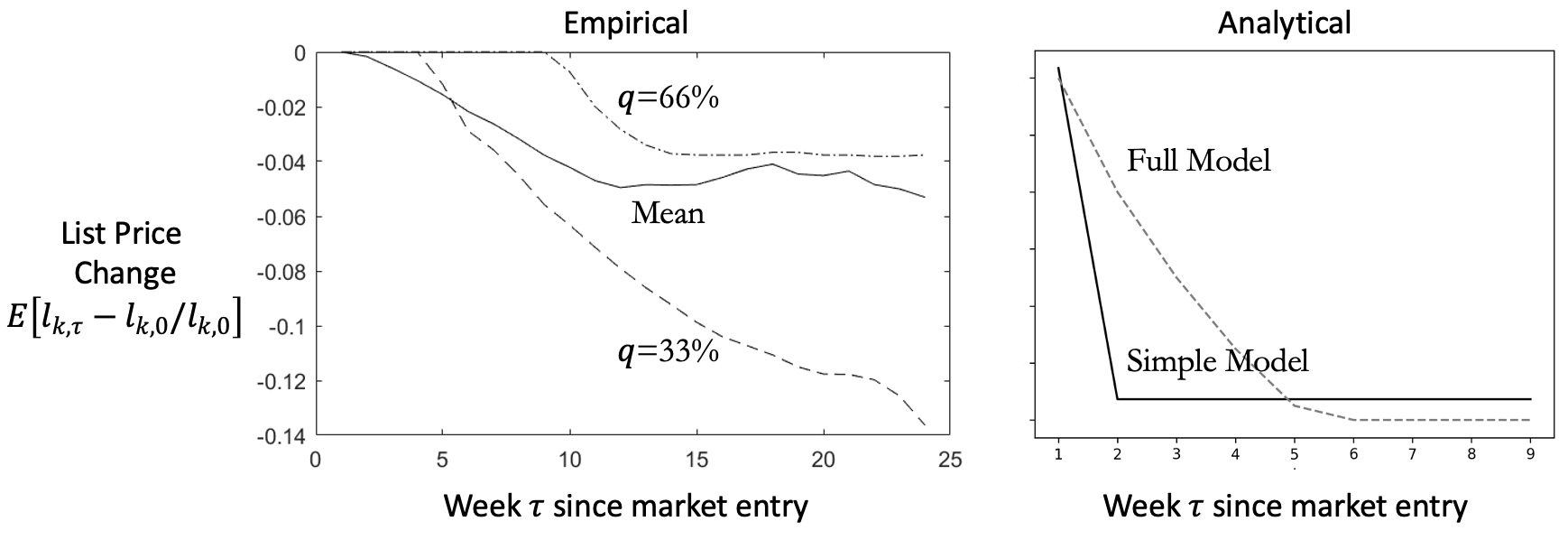}
    \caption{Average list price for home $k$ at week $\tau$ as a percentage of first list price $E[\frac{l_{k,\tau}-l_{k,0}}{l_{k,0}}]$ across - empirical observation (Left Figure), simple (Solid line in Right Figure) and the full analytical model (Dashed line in Right Figure).}
    \label{fig: EmpiricalvAnalytical}
\end{figure}
\begin{figure}
    \centering
    \includegraphics[width=0.7\textwidth]{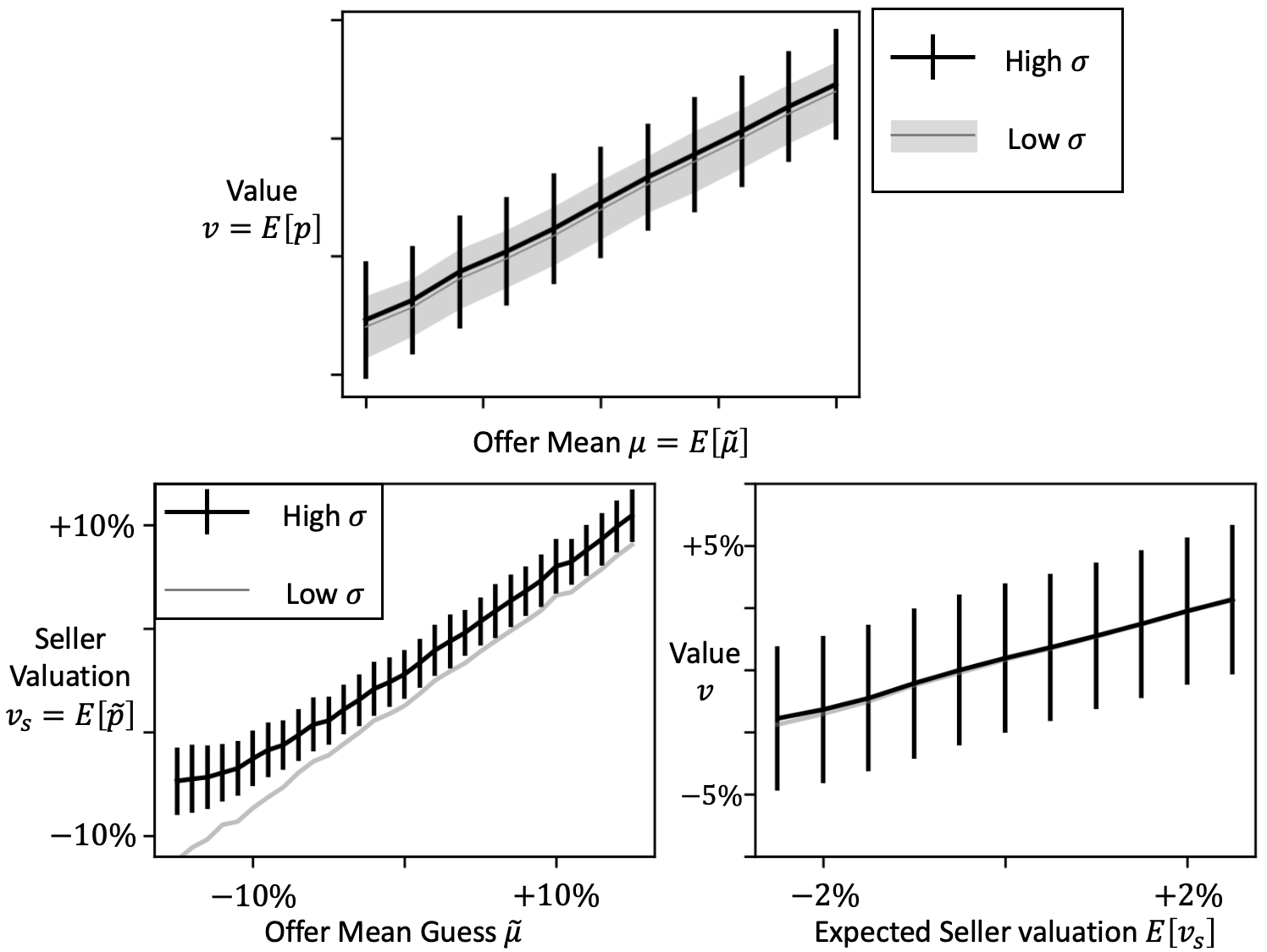}
    \caption{(Top) Home value $v=E[p]$ as a function of offer mean $\mu=E[\tilde{\mu}]$. The standard deviation band of sale prices $S.D.[p]$ is wider under high $\sigma$ (black error bars) compared to low $\sigma$ (grey band). (Bottom Left) The seller's valuation $v_s=E[\tilde{p}]$ as a function of the offer guess $\tilde{\mu}$. (Bottom Right) Home value $v=E[p]$ as a function of the expected seller valuation $E[v_s]$.}
    \label{fig: Lemma1_FullModel}
\end{figure}

Figure \ref{fig: EmpiricalvAnalytical} illustrates the seller choice of list prices $l_{\tau}$ as - (Left) observed empirically, (Right) optimal solution in the simple and full model. On an average, sellers start with a high list price $l_{\tau=0}$ when $\tilde{\sigma}_{\tau = 0}$ is large. The list price choice $l_{\tau}$ decreases as $\tilde{\sigma}_{\tau}$ decreases. This is one example to sanity check that the analytical models capture patterns exhibit by seller choices empirically. Note that the simple model greatly simplifies the learning process using $\sigma_{signal} = 0$ which results in the seller learning the true buyer offer in a single period ($\tilde{\mu}_{\tau = 1} = \mu$ and $\tilde{\sigma}_{\tau = 1} = 0$ ). Consequently, the seller list price $l_{\tau = 1}$ is optimal with respect to the true offer distribution and does not change for any $\tau \geq 1$. In comparison the learning and therefore adjustment of the list prices is more gradual in the full model. Figure \ref{fig: Lemma1_FullModel} exhibits that numerical results from the full analytical model are qualitatively identical to closed form expressions in Lemma 1 from the simple model. First, the expected sale price $E[p]$ (and true home value $v$) is increasing in true best buyer offer mean $\mu$. The variance in sale prices $Var[p]$ is increasing in offer variance $\sigma$. Second, the seller valuation $v_s$ (or $\tilde{v}_s$) is increasing in her belief of buyer offer mean $\tilde{\mu}$. The variance in seller valuation $v_s$ (across sellers) is increasing in variance in belief of buyer offer $\tilde{\sigma}_{\tau = 0}$. Finally, the true value of a home is increasing in expected seller valuations $E[v_s]$ (across sellers). These are all directionally consistent with the expressions in Lemma 1 using the simple model. We skip reporting numerical results to support the expressions in Lemma 2.

\section{Proofs} \label{app: proofs}
\subsection{Proof for Lemmas}
\textbf{Lemma 1}\\
The simple model makes the following assumptions,
\begin{itemize}
	\item (a.1) Best buyer offer $y$ in any period $\tau$ is uniformly distributed between $U[\mu-\sigma,\mu+\sigma]$.
	\item(a.2) Seller guess $P(\tilde{\mu}/\mu)$ has a discrete distribution with equal probability mass over three points $\{\mu-2\sigma,\mu,\mu+2\sigma\}$.
	\item(a.3) Seller perfectly learns the true buyer offer distribution after observing offers in the market for one period i.e., $\sigma_{signal}^2=0$. But they are not forward looking about this learning.
	\item (a.4) The seller's outside option $x$ is low enough that they enter the market and stay in the market at the end of the first period. 
\end{itemize}
These simplifying assumptions help with tractability of results and ease of interpretation. We will first solve for the optimal list price for second period onwards $l_{\tau=2,...}^*$ where the sellers guess of the offer mean is accurate ($\tilde{\mu}_{\tau}=\mu$ $\forall$ $\tau >= 2$) since she has learned from observing offers in the first period. The seller correctly expects the offers to be distributed as $y \sim U[\mu-\sigma,\mu+\sigma]$. Since this offer distribution remains stationary, the sellers' choice to stay in the market, list price and expected payoff remain same for $\tau >= 2$.
\begin{align}
    E[\pi_{\tau}|l_{\tau}] = -c + \tilde{P}(sale|l_{\tau})*l_{\tau} + (1 - \tilde{P}(sale|l_{\tau}))*E[\pi_{\tau}|l_{\tau}] \nonumber \\
    E[\pi_{\tau}|l_{\tau}] = (-c/\tilde{P}(sale|l_{\tau})) + l_{\tau}
\end{align}
The optimal list price that maximizes $E[\pi_{\tau}|l_{\tau}]$ is given by,
\begin{align}
    l_{\tau}^* = \mu + \sigma -\sqrt{2c\sigma} \quad \forall \quad \tau >= 2 \nonumber \\
    E[\pi_{\tau}|l_{\tau}] = \mu + \sigma - 2\sqrt{2c\sigma} \quad ; \quad \tilde{P}(sale|l_{\tau}) = \sqrt{c/2\sigma}
\end{align}
In the first period the seller draws a guess of offers $\tilde{\mu}$ that is equally likely to be - pessimistic $\mu - 2\sigma$, accurate $\mu$ or optimistic $\mu + 2\sigma$. The seller is aware of error in her guess in the first period. She can construct a guess of true offer mean $P(\mu|\tilde{\mu},\sigma_s^2)$ and subsequently the offers as,
\begin{align}
    P(y|\tilde{\mu},\sigma_s^2) = \Sigma_{\mu} P(y|\mu,\sigma_s^2) \times P(\mu|\tilde{\mu},\sigma_s^2) \nonumber \\
    \tilde{y}_1 \sim U[\tilde{\mu} - 3\sigma,\tilde{\mu} + 3\sigma]
\end{align}
\begin{figure}
    \centering
    \includegraphics[width=\textwidth]{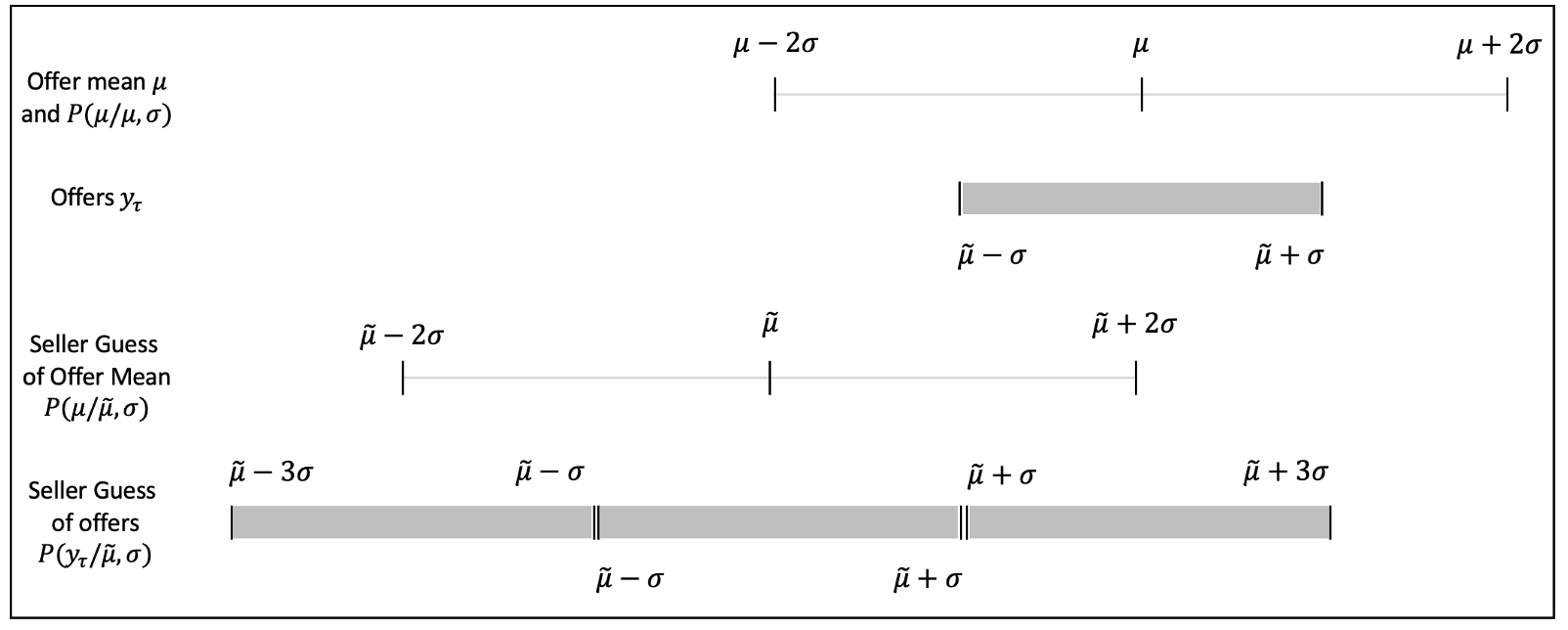}
    \caption{The seller’s guess $\tilde{\mu}$ of offer mean is one out of $\{\mu-2\sigma,\mu,\mu+2\sigma\}$. We consider here a seller who draws a low guess $\tilde{\mu} = \mu - 2\sigma$ The seller with guess $\tilde{\mu}$ does not know that they have drawn a low guess, but they do know that their guess has error $-2\sigma$ or $0$ or $2\sigma$, with probability $1/3$ each. The seller’s unconditional guess of offers is distributed as $U[(\mu - 2\sigma) \pm 3\sigma]$. The actual offer is distributed as $U[\mu \pm \sigma]$.The choice of offer and belief distributions in the simple model avoid overlapping regions and subsequently simplifies the mathematical expressions. Further setting cost as proportional to $\sigma$, for example $c = 2\sigma$, lends to optimal list prices $l_1^*(\tilde{\mu}) = \tilde{\mu} + \sigma$ and $l_2^*(\tilde{\mu}=\mu) = \tilde{\mu}$. }
    \label{fig: Lemma1_FullModel}
\end{figure}
The sellers' optimal list price in the first period is given by,
\begin{align}
    l^*_1 = \tilde{\mu} + 2\sigma - \sqrt{2c\sigma}
\end{align}
Using the optimal listing prices $l^*_1,l^*_{\tau = 2,...}$, we can calculate the expected sale price of a home and equivalently the true home value $v = E[p]$ as,
\begin{align}
E[p|\mu,\sigma] = \Sigma_{\tilde{\mu}} P(\tilde{\mu}/\mu,\sigma) \Big[ P_1(sale | \tilde{\mu},\mu,\sigma)*l_1(\tilde{\mu}) + (1 - P_1(sale | \tilde{\mu},\mu,\sigma))*l_2^*(\mu) \Big] \nonumber \\
E[p|\mu,\sigma] = v = (\mu + 2\sigma/3 - \sqrt{2c \sigma})
\end{align}
Individual sellers' valuation $\tilde{v} = \tilde{E}[p|\tilde{\mu},\sigma]$ is given by,
\begin{align}
\tilde{E}[p|\tilde{\mu},\sigma] = \Sigma_{\mu} P(\mu|\tilde{\mu},\sigma) \times E[p|\mu,\sigma] \nonumber \\
\tilde{v} = \tilde{\mu} + 2\sigma/3 - \sqrt{2c \sigma}
\end{align}
The variance of sellers' valuation is given by,
\begin{align}
Var[\tilde{v}] = E[(\tilde{v} - E[\tilde{v}])^2] = E[(\tilde{\mu} - \mu)^2] \nonumber \\
Var[\tilde{v}] = \sigma_e^2 = 8 \sigma^2/3
\end{align}

\textbf{Proof for Lemma 2}\\
The expected seller payoff is given by,
\begin{align}
\mathbb{E}[\pi] = \Sigma_{\tilde{\mu}} \Big(  P_1(sale | \tilde{\mu},\mu,\sigma)*l^*_1(\tilde{\mu}) + (1 - P_1(sale | \tilde{\mu},\mu,\sigma))*\mathbb{E}[\pi(l^*_{\tau = 2}(\mu))] \Big)  \nonumber \\
\mathbb{E}[\pi(\mu)] = \mu + (2\sigma/3) - c - 5\sqrt{2c\sigma}/3
\end{align}
Using $\sigma = \sqrt{3/8}\sigma_e$ from Lemma 1,
\begin{align}
\mathbb{E}[\pi(\mu)] = \mu - c + \gamma_1\sigma_e - \gamma_2\sqrt{c\sigma_e} \nonumber \\
\text{where } \gamma_1 = (1/3)*(3/2)^{1/2} \text{\quad and \quad}  \gamma_2 = (5/3)*(3/2)^{1/4}
\end{align}
If we use $c = \kappa \sigma_e$ the payoff $\mathbb{E}[\pi(\mu)]$ is decreasing in $\sigma_e$ if, 
\begin{align}
\frac{\partial}{\partial \sigma_e} \mathbb{E}[\pi(\mu)] = \mu - \sigma_e \times (\kappa - \gamma_1 + \gamma_2\sqrt{\kappa}) < 0 \nonumber \\
\kappa - \gamma_1 + \gamma_2\sqrt{\kappa} > 0 \nonumber \\
\kappa > 0.04
\end{align}

\textbf{Proof for Lemma 3}\\
Error in valuations can be broken into two components as,
\begin{align}
E[(\tilde{v} - v)^2] = E[((\tilde{v} - E[\tilde{v}])+ (E[\tilde{v}] - v))^2]
\end{align}
The first component simplifies as,
\begin{align}
\tilde{v} - E[\tilde{v}] = \Big( (1-\alpha)*(v - \mu + \tilde{\mu}) + \alpha * (v + e_z) \Big) - \Big( (1-\alpha)*v + \alpha * (v + e_z) \Big) \nonumber \\
\tilde{v} - E[\tilde{v}] = (1-\alpha)*(\tilde{\mu} - \mu)
\end{align}
The second component simplifies as,
\begin{align}
E[\tilde{v}] - v = \Big( (1-\alpha)*v + \alpha * (v + e_z) \Big) - v = \alpha * e_z
\end{align}
Substituting the simplified first and second component,
\begin{align}
E[(\tilde{v} - v)^2] = E[(((1-\alpha)*(\tilde{\mu} - \mu)) + \alpha * e_z)^2] \nonumber \\
E[(\tilde{v} - v)^2] = (1-\alpha)^2 \sigma_e^2 + \alpha^2 \sigma_z^2(\alpha)
\end{align}
Error in sale prices can be broken into two components and simplified similarly as,
\begin{align}
E[(p - v)^2] = E[((p - E[p])+ (E[p] - v))^2] \nonumber \\
E[(p - v)^2] = \delta \times (1-\alpha)^2 \sigma_e^2 + \alpha^2 \sigma_z^2(\alpha)
\end{align}

The variance in seller payoff is given by,
\begin{align}
Var[\pi] = Var[\pi|z] + E[(E[\pi|z] - E[\pi])^2] \nonumber \\
Var[\pi] = E[(\pi - E[\pi])^2|z] + \alpha^2\sigma^2_z(\alpha)
\end{align}
The variance in payoff for a single realization of ML price $z$ i.e., $Var[\pi|z]$ can be expressed as (we skip realized ML price $z$ from the notation below),
\begin{align}
Var[\pi|z] = E[(\pi - E[\pi])^2] \nonumber \\
= P(\tilde{\mu} = \mu - 2\sigma)\Big( \pi(\tilde{\mu} = \mu - 2\sigma) - E[\pi] \Big)^2 + P(\tilde{\mu} >= \mu) E[(\pi - E[\pi])^2 | \tilde{\mu} >= \mu] 
\end{align}
The first components simplifies as
\begin{align}
\pi(\tilde{\mu} = \mu - 2\sigma) - E[\pi] = (\mu - c - \sqrt{2c\sigma}) - (\mu - c - +\gamma_1\sigma - \gamma_2\sqrt{c\sigma}) \nonumber \\
\pi(\tilde{\mu} = \mu - 2\sigma) - E[\pi] = -\gamma_1 \sigma + (\gamma_2 - \sqrt{2})\sqrt{c\sigma} \nonumber \\
\text{where } \gamma_1 = (1/3)*(3/2)^{1/2} \text{\quad and \quad}  \gamma_2 = (5/3)*(3/2)^{1/4} \quad \text{from Lemma 2}  
\end{align}
The second components simplifies as
\begin{align}
E[(\pi - E[\pi])^2 | \tilde{\mu} >= \mu] = E[(\pi - E[\pi|\tilde{\mu} >= \mu] + E[\pi|\tilde{\mu} >= \mu] - E[\pi])^2 | \tilde{\mu} >= \mu] \nonumber \\
= E[(\pi - E[\pi|\tilde{\mu} >= \mu])^2  | \tilde{\mu} >= \mu] + E[(E[\pi|\tilde{\mu} >= \mu] - E[\pi])^2 | \tilde{\mu} >= \mu] \nonumber \\ \quad + 2*E[(\pi - E[\pi|\tilde{\mu} >= \mu])*(E[\pi|\tilde{\mu} >= \mu] - E[\pi]) | \tilde{\mu} >= \mu]  \nonumber \\
= Var[\pi|\tilde{\mu} >= \mu] + (E[\pi|\tilde{\mu} >= \mu] - E[\pi])^2 + 2*(E[\pi|\tilde{\mu} >= \mu] - E[\pi])*(E[(\pi - E[\pi|\tilde{\mu} >= \mu])]) \nonumber \\
= Var[\pi|\tilde{\mu} >= \mu] + (E[\pi|\tilde{\mu} >= \mu] - E[\pi])^2 \end{align}
Using Lemma 2 we can substitute expressions for $E[\pi|\tilde{\mu} >= \mu]$ and $E[\pi]$ as,
\begin{align}
E[\pi|\tilde{\mu} >= \mu] - E[\pi] = (\mu - c + \sigma - 2\sqrt{2c\sigma}) - (\mu - c - +\gamma_1\sigma - \gamma_2\sqrt{c\sigma}) \nonumber \\
= (1-\gamma_2) - (2\sqrt{2} - \gamma_2)\sqrt{c\sigma}
\end{align}
The distribution of $\pi$ (when $\tilde{\mu} >= \mu$) follows a geometric distribution i.e., probability of sale at $\tau = T$ is given by $(1-P(sale|l_2^*))^{T-2}P(sale|l_2^*)$. The sale price at any $\tau$ remains same while the cost on market adds up to $-c\tau$. Consequently, we can use variance of geometric distribution as,
\begin{align}
Var[\pi|\tilde{\mu} >= \mu] = c^2 \times \frac{1 - P(sale|l_2^*)}{(P(sale|l_2^*))^2} \nonumber \\
= 2c\sigma - c\sqrt{2c\sigma}
\end{align}
We can substitute expressions for $E[\pi|\tilde{\mu} >= \mu] - E[\pi]$ and $Var[\pi|\tilde{\mu} >= \mu]$ as,
\begin{align}
E[(\pi - E[\pi])^2 | \tilde{\mu} >= \mu] = \Big( (1-\gamma_2) - (2\sqrt{2} - \gamma_2)\sqrt{c\sigma} \Big)^2 + 2c\sigma - c\sqrt{2c\sigma} \nonumber \\
(1-\gamma_1)^2 \sigma^2 + 8\gamma_2^2c\sigma - 2\sqrt{2}\gamma_2(1-\gamma1)\sigma\sqrt{c\sigma}-c\sqrt{2c\sigma} + 2c\sigma
\end{align}
Substituting the first and second components we have,
\begin{align}
Var[\pi|z] = (1/3) \times \Big( -\gamma_1 \sigma + (\gamma_2 - \sqrt{2})\sqrt{c\sigma} \Big)^2 \nonumber \\ \quad + (2/3) \times \Big( (1-\gamma_1)^2 \sigma^2 + 8\gamma_2^2c\sigma - 2\sqrt{2}\gamma_2(1-\gamma1)\sigma\sqrt{c\sigma}-c\sqrt{2c\sigma} + 2c\sigma \Big)
\end{align}
While the expression $Var[\pi|z]$ is unwieldy, we numerically verify the conditions where $Var[\pi|z]$ is increasing in $c$. This is trivially satisfied for a wide range $0.01\sigma < c < 100\sigma$. Consider a home values at $\$500,000$. A plausible range of seller uncertainty about offers is $\sigma \in [\$1000,\$50000]$. The per period (say monthly) cost of keeping the home on the market can be in range $c \in [\$500,\$10000]$. $Var[\pi|z]$ is increasing in $c$ for any $(\sigma,c)$ in the plausible range above. This analytical result is consistent with the intuition that as market participation cost $c$ increases, seller's (and equivalently buyers) are less willing to spend time in the market to resolve uncertainty and price accurately. Thus, the realized sale prices and therefore payoffs are more likely to deviate from true value (expected sale price over thousands of hypothetical repetitions).

We can substitute $c=\kappa \sigma_e$ and $Var[\pi|z]$ in expression for $Var[\pi]$ as,
\begin{align}
Var[\pi] = \Omega(\kappa;\beta) (1-\alpha)^2 \sigma_e^2 + \alpha^2 \sigma_z^2(\alpha) \nonumber \\
\small{\text{ where \quad } \Omega(\kappa;\beta) = \beta_0 + \kappa^{1/2} \beta_1 + \kappa \beta_2 + \kappa^{3/2} \beta_3}
\end{align}

\subsection{Proof for Propositions}
\textbf{Proposition 2}\\
We can set estimated ML price error $\hat{\sigma}_z^2 (\alpha)$ expressed in equation \ref{eq:sigma_hat} as less than the true ML price error $\sigma_z^2 (\alpha)$ as 
\begin{align}
    \hat{\sigma}_z^2 (\alpha) = (1 - \alpha)^2 \times \sigma_z^2 (\alpha) + (1 - \alpha)^2 \times \delta \sigma_e^2 < \sigma_z^2 (\alpha) \nonumber \\
    (1 - \alpha)^2 \times \delta \sigma_e^2 < \alpha (2 - \alpha) \times \sigma_z^2 \nonumber \\
    \frac{\sigma_e^2}{\sigma_z^2 (\alpha)} < \frac{1}{\delta} \times \Big(\frac{1}{(1-\alpha)^2} - 1 \Big)    
\end{align}
The LHS is decreasing in $\alpha$ while the RHS takes increasing values $[0,\infty)$ as $\alpha$ increases from $[0,1]$. Therefore a unique $\alpha \in [0,1]$ satisfies this condition.

\textbf{Proposition 3}\\
We can set derivative of estimated ML price error $\hat{\sigma}_z^2 (\alpha)$ expressed in equation \ref{eq:sigma_hat} with respect to $\alpha$ as less than 0 as, 
\begin{align}
    \frac{\partial \hat{\sigma}_z^2 (\alpha)}{\partial \alpha} = \frac{\partial }{\partial \alpha} \Big[ (1 - \alpha)^2 \times \sigma_z^2 (\alpha) + (1 - \alpha)^2 \times \delta \sigma_e^2 \Big] < 0 \nonumber \\
    \frac{\partial }{\partial \alpha} \Big[ (1-\alpha)^2 \times (\sigma_z^2 + \delta \sigma_e^2) + (1-\alpha)^2 (1 - (1 - \alpha)^2) \times \delta Q \sigma^2_e / N \Big] < 0 \nonumber \\
    -2(1-\alpha) \times (\sigma_z^2 + \delta \sigma_e^2) + \Big[ -2(1 - \alpha) + 4(1 - \alpha)^3 \Big] \times \delta Q \sigma_e / N < 0 \nonumber \\
    2(1 - \alpha)^2 \times \delta Q / N < \frac{\sigma_z^2}{\sigma_e^2} + \delta + \delta Q / N \nonumber \\
    (1-\alpha)^2 < \frac{1}{2} + \frac{N}{2Q} + \frac{N}{2\delta Q} \frac{\sigma_z^2}{\sigma_e^2} \nonumber \\
    \alpha > 1 - \sqrt{\frac{1}{2} + \frac{N}{2Q} + \frac{N}{2 Q \delta}\frac{\sigma_z^2}{\sigma_e^2}}
\end{align}
This condition is satisfied for any $\alpha > 0$ if number of traning samples $N$ is greater than two times the number of features $Q$.

\textbf{Proposition 4}\\
We can express equation \ref{eq:sigma_hat} as,
\begin{align}
\hat{\sigma}_z^2 (\alpha) = (1 - \alpha)^2 \times \Big( \sigma_z^2 (\alpha) + \delta \sigma_e^2 \Big) \nonumber \\
\frac{\hat{\sigma}_z^2 (\alpha)}{(1 - \alpha)^2} - \delta \sigma_e^2 = \sigma_z^2 (\alpha)
\end{align}
Using this we can re-express Proposition 2 as,
\begin{align}
\frac{\hat{\sigma}_z^2 (\alpha)}{(1 - \alpha)^2} - \delta \sigma_e^2 = \sigma_z^2 (\alpha) > \hat{\sigma}_z^2 (\alpha)  \nonumber \\
\hat{\sigma}_z^2 (\alpha) \Big( \frac{1}{(1 - \alpha)^2} - 1 \Big) > \delta \sigma_e^2   \nonumber \\
\frac{1 - (1 - \alpha)^2}{(1 - \alpha)^2} > \delta \frac{\sigma_e^2}{\hat{\sigma}_z^2 (\alpha)}
\end{align}
We can express $\sigma_e^2/\hat{\sigma}_z^2 (\alpha)$ in terms of $\alpha$ using equation \ref{eq:alpha} as,
\begin{align}
\alpha = \frac{\sigma_e^2}{\sigma_e^2 + \hat{\sigma}_z^2} \nonumber \\
\frac{\sigma_e^2}{\hat{\sigma}_z^2} = \frac{\alpha}{1-\alpha}
\end{align}
We can now substitute this $\sigma_e^2/\hat{\sigma}_z^2 (\alpha)$ to check if the condition is satisfied at equilibrium,
\begin{align}
\frac{1 - (1 - \alpha)^2}{(1 - \alpha)^2} > \delta \frac{\sigma_e^2}{\hat{\sigma}_z^2 (\alpha)} \nonumber \\
\frac{1 - (1 - \alpha)^2}{(1 - \alpha)^2} > \delta  \frac{\alpha}{1-\alpha}  \nonumber \\
1 - (1 - \alpha)^2 > \delta \alpha (1 - \alpha) \nonumber \\
\alpha (2 - \alpha) > \delta \alpha (1 - \alpha)  \nonumber \\
\alpha < \frac{2}{1+\delta}
\end{align}
Since $\delta < 1$, the RHS is greater than 1. Therefore this condition is always satisfied.

Now consider if estimated ML error is increasing with $\alpha$ at ml feedback loop equilibrium $\alpha^*$. From above we have,
\begin{align}
\frac{1 - (1 - \alpha)^2}{(1 - \alpha)^2} > \delta \frac{\sigma_e^2}{\hat{\sigma}_z^2 (\alpha)} \nonumber \\
\sigma_z^2 (\alpha) > \sigma_e^2 \delta \frac{(1-\alpha)^2}{1 - (1-\alpha)^2} \nonumber \\
\sigma_z^2 + \alpha(2-\alpha) Q \delta \sigma_e^2/N > \sigma_e^2 \delta \frac{(1-\alpha)^2}{1 - (1-\alpha)^2} \nonumber \\
\sigma_z^2 > \Big[ \frac{(1-\alpha)^2}{1 - (1-\alpha)^2} - \frac{1 - (1-\alpha)^2}{n} \Big] \sigma_e^2 Q \delta / N \nonumber \\
\frac{N \sigma_z^2}{Q \delta \sigma_e^2} > \Big[ \frac{(1-\alpha)^2}{1 - (1-\alpha)^2} - \frac{1 - (1-\alpha)^2}{N/Q} \Big]
\end{align}
We can use this lower bound on $N \sigma_z^2/(Q \delta \sigma_e^2)$ into the condition for increasing $\hat{\sigma}_z^2 (\alpha)$ in $\alpha$ from Proposition 3 as,
\begin{align}
\alpha > 1 - \sqrt{\frac{1}{2} + \frac{N}{2Q} + \frac{N}{2 Q \delta}\frac{\sigma_z^2}{\sigma_e^2}} \nonumber \\
(1 - \alpha)^2 < \frac{1}{2} + \frac{N}{2Q} + \frac{N \sigma_z^2}{2Q \delta \sigma_e^2} \nonumber \\
\frac{N \sigma_z^2}{\delta \sigma_e^2} > \Big[ (1-\alpha)^2 - 1/2 - (N/2Q) \Big] \nonumber \\
 \Big[ \frac{(1-\alpha)^2}{1 - (1-\alpha)^2} - \frac{1 - (1-\alpha)^2}{N/Q} \Big] > \Big[ (1-\alpha)^2 - 1/2 - (N/2Q) \Big]
\end{align}
The above condition is satisfied for all $\alpha \in [0,1]$ because,
\begin{align}
\frac{(1-\alpha)^2}{1 - (1-\alpha)^2} > (1 - \alpha)^2 \text{  and  }  
\frac{1 - (1-\alpha)^2}{N/Q} < 1/2
\end{align}
Thus estimated ML error is increasing with $\alpha$ at ml feedback loop equilibrium $\alpha^*$.

\textbf{Proposition 5}

We can reformulate $\alpha$ in equation \ref{eq:alpha} as,
\begin{align}
\alpha = \frac{\sigma_e^2}{\sigma_e^2 + \hat{\sigma}_z^2} \nonumber \\
\frac{\hat{\sigma}_z^2}{\sigma_e^2} = (1/\alpha) - 1
\end{align}
In the expression of $\hat{\sigma}^2_z(\alpha)$ we can reuse above as,
\begin{align}
\hat{\sigma}^2_z(\alpha) = (1-\alpha)^2 \times (\sigma_z^2(\alpha) + \delta\sigma_e^2) \nonumber \\
\hat{\sigma}^2_z(\alpha) = (1-\alpha)^2 \times \Big( \sigma_z^2 + \delta\sigma_e^2(1 + \alpha(2-\alpha)/n) \Big) \nonumber \\
(1/\alpha) - 1 = (1-\alpha)^2 \times \Big( \frac{\sigma_z^2}{\sigma_e^2} + \delta +  \delta \alpha(2 -\alpha)/n \Big) \nonumber \\
(1 - \alpha) \times \Big[ (1 - \alpha) \Big( \frac{\sigma_z^2}{\sigma_e^2} + \delta +  \frac{\delta \alpha(2 -\alpha)}{n} \Big) - \frac{1}{\alpha} \Big] = 0
\end{align}
$\alpha^* = 1$ is one solution for this equation. The remaining solutions must satisfy,
\begin{align}
\Big[ (1 - \alpha) \Big( \frac{\sigma_z^2}{\sigma_e^2} + \delta +  \frac{\delta \alpha(2 -\alpha)}{n} \Big) - \frac{1}{\alpha} \Big] = 0  \nonumber \\
\alpha (1 - \alpha) \Big( \frac{\sigma_z^2}{\sigma_e^2} + \delta +  \frac{\delta \alpha(2 -\alpha)}{n} \Big) = 1  \nonumber \\
\frac{\sigma_z^2}{\sigma_e^2} + \delta +  \frac{\delta \alpha(2 -\alpha)}{n} = \frac{1}{\alpha (1 - \alpha)}
\end{align}
The minimum values of RHS is $4$ and the maximum values of LHS is $(\sigma_z^2/\sigma_e^2) + \delta +  (\delta/n)$. There are no other solutions $\alpha^*$ (except $\alpha^* = 1$) if,
\begin{align}
\frac{\sigma_z^2}{\sigma_e^2} + \delta +  \frac{\delta}{n} < 4 \nonumber \\
\frac{\sigma_z^2}{\sigma_e^2} < 4 - \delta - \frac{1}{n}
\end{align}

\textbf{Proposition 6}\\
We can evaluate derivative of $E[(p - v)^2]$ (from Lemma 3) with $\alpha$ and set greater than zero as,
\begin{align}
\frac{\partial E[(p - v)^2]}{\partial \alpha}  = \frac{\partial }{\partial \alpha} \Big( \delta \times (1-\alpha)^2 \sigma_e^2 + \alpha^2 \sigma_z^2(\alpha) \Big) > 0 \nonumber \\
\frac{\partial }{\partial \alpha} \Big( \delta \times (1-\alpha)^2 \sigma_e^2 + \alpha^2 [\sigma_z^2 + \alpha(2 - \alpha)\delta Q \sigma_e^2/N] \Big) > 0 \nonumber \\
\Big( -2\delta(1-\alpha)\sigma_e^2 \Big) + \Big( 2\alpha\sigma_z^2 \Big) + \Big( (6\alpha^2 - 4\alpha^3) \delta Q \sigma_e^2/N \Big) > 0
\end{align}
In the expression above, the first component is less than 0 while the second and third components are greater than 0 for $\alpha \in [0,1]$. We can ignore the third components ($N/Q >> 1$) to find a conservative upper bound on $\alpha$ that satisfies $\frac{\partial}{\partial \alpha} E[(p - v)^2] > 0$.
\begin{align}
\Big( -2\delta(1-\alpha)\sigma_e^2 \Big) + \Big( 2\alpha\sigma_z^2 \Big) > 0 \implies \alpha > \frac{\sigma_e^2}{\sigma_e^2 + \sigma_z^2/\delta}
\end{align}
If the reliance on ML price $\alpha$ is large enough we can guarantee $\frac{\partial}{\partial \alpha} E[(p - v)^2] > 0$. 
We can substitute for $\alpha$ as,
\begin{align}
\alpha = \frac{\sigma_e^2}{\sigma_e^2 + \hat{\sigma}_z^2} > \frac{\sigma_e^2}{\sigma_e^2 + \sigma_z^2/\delta} \implies \sigma_z^2 > \delta \hat{\sigma}_z^2(\alpha)
\end{align}
Following a similar procedure, $E[(\tilde{v}_i - v)^2]$ is increasing in $\alpha$ if,
\begin{align}
    \alpha > \frac{\sigma_e^2}{\sigma_e^2 + \sigma_z^2} \text{\quad or \quad} \sigma_z^2 > \hat{\sigma}_z^2(\alpha)
\end{align}
$Var[\pi]$ is increasing in $\alpha$ if,
\begin{align}
    \alpha > \frac{\sigma_e^2}{\sigma_e^2 + \sigma_z^2/\Omega(\kappa)} \text{\quad or \quad} \sigma_z^2 > \Omega(\kappa) \hat{\sigma}_z^2(\alpha)
\end{align}

\textbf{Proposition 7}\\
The payoff for risk averse participant is given by,
\begin{align}
\Pi(a) = \mathbb{E}[\pi] - 0.5 a \times \sqrt{Var[\pi]}
\end{align}
The risk neutral payoff $\Pi(a=0)$ is always increasing in reliance $\alpha$ because $\mathbb{E}[\pi]$ is decreasing in $\sigma$ (from Lemma 2), which in turn is decreasing in $\alpha$ (from Lemma 1).

The risk averse payoff $\Pi(a > 0)$ is decreasing in reliance $\alpha$ if,
\begin{align}
    \frac{\partial \Pi}{\partial \alpha} = \frac{\partial}{\partial \alpha} \Big( \mathbb{E}[\pi] - 0.5 a \times \sqrt{Var[\pi]} \Big) < 0 \nonumber \\
    \frac{\partial \mathbb{E}[\pi] }{\partial \alpha} < 0.5 a \frac{\partial \sqrt{Var[\pi]}}{\partial \alpha} \nonumber \\
    \frac{\partial \mathbb{E}[\pi]^2 }{\partial \alpha} < \Big( 0.5a\frac{\mathbb{E}[\pi]}{\sqrt{Var[\pi]}} \Big) \frac{\partial Var[\pi]}{\partial \alpha}
\end{align}
We can formulate the components as,
\begin{align}
    \frac{\partial \mathbb{E}[\pi]^2 }{\partial \alpha} = 2\kappa^2(1-\alpha)\sigma_e^2(\alpha = 0)\text{\quad and \quad} \frac{\partial Var[\pi] }{\partial \alpha} > -2\Omega(1-\alpha)\sigma_e^2(\alpha = 0) + 2\alpha\sigma_z^2
\end{align}
Here $\kappa$ is a constant when cost $c$ is linearly growing with $\sigma$. For example, $\kappa = (1/3)$ when $c = 24c/17$. We can substitute as,
\begin{align}
    2\kappa^2(1-\alpha)\sigma_e^2 < \Big( 0.5a\frac{\mathbb{E}[\pi]}{\sqrt{Var[\pi]}} \Big) \times \Big( -2\Omega(1-\alpha)\sigma_e^2 + 2\alpha\sigma_z^2 \Big) \nonumber \\
    \frac{2\kappa^2 \times \sqrt{Var[\pi]}}{a \times \mathbb{E}[\pi]} +  \Omega < \frac{\alpha}{1-\alpha} \times  \frac{\sigma_z^2}{\sigma_e^2}
\end{align}
This is satisfied if ML error underestimation over the feedback loop is severe enough as,
\begin{align}
    \frac{\sigma_z^2}{\hat{\sigma}_z^2} > \frac{1}{a} \Big(\frac{2\kappa^2 \times \sqrt{Var[\pi]}}{\mathbb{E}[\pi]} \Big) +  \Omega
\end{align}

Finally, consider risk averse payoff at equilibrium $\alpha^* = 1$ compared with no machine learning i.e., $\alpha = 0$.
\begin{align}
    \Pi(a,\alpha^* = 1) < \Pi(a,\alpha = 0) \nonumber \\
    \mu - \frac{a}{2}\sqrt{\sigma_z^2(\alpha = 1)} < \Big( \mu - \Gamma(\kappa) \sigma_e \Big) - \frac{a}{2}\sqrt{\Omega \sigma_e^2} \nonumber \\
    \sigma_z(\alpha = 1) > \frac{2\Gamma(\kappa)}{a} \sigma_e + \sqrt{\Omega} \sigma_e  \nonumber \\
    \frac{\sigma_z(\alpha)}{\sigma_e} > \sqrt{\Omega} + 2\Gamma(\kappa)/a  \nonumber \\
    \frac{\sigma^2_z}{\sigma^2_e} > \Big( \sqrt{\Omega} + \frac{2\Gamma(\kappa)}{a} \Big)^2 - \frac{\delta Q}{N}
\end{align}
From Proposition 5, $\alpha^* = 1$ is the only equilibrium if 
\begin{align}
    \frac{\sigma^2_z}{\sigma^2_e} < 4 - \delta - \frac{\delta Q}{N}
\end{align}
Thus risk averse payoff at the only equilibrium is worse than no machine learning if,
\begin{align}
   \Big( \sqrt{\Omega} + \frac{2\Gamma(\kappa)}{a} \Big)^2 - \frac{\delta Q}{N} < \frac{\sigma^2_z}{\sigma^2_e} < 4 - \delta - \frac{\delta Q}{N}
\end{align}
This is viable if,
\begin{align}
   \Big( \sqrt{\Omega} + \frac{2\Gamma(\kappa)}{a} \Big)^2 - \frac{\delta Q}{N} < 4 - \delta - \frac{\delta Q}{N} \nonumber \\
   \Big( \sqrt{\Omega} + \frac{2\Gamma(\kappa)}{a} \Big)^2 + \delta < 4
\end{align}

\end{appendices}
	
	\newpage

\end{document}